\pdfoutput=1

\def\editmode{0}
\def\reportmode{0}
\def\bibfilenames{WISENET}

\if\reportmode1
\documentclass[10pt,final,onecolumn]{IEEEtran}
\else
\documentclass[journal]{IEEEtran}
\fi

\usepackage{dsfont,enumerate}
\if\editmode1
\usepackage{cancel}
\usepackage{mathrsfs}
\usepackage{graphicx}
\usepackage{subcaption,caption}
\usepackage{amssymb, amsthm}
\usepackage[utf8]{inputenc}
\usepackage{algorithm}
\usepackage[noend]{algpseudocode}
\floatname{algorithm}{Procedure}
\usepackage[backend=bibtex,style=alphabetic,sorting=debug]{biblatex}
\DeclareFieldFormat{labelalpha}{\thefield{entrykey}}
\DeclareFieldFormat{extraalpha}{}
\bibliography{\bibfilenames}
\newcommand{\cmt}[1]{\noindent\textcolor{lightgreen}{\underline{[#1]}}} 
\newcommand{\hc}[1]{\textcolor{magenta}{#1}} 
\newenvironment{myitemize}{\begin{itemize}}{\end{itemize}}
\newcommand{\myitem}{\item}

\else
\usepackage{graphicx}
\usepackage{subcaption}
\usepackage{mathrsfs}
\usepackage{float}
\usepackage{amsthm, amssymb}
\usepackage[utf8]{inputenc}
\usepackage{algorithm}
\usepackage[noend]{algpseudocode}
\floatname{algorithm}{Procedure}
\usepackage{cite}
\newcommand{\cmt}[1]{} 
\newcommand{\hc}[1]{\textcolor{black}{#1}} 
\newenvironment{myitemize}{}{}
\newcommand{\myitem}{}

\fi 

\usepackage{fixltx2e}

\usepackage{graphicx}

\usepackage{subcaption}              

\usepackage[utf8]{inputenc}

\usepackage{amsfonts}
\usepackage{amsmath}
\usepackage{mathtools}

\usepackage{amssymb}

\usepackage{bm}

\usepackage{color,verbatim}
\usepackage{multirow}
\usepackage{accents}

\usepackage{theoremref}

\usepackage{url}

\newtheorem{myauxproblem}{Problem}

\newcounter{rulecounter}
\newcommand{\resetrule}{ \setcounter{rulecounter}{0}}
\resetrule

\newsavebox{\selvestebox}
\newenvironment{colbox}[1]
  {\newcommand\colboxcolor{#1}%
   \begin{lrbox}{\selvestebox}%
   \begin{minipage}{\dimexpr\columnwidth-2\fboxsep\relax}}
  {\end{minipage}\end{lrbox}%
   \begin{center}
   \colorbox{\colboxcolor}{\usebox{\selvestebox}}
   \end{center}}

\definecolor{orange}{rgb}{1,0.8,0}
\definecolor{gray}{rgb}{.9,0.9,0.9}
\definecolor{darkgray}{rgb}{.3,0.3,0.3}
\definecolor{darkblue}{rgb}{.1,0.0,0.3}
\definecolor{lightblue}{rgb}{0.7,0.7,1}
\definecolor{lightred}{rgb}{1,0.7,.7}
\definecolor{purple}{RGB}{204,153,255}
\definecolor{lightgray}{rgb}{.95,0.95,0.95}
\definecolor{lightgreen}{rgb}{0.6,0.8,0.6}
\definecolor{darkgreen}{rgb}{0.05,0.3,0.05}
\definecolor{pistachio}{RGB}{204,255,153}
\definecolor{paleturquoise}{RGB}{175,238,238}
\definecolor{yellow}{RGB}{255,255,153}

\newcommand{\tbm}[1]{{\tilde{\bm #1}}}

\newcommand{\hbm}[1]{{\hat{\bm #1}}}

\newcommand{\rfield}{\mathbb{R}}

\newcommand{\transpose}{^T}
 \newcommand{\define}{\triangleq}

\newcommand{\minimize}{\mathop{\text{minimize}}}

\DeclareMathOperator*{\argmin}{arg\,min}

\newtheorem{myproposition}{Proposition}
\newtheorem{myquestion}{Question}
\newtheorem{myquiz}{Quiz}
\newtheorem{myremark}{Remark}
\newtheorem{myproblemstatement}{Problem Statement}
\newtheorem{mylemma}{Lemma}
\newtheorem{mytheorem}{Theorem}
\newtheorem{mydefinition}{Definition}
\newtheorem{mycorollary}{Corollary}
\newtheorem{myexample}{Example}

\newcommand{\acom}[1]{\noindent\textcolor{red}{{[#1]}}} 
\renewcommand{\acom}[1]{}
\newcommand{\rev}{} 
\newcommand{\revv}{}

\newcommand{\nextv}{}
\newcommand{\nextvv}[1]{}

\renewcommand{\transpose}{^\top }
\newcommand \auxSgamma {\theta }

\newcommand{\tisohindsightobj}{\hc{C}}
\newcommand{\tirsohindsightobj}{\hc{\tilde C}}
\newcommand{\energybound}{\hc{B}_y}

\newcommand{\trho}{\tilde{\rho}}
\newcommand{\ftn}{f_t^{(n)}}
\newcommand{\nReg}{\rev {\Omega}^{(n)}}
\newcommand{\strongcvxparf}{\beta_{\tilde{\ell}}}
\newcommand{\gradboundf}{G_{\tilde{\ell}}}
\newcommand{\gradf}{\bm g_t^{\tilde {\ell}}}
\newcommand{\gradreg}{\bm g_{t+1}^\Omega }

\newcommand{\ltn}{\tilde \ell_t^{(n)}}
\newcommand{\timevarhindsight}{\tbm a_n^\circ}

\newtheorem{theorem}{Theorem}

\newcommand{\winnot}[1]{_{[#1]}}

\usepackage{theoremref}
\usepackage{xr,xcite}
\externalcitedocument{SupplementaryMaterial}

\begin{document}
	\title{Online Topology Identification  from Vector Autoregressive Time Series }

	\if\reportmode1
 \author{Bakht Zaman,~\IEEEmembership{Student Member,~IEEE,} Luis Miguel Lopez Ramos,~\IEEEmembership{Member,~IEEE,}  \\ Daniel Romero,~\IEEEmembership{Member,~IEEE,}
  and Baltasar Beferull-Lozano,~\IEEEmembership{Senior Member,~IEEE}
\thanks{The work in this paper was supported by the SFI Offshore Mechatronics grant 237896/E30, the PETROMAKS Smart-Rig grant 244205 and the IKTPLUSS Indurb grant 270730/O70 from the Research Council of Norway.
 }
\thanks{The authors  are with the WISENET Lab, Dept. of ICT,
   University of Agder, Jon Lilletunsvei 3, Grimstad, 4879 Norway. E-mails:\{bakht.zaman, luismiguel.lopez, daniel.romero, baltasar.beferull\}@uia.no.}
\thanks{The material in this work was presented, in part, at CAMSAP 2017 \cite{zaman2017onlinetopology}.}
}%
\else
\author{Bakht Zaman,~\IEEEmembership{Student Member,~IEEE,} Luis Miguel Lopez Ramos,~\IEEEmembership{Member,~IEEE,}\\ Daniel Romero,~\IEEEmembership{Member,~IEEE,}
  and Baltasar Beferull-Lozano,~\IEEEmembership{Senior Member,~IEEE}
\thanks{The work in this paper was supported by the SFI Offshore Mechatronics grant 237896/E30, the PETROMAKS Smart-Rig grant 244205 and the IKTPLUSS Indurb grant 270730/O70 from the Research Council of Norway. }
\thanks{The authors  are with the WISENET Lab, Dept. of ICT,
   University of Agder, Jon Lilletunsvei 3, Grimstad, 4879 Norway. E-mails:\{bakht.zaman, luismiguel.lopez, daniel.romero, baltasar.beferull\}@uia.no.}
\thanks{The material in this work was presented, in part, at CAMSAP 2017 \cite{zaman2017onlinetopology}.}
}%
\fi
\maketitle \begin{abstract} Causality graphs are
	routinely estimated in social sciences, natural sciences, and
	engineering due to their capacity to efficiently represent the
	spatiotemporal structure of multi-variate data sets in a
	format amenable for human interpretation, forecasting, and
	anomaly detection. A popular approach to mathematically
	formalize causality is based on vector autoregressive (VAR)
	models and constitutes an alternative to the well-known, yet
	usually intractable, Granger causality. Relying on such a VAR
	causality notion, this paper develops two algorithms with
	complementary benefits to track time-varying causality graphs
	in an online fashion.  Their constant complexity per update
	also renders these algorithms appealing for big-data
	scenarios. Despite using data sequentially, both algorithms
	are shown to asymptotically attain the same average
	performance as a batch estimator which uses the entire data
	set at once. \rev{To this end, sublinear (static) regret
	bounds are established. Performance is also characterized in
	time-varying setups by means of dynamic regret
	analysis}. \rev{Numerical results with real and synthetic data
	further support the merits of the proposed algorithms in static
	and dynamic scenarios.}  \end{abstract}
\section{Introduction}	 	\label{s:intro}
		

 \cmt{Motivation Online Topology Id}%
\begin{myitemize}%
\myitem\cmt{causality relations}Inferring causal relations among time
series finds countless applications in social sciences, natural
sciences, and engineering.
\myitem\cmt{causality graphs}\begin{myitemize}%
\myitem\cmt{def}These relations are typically encoded as
the edges of a causality graph, where each node corresponds to a time
series, and
\myitem\cmt{motivation}oftentimes reveal  the
topology of e.g. an underlying social, biological, or brain network~\cite{kolaczyck2009}. Causality graphs may also offer
valuable insights into the spatio-temporal structure of time series and
 assist data processing tasks such as
forecasting~\cite{isufi2018forecasting}, signal
reconstruction~\cite{lorenzo2016lms}, anomaly
detection~\cite{liu2016unsupervised}, and
dimensionality
reduction~\cite{shen2017dimensionalityreduction}. \rev{In some applications,  graphs capturing different forms of causality can be constructed based on domain knowledge; see e.g.~\cite[Ch. 8]{bishop2006}. However, this approach is often impractical in the aforementioned applications due to the  large dimension of the data or because such  prior knowledge is unavailable. Instead, causality graphs need to be inferred from data in these situations.
\myitem\cmt{online estimation}This paper accomplishes this task  in an online fashion.}
\end{myitemize}%
\end{myitemize}%

\cmt{Literature on batch topology estimation}Identifying
graphs capturing the spatiotemporal
 ``interactions'' among time series has attracted great
 attention \cite{kolaczyck2009,mateos2018connecting}.
\begin{myitemize}%
  \myitem\cmt{Memoryless interactions}%
  Some approaches focus on instantaneous interactions, i.e.,
  they disregard the temporal structure.
\begin{myitemize}%
\myitem\cmt{undirected graphs}%
\begin{myitemize}%
\myitem\cmt{refs}%
\begin{myitemize}%
\myitem\cmt{correlation}The simplest one is to connect two nodes if the sample
correlation between the associated time series exceeds a certain
threshold~\cite{kolaczyck2009}.
\myitem\cmt{partial correlations}To distinguish mediated from
unmediated  interactions~\cite[Sec. 7.3.2]{kolaczyck2009}, one may
resort to conditional independence, partial correlations,  Markov random
fields, or other approaches in graph signal processing; see
e.g.~\cite{angelosante2011graphical,friedman2008sparse,lauritzen1996graphical,bishop2006,dong2016learninglaplacian,segarra2017templates}.
\end{myitemize}%
\myitem\cmt{limitations}For directed interactions,
\end{myitemize}%
\myitem\cmt{directed graphs}%
\begin{myitemize}%
\myitem\cmt{structural equation models (SEMs)}one may employ
structural equation models (SEM)~\cite{kline2015} (see also \cite{shen2017tensor} and references
	therein)
\myitem\cmt{Bayesian networks}or Bayesian networks~\cite[Sec.~8.1]{bishop2006}.
\end{myitemize}%
\myitem\cmt{limitations}However, these methods account
only for \emph{memoryless} interactions, i.e., they cannot
accommodate delayed 
interactions where the value of a time series at
a given time instant is related to the past values of other time
series.
\end{myitemize}%
\par 
\myitem\cmt{Memory based approaches}%
\begin{myitemize}%
\myitem\cmt{Granger causality notion}%
\begin{myitemize}%
\myitem\cmt{def}The earliest effort to formalize
the notion of causality among time series is due to
Granger~\cite{granger1988causality} and relies on the
 rationale that \emph{the cause precedes
the effect}. A time series is said to be
Granger-caused by another if the optimal prediction error of the former is
decreased when the past of the latter is taken into
account.
\myitem\cmt{limitations:Elegant but
practically limited}Albeit elegant, this  definition  is
generally impractical since the \emph{optimal} prediction error is difficult
to determine \cite[p. 33]{zellner1979causality}, \cite{kay1}.
\end{myitemize}%
\myitem\cmt{VAR}%
\begin{myitemize}%
\myitem\cmt{VAR causality}Thus, alternative causality definitions
 based on vector autoregressive (VAR) models  are typically
preferred~\cite{goebel2003varcausality,basu2015granger,tank2018neural}.
\begin{myitemize}%
  \myitem\cmt{Granger}VAR causality is determined from the support of
  VAR matrix parameters and is equivalent to Granger
  causality~\cite[Chap. 2]{lutkepohl2005} in certain cases (yet
  sometimes treated as
  equivalent~\cite{basu2015granger,tank2018neural}).
  \myitem\cmt{generality}VAR causality is further motivated by the
  widespread usage of VAR models to
  approximate the response of systems of linear partial differential
  equations~\cite{lions1971} and, more generally, in disciplines such as  \myitem\cmt{popularity}econometrics, bio-informatics,  neuroscience, and engineering   \cite{lutkepohl2004applied,fujita2007modeling,valdes2005estimating}.
\end{myitemize}%
\myitem\cmt{Gaussian and stationary}VAR topologies  are estimated
 assuming Gaussianity
 and stationarity in~\cite{bach2004learning,songsiri2010selection} and
\myitem\cmt{sparsity}%
\begin{myitemize}%
	\myitem \cmt{Group-Lasso Estimator}assuming sparsity in~\rev{\cite{bolstad2011groupsparse,songsiri2013vargranger,mei2017causal,ayazoglu2011blind}}.
\end{myitemize}%
\end{myitemize}%
\end{myitemize}%
\end{myitemize}%
\cmt{Literature on time-varying topology ID}%
\begin{myitemize}%
\myitem\cmt{motivation}All these approaches assume that the graph does not
 change over time. Since this is not the case in many applications,
\myitem\cmt{references}approaches have been devised to identify
 undirected time-varying
 topologies~\cite{kolar2010estimating,lee2017dynamicrandomgraph} \rev{and directed piecewise-constant time-varying topologies \cite{lopez2018dynamic}}.
\end{myitemize}%
\cmt{Literature on Online topology est}%
\begin{myitemize}%
\par 
        \myitem\cmt{motivation}The complexity of all
previously discussed approaches becomes
prohibitive for long observation windows since they process
the entire data set at once and cannot accommodate data arriving sequentially.  The modern approach to tackle these
issues is
 \emph{online} optimization, where an estimate is refined with
every new data instance.
	\myitem	\cmt{Refs }%
        \begin{myitemize}%
        \myitem\cmt{no memory}%
        \begin{myitemize}%
	\myitem \cmt{graphical + SEM}Existing online topology
        identification algorithms include \cite {hallac2017network,shen2017tensor},\rev{\cite{baingana2014trackingcascades,rodriguez2013structure,zaman2020dynamic}},
        \rev {and
        \myitem \cmt{Undirected Online TopID}\cite{shafipour2019onlinetopology}, but they only account for memoryless interactions.}
\end{myitemize}%
\end{myitemize}%
\end{myitemize}%
\par 
\cmt{contribution}\rev{ The present work is the first to propose  online algorithms to estimate  the \emph{memory-aware} causality graphs associated with   a collection of time series}\revv{\footnote{\revv{The related work in \cite{shen2018online} was run in parallel and published after the conference version \cite{zaman2017onlinetopology} of this work.}}.}
\revv{We take as a starting point an online algorithm for estimating directed VAR causality graphs which basically minimizes a sequential, sparse topology identification criterion 
by means of a composite-objective
iteration~\cite{duchi2010comid}. 
This procedure, which we termed TISO (Topology
Identification via Sparse Online learning) throughout the paper, promotes sparse updates and }enjoys constant computational complexity
  and memory requirements per iteration, which renders it suitable for sequential and big-data
  scenarios. \revv{Building upon this basic algorithm, the contributions of the present paper include the derivation of a more advanced algorithm, theoretical results that characterize the performance of both algorithms, and empirical validation of their performance through extensive experiments with synthetic and real data sets.}

\myitem\cmt{TIRSO} \revv{The proposed algorithm is named} \emph{Topology
Identification via Recursive Sparse Online learning} (TIRSO), which
\begin{myitemize}%
\myitem\cmt{what} substantially improves the tracking performance of TISO and robustness to input variability by
\myitem\cmt{how}minimizing a novel estimation
criterion inspired by  \emph{recursive least
squares} (RLS) where the instantaneous loss function
accounts for past samples.
\myitem\cmt{features}%
\begin{myitemize}%
\myitem\cmt{all of TISO}TIRSO inherits  certain
benefits of TISO  but incurs
\myitem\cmt{increased complexity}a moderate increase in  computational
complexity, which is still constant per iteration.
\end{myitemize}%
\end{myitemize}%

\revv{ We summarize our theoretical results, which constitute the main contribution of our paper: (R1)} it is established that the hindsight solution of TISO and TIRSO are asymptotically the same; \revv{(R2)} \revv{The performance of TISO and TIRSO is analyzed in terms of static regret bounds, which are sublinear and suggest that TIRSO outperforms TISO.} Hence, in the long run, these algorithms perform as well as the best (batch) predictor in hindsight, which supports their adoption for online topology identification.
	\revv{The static regret analysis goes beyond simply stating that the regret is sublinear (which is a direct consequence of applying the algorithm in \cite{duchi2010comid} to the aforementioned criterion), but rather establishes a bound based on properties of the time series that can be checked in practice;}
	\revv{(R3)} A logarithmic regret bound is proved for TIRSO \revv{(such a bound has been proven for TIRSO and could not be proven for TISO thanks to the strong convexity of the loss function).}
	\revv{(R4)} To analyze the performance of TIRSO when the topology is time-varying, a dynamic regret bound is derived. Moreover, the steady-state error of TIRSO in time-varying scenarios is quantified in terms of the data properties.
	Remarkably, the performance (regret)
	analysis does not require probabilistic assumptions, which endows the
	developed approaches with high generality.
\par
\revv{The conference version \cite{zaman2017onlinetopology} of this work presents two online algorithms that are different from the algorithms presented here. One is based on the subgradient approximation for regularized RLS proposed in \cite{eksioglu2011rls} and has computational complexity comparable to that of TIRSO, and the other one is based on a block coordinate minimization via Newton's method and has lower computational complexity for large networks with small process order. In addition, no convergence guarantees were provided.}

        \cmt{Paper Structure}The rest of the paper is organized as
        follows: Sec.~\ref{s:preliminaries} presents the model, a
        batch estimation criterion, and background on online
        optimization. Sec.~\ref{sec:sol} develops TISO and TIRSO. \rev
        { Sec.~\ref{sec:theoresults} \revv{and} Sec.~\ref{sec:numericalresults}
        respectively assess performance analytically and via
        simulations, whereas Sec.~\ref{s:concl} concludes the paper.}
\rev{All code will be made public at the authors' websites.}

\cmt{Notation}\textbf{Notation.} Bold lowercase (uppercase) letters
denote column vectors (matrices). Operators $\mathbb E[\cdot]$,
$\nabla$, $\tilde \nabla$, \rev{$\partial$},
$(\cdot)^\top $, $\mathrm{vec}(\cdot)$,
$\lambda_{\mathrm{max}}(\cdot)$, $\mathcal{R}(\cdot)$, $(\cdot)^\dagger$, and
$\mathrm {diag}(\cdot)$ respectively denote expectation, gradient, subgradient, \rev{sub-differential}, matrix
transpose, vectorization, maximum eigenvalue, range or column space, pseudo-inverse, and
diagonal of a matrix.  Symbols $\bm 0_N$, $\bm 1_N$, $\bm 0_{N\times N}$, and
$\bm I_{N}$ respectively represent the  all-zero vector of size $N$, the all-ones vector of size $N$, the all-zero matrix of size $N\times N$, and
the size-$N$ identity matrix. Also, $[\cdot]_+=\mathrm {max} (\cdot, 0)$. For functions $f(x)$ and $g(x)$, the
notation $f(x)\propto g(x)$ means  $\exists a>0,b:f(x) = a g(x) +b$. The operator $\mathds{1}$ is the indicator satisfying $\mathds 1 \{x\}=1$ if $x$ is true and
    $\mathds 1 \{x\}=0$ otherwise. \revv{Finally, for time series, the notation $\{y_n[t]\}_t$ corresponds to $\{y_n[t]\}_{t\in\mathbb{Z}}$.}

\section{Preliminaries}  \label{s:preliminaries}
After outlining the notion of directed causality graphs, this section
reviews how these graphs can be identified in a batch fashion. Later,
the basics of online optimization are described.
\vspace{-4mm}
\subsection{Directed Causality Graphs}\label{sec:directedcausalitygraphs}
\begin{myitemize}%
  \myitem \cmt{Time series and causality notion:}
  \begin{myitemize}%
    \myitem \cmt{Time series}Consider a collection of $N$ time
    series \rev{$\{y_n[t]\}_{t}$, $n=1,...,N$,} where $y_n[t]$ denotes the value of the $n$-th time series at
    time $t$.
    \myitem\cmt{Goal: directed causal
      graphs}A causality graph   $\mathcal
    G\!\triangleq\!(\mathcal V, \mathcal E)$ is a graph
    where
    \begin{myitemize}%
      \myitem\cmt{vertices}the $n$-th vertex in $\mathcal V\!=\!\{
      1,\ldots,N\}$  is identified with the $n$-th time
      series \rev{$\{y_n[t]\}_t$} and
      \myitem\cmt{edges}there is an edge (or
      arc) from $n'$ to $n$ (i.e. $(n,n')\in  \mathcal E$) if and only if (iff)
      \rev{$\{y_{n'}[t]\}_t$} \emph{causes} \rev{$\{y_{n}[t]\}_t$} according to a certain
      causality notion.                                \end{myitemize}%
  \end{myitemize}%
\end{myitemize}%
\begin{myitemize}%
  \myitem \cmt{Model:VAR}%
  For the reasons
  outlined in Sec.~\ref{s:intro}, a prominent notion of
  causality \rev{described later in this section} can be defined using  VAR models.
  \begin{myitemize}%
    \myitem\cmt{def}To this end, \rev{let $\bm y[t]\!\triangleq\!
  [y_1[t], \ldots, y_N[t] ]^\top$ and define a VAR time series $\{\bm y[t]  \}_t$  as a sequence generated by}  the  order-$P$ VAR model\rev{\cite {lutkepohl2005}}
    \begin{equation}\label{eq:model}
      \bm y[t] =\textstyle{\sum_{p=1}^{P}}\bm A_p \bm y[t-p] +\bm u[t],
    \end{equation}
    where
    \begin{myitemize}%
      \myitem$\bm A_p \!\in \!\mathbb R
      ^{N \times N}, p=~\!1, \ldots, P$, are the
      VAR parameters\footnote{\rev{For the sake of clarity,
        matrices $\{\bm A_p\}_{p=1}^P$ are deemed constant
        throughout this section. However, all the notions explained
        here can be easy generalized to  time-varying
        scenarios, as detailed in subsequent sections.}}
      and
      \myitem$\bm u[t]\!\triangleq   \![
      u_1[t],\ldots,u_N[t] ]^\top$ is the \emph{innovation
  process}. This process is generally assumed to be a \revv{temporally} white, zero-mean stochastic process, i.e.,
      $\mathbb E [\bm u[t]  ]=\bm 0_N$ and $\mathbb E [\bm
      u[t]\bm u^\top[\tau]  ]\!=\!\bm 0_{N\times N}$ for $t \! \ne \! \tau$. 
     \rev{Yet, the present work   does  not even need to assume that
   $\bm u[t]$ is random, \nextv{which benefits its generality}; see the remark at the end of Sec.~\ref{sec:theoresults}.}
    \end{myitemize}%
    \myitem\cmt{LTI interpr}With $a_{n,n'}^{(p)}$ the
  $n,n'$-th entry of $\bm A_p$, expression \eqref{eq:model} becomes
    \begin{align} \nonumber
      y_n[t]&=\textstyle{\sum_{n'=1}^{N}\sum_{p=1}^P}a_{n,n'}^{(p)}y_{n'}[t-p]+u_n[t]
      \\ \label{eq:fnE}
            &= \textstyle{\sum_{n'\in \mathcal{N}(n)}\sum_{p=1}^P} a_{n,n'}^{(p)}y_{n'}[t-p]+ u_n[t]
    \end{align}
 for $n=1, \ldots, N$,
    where $\mathcal{N}(n) \!\triangleq\! \{ n'\!:\! \bm a_{n,n'}\neq \bm 0_P \}$ and
    $\bm a_{n,n'}\!\define\!
    [a_{n,n'}^{(1)},\ldots,a_{n,n'}^{(\rev {P})}]\transpose$. Recognizing the
    convolution operation in the right-hand side enables one to
    express
    \eqref{eq:fnE} as $
      y_n[t] =\textstyle{\sum_{n'\in
        \mathcal{N}(n)} }a_{n,n'}^{(t)}\ast
        y_{n'}[t]\!+ \!u_n[t]$
    in signal processing notation. Thus, in a VAR model,  $y_n[t]$ equals the
    sum of  noise and the  output of
    $|\mathcal{N}(n)|$ linear time-invariant
    filters \rev{ where the $n,n'$-th filter has input
    $\{y_{n'}[t]\}_{t}$ and coefficients
    $\{a_{n,n'}^{(p)}\}_{p=1}^P$}.
  \end{myitemize}%
  \myitem\cmt{VAR causality}%
  \begin{myitemize}%
  
  	 \myitem\cmt{intuition}When
    $\bm u[t]$ is a zero-mean and temporally
    white stochastic process,  the term
    $\hat
    y_n[t]\!\define\! \sum_{n'\in \mathcal{N}(n)}\sum_{p=1}^P
    a_{n,n'}^{(p)}y_{n'}[t-p]$
    in \eqref{eq:fnE} is the \emph{minimum
      mean square error estimator} of $y_n[t]$
    given the previous values of all time
    series
    $\{y_{n'}[\tau],n'\!=\!1,...,N,\tau <\!
    t\}$; see e.g.~\cite[Sec. 12.7]{kay1}.  The set
    $\mathcal{N}(n)$ therefore collects
    the indices of those time series that
    participate in this optimal
    predictor of $y_n[t]$ or,
    alternatively, the information
    provided by time series \rev {$\{y_{n'}[\tau]\}_{\tau<t}$}
    with $n'\!\notin \!\mathcal{N}(n)$ is not
    informative to predict
    $y_n[t]$.  \myitem\cmt{def}This motivates the
    following definition of causality\nextv{, which
    embodies the spirit of Granger
    causality (see Sec.~\ref{s:intro})}:
    \rev {$\{y_{n'}[t]\}_t$} \emph{VAR-causes}
    \rev {$\{y_{n}[t]\}_t$} whenever
    $n'\!\in \!\mathcal{N}(n)$. Equivalently,
   	\rev {$\{y_{n'}[t]\}_t$} \emph{VAR-causes}
    \rev {$\{y_{n}[t]\}_t$} if $\bm a_{n,n'} \!\neq \! \bm 0_P$.
    \myitem\cmt{relation to Granger}%
    \revv{A detailed comparison  with Granger causality
    lies out of scope, yet
    it is worth mentioning that the main
    distinction lies in the prediction
    horizon\footnote{\revv{Whereas VAR causality just
    pertains to prediction 1 time instant
    ahead, Granger causality involves
    prediction of all future samples
    $y_n[t'],~t'\geq t$, given the  ones up to
    a certain time instant
    $\{y_{n'}[\tau],~n'=1,\ldots,N,~\tau <
    t\}$. Therefore VAR causality implies
    Granger causality, but the converse is
    false.}}; see \cite[Sec. 2.3.1]{lutkepohl2005}
    for a more detailed comparison.}
    \myitem\cmt{graph}VAR causality relations among
    the $N$ time series can be represented  using a
    causality graph where
    \begin{myitemize}%
      \myitem\cmt{edges}$\mathcal{E}\triangleq \{(n,n')\!:\!\bm
      a_{n,n'}\!\neq\! \bm 0_P\}$.
      \myitem\cmt{neighborhood}Clearly, in
      such a graph, $\mathcal{N}(n)$ is the
      in-neighborhood of node $n$.
      \myitem\cmt{weights}To
      quantify the strength of these
      causality relations, a weighted graph
      can be constructed by assigning e.g. the weight $\|\bm a_{n,n'}\|_2$
      to the edge~$(n,n')$.%
    \end{myitemize}%
    \begin{figure}
      \centering
      \includegraphics[width=0.4\textwidth]{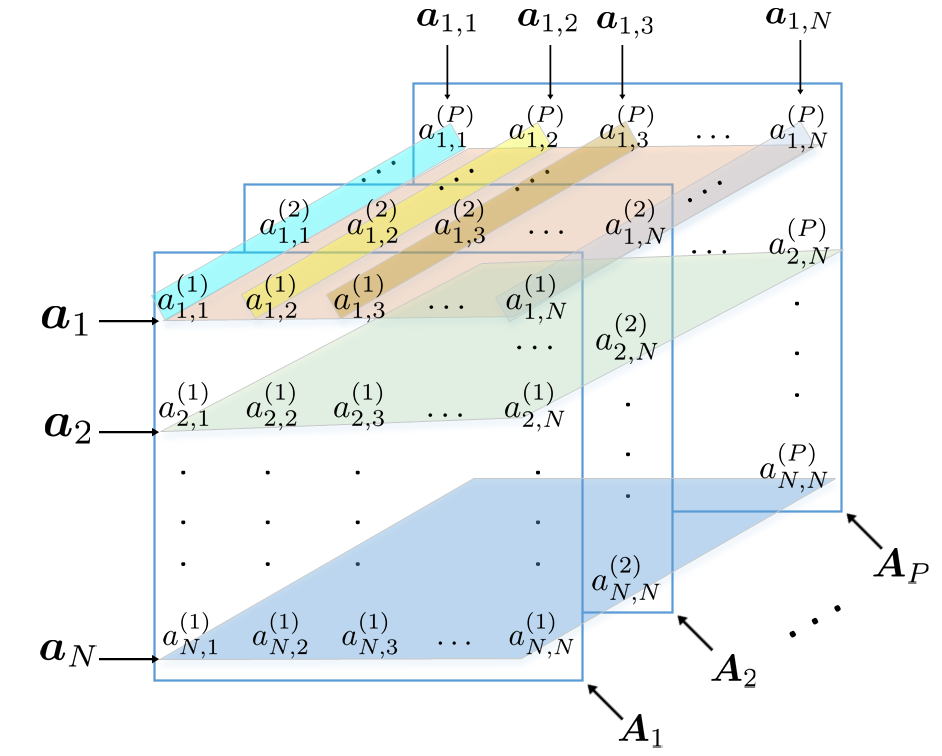}
      \caption{\rev{\small Tensor $\mathcal{A}$ collecting the VAR parameter matrices. }}
      \label{fig:structureofa}
    \end{figure}
  \end{myitemize}%
\par
  \myitem\cmt{Batch problem Statement:}With these definitions,
  the \emph{batch} problem of identifying a VAR causality
  graph reduces to estimating the VAR coefficient matrices
  $\{ \bm A_p\}_{p=1}^P$ given $P$ and the observations
  $\{\bm y[t]\}_{t=0}^{T-1}$. To simplify
  notation, form the tensor $\mathcal{A}$ by stacking the matrices
  $\{\bm A_p\}_{p=1}^P$ along the third dimension \rev{ as shown in Fig.~\ref{fig:structureofa}}.%
\end{myitemize}%
\subsection{Batch Estimation Criterion for Topology Identification} \label{s:online}
This section presents an estimation criterion to address the batch  problem
formulated in Sec.~\ref{sec:directedcausalitygraphs}.
\begin{myitemize}%
  \myitem \cmt{Least squares}A natural estimate could be
  pursued through  \emph{least-squares}
by  minimizing~\cite{lutkepohl2005}
  \begin{eqnarray}
    \mathcal {L}\left (\mathcal{A}\right)\triangleq \frac{1}{\,2(T-P)}  \sum_{\tau=P}^{T-1}\Big \lVert \bm y[\tau]-\sum_{p=1}^{P} \bm A_p \, \bm y[\tau -p]\Big \lVert_2^2 \nonumber\\
    =\frac{1}{2\, (T-P)}\sum_{n=1}^{N}\sum_{\tau=P}^{T-1}\Big [ y_n[\tau]-\sum_{n'=1}^{N}\sum_{p=1}^{P}  a_{n,n'}^{(p)} \, y_{n'}[\tau-p]\Big ]^2 \nonumber.
  \end{eqnarray}
  This estimation task becomes underdetermined  unless
  the number $NT$ of available data samples meaningfully exceeds the
  number of  unknowns $ PN^2$, \nextv{ or, equivalently, $T\geq PN+P$. Even more, to obtain a
  reasonable performance, one requires $T\gg PN+P$ which may not be
  possible in practice, especially if the parameters $\{ \bm
  A_p\}_{p=1}^P$ remain constant only for short periods of time. }%
  \myitem{}To circumvent this limitation, \nextv{one may note that most
  causality relations between two time series will be \emph{mediated}
  by one or more time series. This means that the causality graph
  introduced in Sec.~\ref{sec:directedcausalitygraphs} is expected to be sparse, meaning
  that many of the vectors $\bm a_{n,n'}$ equal zero. Such a sparsity
  structure can be promoted by properly regularizing the aforementioned
  least squares objective.} \nextvv{Thus, the estimator becomes
  \begin{eqnarray}\label{eq:prob}
    \underset{ \mathcal{A}}{\arg\min} ~ \mathcal L\left (
    \mathcal{A}\right) \!+\! \lambda \sum_{n=1}^{N}\sum_{\substack {
    n'=1\\ n' \neq n}}^{N} \mathds{1}  \bigg \{ \lVert \bm a_{n,n'} \rVert_2   \bigg \},
  \end{eqnarray}
  where
  \begin{myitemize}%
\myitem \cmt{Sparsity-promoting
      Regularizer}The second term in \eqref{eq:prob} equals the
    cardinality of $\mathcal E$ (i.e., the number of edges) times the
    regularization parameter $\lambda$.   \myitem
    \cmt{Self-loops not regularized}
  \end{myitemize}%
  \myitem \cmt{Convex approximation}Problem \eqref{eq:prob} is not
  convex due to the regularizer, so one may instead replace the
  indicator function, also known as zero norm, with its convex
  surrogate, the $\ell_1$-norm.}%
To this end, the following criterion has been proposed in~\cite{bolstad2011groupsparse}:
  \begin{eqnarray}\textstyle \label{eq:probl1}
    \underset{ \mathcal{A}}{\arg\min} ~ \mathcal L\left (
    \mathcal{A}\right) \!+\! \lambda \sum_{n=1}^{N}\sum_{\substack {
    n'=1, n' \neq n}}^{N} \left \lVert \bm a_{n,n'} \right \rVert_2,
  \end{eqnarray}
where $\lambda>0$ is a regularization parameter that can be adjusted e.g. via cross-validation
    \cite[Ch. 1]{bishop2006}.
  The second term in   \eqref{eq:probl1} is conventionally referred to as 
  a \emph{group-lasso}\revv{\footnote{\revv{Although other norms (such as the sum of infinity norms) can be used to enforce group sparsity, recoverability results associated with this norm are provided in \cite{bolstad2011groupsparse}.}}} regularizer and the solution to
  \eqref{eq:probl1} as a  \emph{group-lasso} estimate~\cite{yuan2006grouplasso}. This promotes a \emph{group-sparse
    structure} in $\{ \bm A_p\}_{p=1}^P$ to exploit the information
    that the number of edges in $\mathcal E$ is typically small. Self-connections  ($\bm a_{n,n}$,\! $n=1,...,N$) are excluded from the
    regularization term so that the inferred causal relations
    pertain to the component of each time series that cannot be
    predicted using its own past\nextv{. This is motivated by the improvement
    in consistency reported in}~\cite{bolstad2011groupsparse}.
\nextv{The criterion \eqref{eq:probl1} can be further motivated on the grounds of the consistency of group-lasso estimators~\cite{liu2009estimation}.}
\par
\myitem\cmt{separability}Remarkably, \eqref{eq:probl1} separates
along $n$. To see this, let
$    \bm a_{n}\triangleq[\bm a_{n,1}^\top,\bm a_{n,2}^\top,..., \, \bm
a_{n,N}^\top]^\top \! \in \! \mathbb R^ {NP}$ and
\begin{align}
    \bm g[t]\triangleq \mathrm{vec}\big (\left [\bm y[t-1], \ldots, \bm y [t -P] \right ]^\top \!\big) \in \mathbb R^ {NP}, \label{eq:g}
\end{align}
and express $\mathcal L(\mathcal{A})$ as $ \mathcal
L(\mathcal{A})\!=\!{\sum_{n=1}^N {\ell} ^{(n)}(\bm a_n)}$, where ${\ell}^{(n)}(\bm a_n)\!\triangleq\!{1}/(T-P)\sum_{t=P}^{T-1}\ell_t^{(n)}(\bm a_n)$ and $\ell_t^{(n)}(\bm a_n)\!\triangleq\! {1}/{2}  (y_n[t]- \bm g^\top[t] \bm a_n  )^2$.
Then, \eqref{eq:probl1} becomes
  $  \{ {\bm a}_n^*\}_{n=1}^N\!=\!\argmin_{ \{\bm a_n\}_{n=1}^N}  \!\sum_{n=1}^{N} \![ {\ell}^{(n)}(\bm a_n)\!\!+\!\!{\lambda} \! \sum_{\substack { n'=1, n' \neq n}}^{N}\!\left \lVert \bm a_{n,n'}\right\rVert_2   ],$
with
    \begin{equation} \textstyle \label{eq:prob-sep}
      {\bm a}_n^*=  \underset{ \bm a_n}{\arg\min} ~ {\ell}^{\left (n
        \right )}(\bm a_n)+{\lambda} \sum_{\substack { n'=1, n' \neq n}}^{N} \left \lVert \bm a_{n,n'}\right\rVert_2
    \end{equation}
for $n=1,\ldots,N$. Thus, the VAR causality graph can  be identified by separately
estimating the VAR coefficients, and hence incoming edge weights, for each node. 

\myitem\cmt{Limitations of batch}The batch estimation criterion in
\eqref{eq:prob-sep} requires all data $ \{\bm y[t]\}_{t=0}^{T-1}$
before an estimate can be obtained and cannot track changes. Furthermore, 
solving  \eqref{eq:prob-sep} eventually becomes prohibitively complex for sufficiently large $T$. To address these challenges, this paper
adopts the framework of online optimization, \rev{which is reviewed in
the following subsection.}
\end{myitemize}%
\par
\revv{\textbf{Remark:} As seen in \eqref{eq:probl1}, $\lambda$ is the
      same for all candidate edges $(n,n')$. This can be readily replaced with
an edge-dependent regularization parameter $\lambda_{n,n^\prime} $
without any complexity increase  to exploit possibly available
prior-information about edges.}
\subsection{Background on Online Optimization}
\label{sec:onlineopt}
\cmt{Overview}This section reviews the fundamental notions of online
optimization from a general perspective, not necessarily applied to
the problem of topology identification.
\begin{myitemize}%
  \myitem \cmt{Batch vs Online}%
  \begin{myitemize}%
    \myitem\cmt{generic problem}To this end, consider the generic unconstrained optimization problem
    \begin{align}
      \label{eq:genericproblem}
 \minimize_{\bm a}~ \frac{1}{T_0}\sum_{t=0}^{T_0-1}~h_t(\bm a),
    \end{align}
    where $h_t(\bm a)$ is a convex function, which in many applications  depends on the data received at time $t$. For example, in least squares  $h_t(\bm a) \!= \!\lVert \bm X[t] \bm a - \bm
    y[t]\rVert_2^2$, where $\bm y[t]$ \rev{and $\bm X[t]$ are the data
  vector and matrix made available at time $t$.}
        \myitem\cmt{batch approaches}%
    \begin{myitemize}%
      \myitem\cmt{def}To solve \eqref{eq:genericproblem}, it is necessary that      all  $\{h_t(\bm a)\}_{t=0}^{T_0-1}$  be available. Approaches
      that process all data at once are termed \emph{batch}
      \myitem\cmt{limitations}%
      \begin{myitemize}%
        \myitem\cmt{latency}and, hence,  suffer from potentially long
        waiting times, which generally render them inappropriate for real-time
        operation.
        \myitem\cmt{complexity}Besides, computational complexity and
        memory generally grow
        super-linearly with $T_0$, which eventually becomes prohibitive.
      \end{myitemize}%
    \end{myitemize}%
    \par
    \myitem\cmt{online approaches}%
    \begin{myitemize}%
      \myitem\cmt{def}Online algorithms  alleviate these limitations.
      \begin{myitemize}%
        \myitem\cmt{sequential}\rev{Let $\bm a[t+1]$ denote an
  estimate of the solution to \eqref{eq:genericproblem} at time $t$ produced by an online algorithm. Online algorithms} compute a new  $\bm a[t+1]$ every time a new
         $(\bm X[t],\bm y[t])$ data element (or,  more generally, a new  $h_t(\bm a)$) is processed.
        \myitem\cmt{bounded complexity}At every iteration,
        also known as \emph{update}, $\bm a[t+1]$ is obtained from $\bm
        a[t]$, $\bm y[t]$, $\bm X[t]$, and possibly some additional information
        carried from each update to the next. The memory requirements and number of arithmetic
        operations  per
        iteration  must not
        grow unbounded for increasing $t$. \rev{This requirement rules out
        approaches involving   solving
        \eqref{eq:genericproblem} as a batch problem per update or
        carrying  all the past data $\{(\bm X[\tau],\bm
        y[\tau])\}_{\tau=0}^{t-1}$ from the $(t\!-\!1)$-th update to the
        $t$-th update.} %
      \end{myitemize}%
      \myitem\cmt{strengths}%
      \begin{myitemize}%
        \myitem\cmt{sequential}Thus,  online algorithms are especially
        appealing when data vectors are received
        sequentially
        \myitem\cmt{complexity}or  $T_0$ is so large that batch
        solvers are not
        {computationally} affordable.
        \myitem\cmt{tracking}Additionally, online algorithms can track
        changes in the underlying model.
\nextv{  When $\bm a$ represents a
         probabilistic model parameter  that must be estimated, the estimate obtained
        through an online method is generally capable of tracking
        variations in $\bm a$ so long as they do not occur too rapidly.}
      \end{myitemize}%
    \end{myitemize}%
  \end{myitemize}%
  \par
  \myitem \cmt{Static Regret}The most common performance metric  to evaluate online algorithms is the \textit{regret},
  which quantifies the cumulative loss incurred by an
  online algorithm relative to the loss corresponding to the optimal constant solution in
  hindsight. 
  \begin{myitemize}%
    \myitem\cmt{def}Formally, the (static) regret\rev{\footnote{\rev{The static regret is known simply as regret in earlier works, e.g. \cite{shalev2011online}, and different types of regret were formalized later, see e.g. \cite{hall2015dynamicregret}.}}} at iteration  $T_0-1$ is
    given by \cite{shalev2011online}:
    \begin{align}  \label{eq:regret_def}
      \begin{aligned}
        R\rev{_{s}[T_0]}\triangleq&\sum_{t=0}^{T_0-1} [h_t \left ( \bm a[t] \right)-h_t\left (\bm a^\ast [T_0] \right ) ],
      \end{aligned}
    \end{align}
    where $
      \bm a ^\ast [T_0]\triangleq\argmin_{\bm a}~ ({1}/{T_0})\sum_{t=0}^{T_0-1}  h_t(\bm a)$
    is the optimal \emph{constant} hindsight solution, i.e., the batch
        solution after $T_0$ data vectors have been
        processed.
    \myitem\cmt{can be negative}\nextv{Observe that the regret in \eqref{eq:regret_def} may be negative since the estimates $\{\bm a[t]\}_t$ are allowed to depend on $t$ and, hence, it may hold that
    $h_t  ( \bm a[t]  )\leq h_t (\bm a ^\ast [T_0])$ for multiple
    (potentially all) values of $t$.
    \myitem\cmt{must be sublinear}In practice, nevertheless, the regret
    will typically be positive and increase with $T_0$.} To be deemed
    admissible, online algorithms must yield a \emph{sublinear regret},
    i.e., $R\rev{_{s}[T_0]}/T_0\!\rightarrow\! 0$ as
        $T_0 \!\rightarrow\! \infty$. Thus, online algorithm with sublinear regret perform
    asymptotically as well as
    the batch solution \emph{on average}.
    \myitem\cmt{no data assumptions}\rev{It is worth noting that the online learning framework does not involve statistical assumptions on the data, which can even be generated  by an ``adversary'' \cite{hazan2016online}.}
  \end{myitemize}%
\par
\rev{
\myitem \cmt{Dynamic Regret}%
\begin{myitemize}%
	\myitem \cmt{dynamic environment}In dynamic settings where the
	parameters of the data generating process vary over time, $ \bm a ^\ast
	[T_0]$ may not be a suitable reference since its computation
	involves potentially very old data, namely $\{h_t\}_{t\ll T_0}$,
	which is informative about old values of the true parameters but
	not about the new values. In
	those cases, it is customary to compare against the
	instantaneous minimizer  $\bm a ^\circ
	[t]\! \define \! \argmin_{\bm a}  h_t(\bm a)$ by means of the
	so-called dynamic regret \cite{hall2015dynamicregret}, \cite{mokhtari2016onlineoptimization}:
\begin{align*}  
\begin{aligned}
R_{d}[T_0]\triangleq&\sum_{t=0}^{T_0-1} [h_t \left ( \bm a[t] \right)-h_t\left (\bm a^{\circ} [t] \right ) ].
\end{aligned}
\end{align*}
\myitem \cmt{applications}
More details about the dynamic regret are given in Sec.~\ref{sec:dynamicregret}.
\end{myitemize}}%
\end{myitemize}%
\section{Online Topology Identification} \label{sec:sol}
\cmt{overview}This section develops online algorithms \rev {for the
considered problem of topology identification from time series.} To
this end,  cast  \eqref{eq:prob-sep} for the $n$-th node in the form \eqref{eq:genericproblem} by setting
\begin{equation} \textstyle \label{eq:hsetting}
  h_{t}(\bm a_n) =\ell^{(n)}_{t+P}(\bm a_n) + \lambda\sum_{\substack { n'=1, n' \neq n}}^{N} \left \lVert \bm a_{n,n'}\right\rVert_2,
\end{equation}
for $t=0,...,T-P-1$.
 \cmt{ online subgradient descent}%
  \begin{myitemize}%
    \myitem\cmt{Application of OSGD}The most immediate approach to
    solve \eqref{eq:genericproblem} 
    would be applying \emph{online subgradient descent} (OSGD), whose 
    updates are given by
    $\bm a_n[t+1] = \bm a_n[t] -\alpha_t \tilde{\mathbf{w}}_n[t]$ with
    $\tilde{\mathbf{w}}_n[t]$  a subgradient of $h_t$ at $\bm a_n[t]$
  and $\alpha_t$ the step size at time $t$.
    From \eqref {eq:hsetting}, $\tilde{\mathbf{w}}_n[t]$ equals
    $\nabla\ell^{(n)}_{t+P}(\bm a_n[t])$
 plus \rev{$\lambda$ times} a valid subgradient of the form
    \begin{multline*} \textstyle 
      \hspace{-5mm}
      \tilde {\nabla}_{\bm a_n} \!\!\sum_{\substack { n'=1\\ n' \neq n}}^{N} \!\left \lVert \bm a_{n,n'}\right\rVert_2  \!\triangleq\!  [ \tilde { \nabla}_{\bm a_{n,1}} ^{\top} \! \left \lVert \bm a_{n,1}\right\rVert_2, \ldots,\tilde { \nabla}_{\bm a_{n,n-1}}^{\top}  \!\left \lVert \bm a_{n,n-1}\right\rVert_2 , \\
      \bm 0_P, \tilde {\nabla}_{\bm a_{n,n+1}} ^{\top} \left \lVert \bm a_{n,n+1}\right\rVert_2, \ldots, \tilde {\nabla}_{\bm a_{n,N}} ^{\top} \left \lVert \bm a_{n,N}\right\rVert_2]^\top,
    \end{multline*}
evaluated at $\bm a_n[t]$. \rev{For example,  for $\bm x \in \mathbb{R}^P$, set $\tilde  {\nabla}_{\bm x}  \lVert \bm x \rVert_2\!=\! \bm x / \lVert \bm
  x  \rVert_2$ for $ \bm x \! \ne \! \bm 0_P $~and $\tilde {\nabla}_{\bm
  x}  \lVert \bm x \rVert_2\!=\!\bm 0_P$ for $\bm x= \bm 0_P$.} 
    \myitem\cmt{online subgrad descent does not yield
      sparsity}\rev{It is easy to see that the resulting iterates $\bm a_n[t]$ are not necessarily sparse; see also \cite{duchi2010comid}.
    Since the solution to the batch problem is indeed sparse \revv{for a properly selected $\lambda$}, alternative approaches are
    required} 
  \end{myitemize}%
\par
 \cmt{Composite approaches}
  \begin{myitemize}%
\myitem\cmt{reason OSGD fails}To this end, note that
    OSGD fails to provide sparse iterates because it  implicitly linearizes the 
   instantaneous objective $h_t(\bm a_n)$. Since the regularizer  (last
   term in \eqref{eq:hsetting}) is
   not differentiable, it is not well approximated by a linear
   function and, as a result, it fails to promote sparsity.
\myitem\cmt{composite better}To address this issue, \emph{composite} algorithms  decompose
  $h_t(\bm a_n)$  as $h_t(\bm a_n)\! = \!\rev{ \ftn }(\bm a_n)+\rev {\nReg}(\bm a_n)$, where $
  \rev{ \ftn }(\bm a_n)$ is a convex loss function and  $\rev {\nReg}(\bm a_n)$ is a
  convex regularizer, and linearize only $\rev{ \ftn }(\bm a_n)$.
    \myitem\cmt{oveview}Algorithms of this family, which include
    \emph{regularized dual averaging} (RDA) \cite{xiao2010dualaveraging} and
    \emph{composite objective mirror descent} (COMID)
    \cite{duchi2010comid}, solve the generic problem
  \begin{equation} \label{eq:compositeObjective}
    \underset{\bm a_n }{\minimize} ~\frac{1}{T_0} \sum_{t=0}^{T_0-1} \big [	\rev{ \ftn }	(\bm a_n)+ \rev {\nReg}(\bm a_n) \big ],
  \end{equation}
  by linearizing $\rev{ \ftn }(\bm a_n)$ \rev {but not $\nReg(\bm a_n)$}. 
\revv{This work  focuses on COMID since, unlike RDA, there exist
bounds for its regret for constant step size when the regularizer is
not strongly convex. The COMID update is }
\myitem\cmt{COMID description}%
\begin{myitemize}%
\myitem\cmt{update}%

\vspace{-2mm}
  \begin{multline} \label{eq:comid}
  \textstyle 
    {\bm a_n}[t+1]=\underset{\bm a_n}{\arg \min} \big [
                  \rev {\alpha_t} \tilde {\nabla} \rev{ \ftn }\transpose(\bm a_n[t])\left ( \bm a_n- \bm a_n[t]\right ) 
                \\+  B_\psi \left (\bm a_n, \bm a_n[t] \right ) 
                +\rev {\alpha_t \nReg}( \bm a_n)\big], 
  \end{multline}
   where
  \begin{myitemize}%
    \myitem \cmt{Subgradient}$\tilde {\nabla} \rev{ \ftn }(\bm
    a_n[t])$ is a  subgradient of $\rev{ \ftn }$ at point $\bm
    a_n[t]$ \rev {(that is, $\tilde {\nabla} \rev{ \ftn }(\bm
    	a[t])\! \in \! \partial \ftn (\bm a_n[t])$)},
    \myitem \cmt{Stepsize}$\alpha_t\!> \!0$ is a step size, and
    \myitem \cmt{Bregman divergence}%
    $
      B_\psi(\bm w, \bm v)\triangleq \psi(\bm w)-\psi (\bm v) - \nabla \psi\transpose \left(\bm v \right)(\bm w-\bm v)
    $
 is the so-called Bregman divergence associated with a
 $\zeta$-strongly convex and continuously differentiable function
 $\psi $. The strong convexity condition   means that  $B_\psi(\bm w, \bm v) \geq
({\zeta}/{2}) \lVert \bm w -\bm v \rVert ^2$, which motivates using
$B_\psi(\bm w, \bm v)$ as a surrogate of a distance between $\bm w$
and $\bm v$. Thus, the Bregman divergence in \eqref{eq:comid} penalizes updates $\bm a_n[t+1]$ lying far from the previous one $\bm a_n[t]$, which
essentially smoothes the sequence of iterates.
  \end{myitemize}%
  \myitem\cmt{explanation}%

Relative to each term in \eqref{eq:compositeObjective},
    the loss  $\rev{ \ftn }$ in  \eqref{eq:comid} has been linearized but
    the regularizer $\rev {\nReg}(\bm a_n)$ has been kept intact. When
    $\rev {\nReg}(\bm a_n)$ is a sparsity-promoting regularizer, then the  online estimate $\bm
    a_n[t+1]$~is therefore expected to be sparse.
\end{myitemize}%
\end{myitemize}%
\par
In view of these appealing features, the algorithm proposed in Sec.~\ref{ss:tiso} builds upon
COMID to address the problem of online causality
graph identification \rev {from time series}.%
\vspace{-3mm}
\subsection{Topology Identification via Sparse Online optimization} \label{ss:tiso}
\cmt{Overview}This section proposes \emph{topology
  identification via sparse online optimization}  (TISO), an online
algorithm for the problem in Sec.~\ref{s:online} that provides a causality graph estimate every time a new
 $\bm y[t]$ is processed. The key idea of this first algorithm
is to refine the previous topology estimate with the information
provided by the new data vector by means of a  COMID update.
\par
\cmt{Application of COMID}To this end, 
\begin{myitemize}%
  \myitem\cmt{decomposition}express  $h_t$ in \eqref{eq:hsetting}  in the form
$h_t(\bm a_n) = \rev{ \ftn }(\bm a_n)+\rev {\nReg}(\bm a_n)$ by setting
\begin{subequations}
	\vspace{-1mm}
 \label{eq:comidlossdef}
\begin{align}
\textstyle
  \rev{ \ftn }(\bm a_n) &= \ell_{t+P}^{(n)}(\bm a_n),\\
 \label{eq:comidlossdefreg}
  \rev {\nReg}(\bm a_n)&= \lambda \textstyle \sum_{\substack { n'=1,
  n' \neq n}}^{N} \left \lVert \bm a_{n,n'}\right\rVert_2 ,
\end{align}
\end{subequations}
for $t\!=\!0,...,T-P-1.$ 
\myitem\cmt{Bregman}
To choose $ B_\psi(\bm w, \bm v)$, note that \eqref{eq:comid} with $\rev{ \ftn }(\bm a_n)$ and $\rev
{\nReg}(\bm a_n)$ given by
\eqref{eq:comidlossdef} can be solved in closed form when   $\psi(\cdot) =
1/2\lVert \cdot \rVert_2^2$. In that case,
 $	B_\psi(\bm w, \bm v) \!=\! 1/2 \lVert \bm w- \bm
 v\rVert_2^2 $ and $\bm a_n[t+1]$ can be found via a
 modified \emph{group soft-thresholding} operator, as
 detailed next.
  \myitem \cmt {TISO update expression:}With these expressions, the
  TISO  update after processing  $\{\bm
  y[\tau]\}_{\tau=0}^t$  is
  \begin{align}  \label{eq:comid2}
  \textstyle
    {\bm a}_n[t+1]&=\underset{\bm a_n }{\arg \min} ~ J\rev{_{t}^{(n)}}(\bm a_n),
  \end{align}
  where
  \vspace{-1mm}
  \begin{multline} \label{eq:tisoupdateobj}
    J\rev{_{t}^{(n)}}(\bm a_n)\define\bm v_n^\top[t] (\bm a_n-\bm a_n[t])+  \frac{1}{2\alpha_t} \left \lVert \bm a_n- \bm a_n[t] \right \rVert_2^2  \\
               +\lambda\textstyle \sum_{\substack { n'=1, n' \neq n}}^{N} \left \lVert \bm a_{n,n'}\right\rVert_2		
  \end{multline}
  \begin{myitemize}%
    \myitem \cmt {subgradient:}
    and 
    \revv{(using the vector $\bm g[t]$ defined in \eqref{eq:g})}
    \begin{equation} \label{eq:instgrad}
        \vspace{-1mm}
      \bm v_n[t]\define \nabla {\ell}_t^{(n)}(\bm a_n[t])=   \bm g[t]\,( \bm g^\top[t]\, \bm a_n[t]-  y_n[t] ).
    \end{equation}
\par
    \myitem \cmt{Solving \eqref{eq:comid2}:}To solve 
    \eqref{eq:comid2} in closed form,
    \begin{myitemize}%
    \myitem \cmt{Rewriting the cost function}expand the
    squared norm in \eqref{eq:tisoupdateobj} to obtain
    \vspace{-1mm}
    \begin{align}
      J\rev{_{t}^{(n)}}(\bm a_n)& \propto\! \frac{\|\bm a_n\|_2^2}{2 \alpha_t} \!+\! \bm a_n^\top \big (\bm v_n[t]\!-\!\frac{1}{ \alpha_t}\bm a_n[t] \big )\!+\!\lambda\textstyle \sum_{\substack { n'=1\\ n' \neq n}}^{N} \left \lVert \bm a_{n,n'}\right\rVert_2 \nonumber \\
                &= \sum_{n'=1}^N \Big[\frac{1}{2 \alpha_t} \|\bm a_{n,n'}\|_2^2\!+\!\bm a_{n,n'}^\top \big (\bm v_{n,n'}[t]\!-\!\frac{1}{\alpha_t} \,\bm a_{n,n'}[t] \big ) \nonumber \\
                & \quad +\lambda \left \lVert \bm a_{n,n'}\right\rVert_2 \mathds{1} \{ n' \neq n\} \Big], \label {eq:solcomid}
    \end{align}
    where  
     $\bm v_n[t]  \triangleq [\, \bm v_{n,1}^\top[t],
  ..., \bm v_{n,N}^\top[t] \,]^\top$ \rev{and $\bm
    v_{n,n'}[t] \!\in \!\mathbb{R}^P~\forall n'$.}
    \myitem  \cmt{separability across groups}From \eqref{eq:solcomid},
    it can be observed that the updates in \eqref{eq:comid2} can be
    computed separately for each group $n'=1,..., N$. 

      \myitem \cmt{Solution when not self-loops}For $n'\neq n$, the
      $n'$-th subvector of $\bm a_n[t+1]$ (or $n'$-th \emph{group}) can be
      expressed in terms of the so-called multidimensional shrinkage-thresholding
      operator  \cite{puig2011multidimensional} as:
      \vspace{-1mm}
      \begin{equation} \label{eq:comidsol1}
        {\bm a}_{n,n'}\left [t+1 \right ]=   \bm
      a_{n,n'}^\text{f}\left [t\right] \left
      [1-\frac{\alpha_t \, \lambda}{\textstyle \left \lVert \bm a_{n,n'}^\text{f}\left [t\right] \right\rVert_2}\right ]_+,
      \end{equation}
      where
      \begin{myitemize}%
        \myitem $\bm a_{n,n'}^\text{f}\left [t\right] \!\triangleq\! \bm
        a_{n,n'}[t]\! - \! \alpha_t \bm v_{n,n'}[t]$.
      \end{myitemize}%
      \cmt{interpretation}Expression \eqref{eq:comidsol1} is composed
      of two terms: whereas $\bm
      a_{n,n'}^\text{f}[t]$ is the result of performing a
      gradient-descent step in a direction that decreases the
      instantaneous loss ${\ell}_t^{(n)}(\bm
      a_n)$, the second term promotes \emph{group sparsity} by setting $\bm a_{n,n'}[t+1]\!=\!\bm 0_P$ for those groups $n'$ with $\lVert \bm a_{n,n'}^\text{f} [t] \rVert_2\leq\alpha_t \, \lambda $. Recalling that each vector $\bm a_{n,n'}$ corresponds to an edge in the estimated causality
      graph (see Sec.~\ref{sec:directedcausalitygraphs}), expression
      \eqref{eq:comidsol1} indicates that only the relatively strong
      edges (i.e. causality relations) survive. In view of such a shrinkage
      operation, a larger 
      $\lambda$ will result in sparser estimates. 
\par
      \myitem \cmt{Solution when self-loops}On the other hand, when $n'\!=\! n$, the
      $n'$-th subvector of $\bm a_n[t+1]$ in \eqref{eq:comid2} is given by:
      \begin{equation}\label{eq:comidsol2}
        {\bm a}_{n,n'}[t+1]=  \bm a_{n,n'}[t]-\alpha_t \bm v_{n,n'}[t]
        = \bm a_{n,n'}^\text{f}[t]
      \end{equation}
      and, as intended, no sparsity is promoted on self-connections; see Sec.~\ref{s:online}.
      \myitem \cmt{Combined solution}Combining \eqref{eq:comidsol1}
      and \eqref{eq:comidsol2},  the estimate of the $n'$-th  group at time $t+1$ is given by:
      \begin{equation}  \label{eq:comidgroupsol}
        {\bm a}_{n,n'}[t+1]= \bm a_{n,n'}^\text{f}\left [t\right]\left  [1-\frac{\alpha_t \lambda ~ \mathds 1 \{ n \neq n'\}}{\textstyle\left \lVert\bm a_{n,n'}^\text{f}\left [t\right] \right\rVert_2}\right ]_+.
      \end{equation}
    \end{myitemize}%
  \end{myitemize}%
  \par
  \myitem\cmt{Step size}The performance of TISO depends on the choice
  of the step-size sequence $\{\alpha_t\}_t$, \rev{as discussed in Sec.~\ref{sec:theoresults}.}
\end{myitemize}%
\cmt{Overall algorithm}The overall TISO algorithm is
listed as \textbf{Procedure~\ref{alg:TISO}}.
\begin{myitemize}%
  \myitem \cmt{Computational Complexity}%
\begin{myitemize}%
  \myitem\cmt{Memory requirement}It only requires $\mathcal O(N^2P)$
  memory entries to  store the last $P$ data vectors and the last estimate.
  \myitem\cmt{arithmetic operations}On the other hand,
  each update requires $\mathcal O(N^2P)$ arithmetic operations, which
  is in the same order as the number of parameters to be estimated. Thus, TISO can arguably be deemed
  a low-complexity algorithm.
\end{myitemize}%
\par
\myitem\cmt{relation to other algorithms}
\begin{myitemize}%
\end{myitemize}%
  \begin{algorithm}
    \caption{Topology Identification via Sparse Online optimization (TISO)}\label{alg:TISO}
    \textbf{Input:} $\lambda, \{\alpha_t\}_t, \{\bm y[\tau]\}_{\tau =0}^{P-1} $ \\
    \textbf{Output:} $\{  {\bm a}_n[\tau]\}_{n=1}^N, \tau= P+1, \ldots$ \\
    \textbf{Initialization:} $  {\bm a}_n[P]=\bm 0_{NP}, n=1,\ldots, N$
    \begin{algorithmic}[1] %
      \For {$t=P,P+1, \ldots $} 
      \State {Receive data vector $\bm y[t]$}
      \State {Form $\bm g[t] $} via \eqref{eq:g} 
      \For {$n=1,2, \ldots ,N $ } 
      \State $\bm v_n[t]=( \bm g^\top[t] \,  {\bm a}_n[t]-y_n[t])\bm g[t]$ 
      \For {$n'=1,2, \ldots ,N $ }
      \State $\bm a_{n,n'}^\text{f}[t] =   \bm a_{n,n'}[t]-\alpha_t \bm v_{n,n'}[t]$
      \State {Compute $ {\bm a}_{n,n'}[t+1]$ via \eqref {eq:comidgroupsol} }
      \EndFor
      \State \textbf{end for}
      \EndFor
      \State \textbf{end for}
      \EndFor
      \State \textbf{end for}
    \end{algorithmic}
  \end{algorithm}%
\end{myitemize}%
  \par
\cmt{overview rest}
The next section will
build upon TISO to develop an algorithm with increased
robustness to input variability.

\subsection{Topology Identification via Recursive Sparse Online optimization} \label{ss:tirso}
\cmt{Overview}As seen in Sec.~\ref{ss:tiso}, each update of TISO
depends on the data through the \emph{instantaneous} loss
$ \ell_t^{(n)}(\bm a_n[t])$, 
which quantifies the prediction
error of the newly received vector $\bm y[t]$ when the VAR parameters
$\mathcal{A}$ are given by the previous estimate $\bm a_n[t]$. Thus,
the residual of predicting each data vector is used only in a single
TISO update. Although this renders TISO a
computationally efficient algorithm for online topology identification, it also increases
 sensitivity to noise and input variability. To this end, this section pursues an alternative approach 
at the expense of a moderate increase in computational complexity and memory requirements.

\cmt{Motivation TIRSO loss}%
  \begin{myitemize}%
    \myitem\cmt{reduce output variability of TISO thru step-size}It
    is clear from \eqref{eq:comid2} that $\bm a_n[t+1]$ is
    determined by  $\bm a_n[t]$ and $\bm v_n[t]$. The latter incorporates the residual only at time $t$.
    The step size $\alpha_t$ controls how much variability in the
    input data  propagates to the estimates $\{\bm a_n[t]\}_t$. 
\begin{myitemize}%
  \myitem\cmt{diminishing step-size}When a diminishing step-size
  sequence is adopted, the influence of each new $\bm y[t]$ on the
  estimate becomes arbitrarily small, and the variability of the estimates fades
  away. However, decreasing sequences cannot be utilized when the
  application at hand demands tracking changes in the
  coefficients $\mathcal{A}$.
\myitem\cmt{constant step-size}In these settings,
  a constant 
  step size   $\alpha_t\!=\!\alpha$ is preferable.
\cmt{smaller step-size}In
  such a scenario, a desire to reduce output variability would
  therefore force one to adopt a small $\alpha$, but this
  would hinder TISO from tracking changes in the topology. 
\end{myitemize}%
\par
\myitem\cmt{reduce output variability by changing loss}
\begin{myitemize}%
  \myitem\cmt{inspiration: RLS}\rev{An approach to reduce output
  variability without sacrificing tracking capability }will be
  developed next by drawing inspiration from the connections between
  TISO, the \emph{least mean squares} (LMS) algorithm, and the \emph{recursive least squares} (RLS)
  algorithm \cite{sayed2003}. Indeed, observe that  TISO
  generalizes LMS, which is recovered for $\lambda\!=\!0$. To speed up
  convergence and reduce variability in the output of LMS, it is
  customary to resort to RLS, which accommodates the received data in
  a more sophisticated fashion, allowing to control the influence of
  each data vector on future estimates through forgetting factors.
\par
  \myitem\cmt{average loss function}Along these lines, the
  trick is to replace the \emph{instantaneous} loss
  $\ell_t^{(n)}(\bm a_n)$ in \eqref{eq:comidlossdef} with a
  \emph{running average} loss.  To maintain tracking capabilities,  a
  heavier weight is assigned to recent data
  using the exponential window customarily adopted by
  RLS. Specifically, consider setting
  $ \rev {\ftn}(\bm a_n) \!= \!\ltn(\bm a_n)$ in
  \eqref{eq:comidlossdef} with
  \begin{equation} \textstyle \label {eq:recloss}
    \ltn(\bm a_n) \define\mu\sum_{\tau=P}^{t}  \gamma^{t-\tau}\ell_{\tau}^{(n)}(\bm a_n),
  \end{equation}
  where
  \begin{myitemize}%
    \myitem $\gamma\! \in \!(0,1)$ is the user-selected forgetting factor
    and $\mu\! = \!1- \gamma$  is set to
    normalize the exponential weighting window, i.e.,
    $\mu\sum_{\tau=0}^\infty  \gamma ^\tau\! = \!1$.
  \end{myitemize}%
\end{myitemize}%
\end{myitemize}%
\par
\cmt{Derivation TIRSO update}Having specified a loss function, the
next step is to derive the update equation.
\begin{myitemize}%
  \myitem\cmt{raw application of COMID}In a direct
  application of COMID to solve \eqref{eq:compositeObjective} with $\rev{ \ftn }(\bm a_n)\!  = \!\ltn(\bm a_n)$, each iteration would involve the
  evaluation of the gradient of the $t\!-\!P\!+\!1$ terms  of $\ltn$. The computational complexity per iteration would grow with $t$ and, therefore, the resulting updates would
  not make up a truly online algorithm according to the requirements expressed in Sec.~\ref{sec:onlineopt}.
  \myitem \cmt{Introduction of $\bm \Phi, \bm r$}To remedy this issue, the structure of \eqref{eq:recloss} will be exploited next to develop an  algorithm with
  constant memory and complexity per iteration. To this end,  expand
  and rewrite \eqref{eq:recloss} to obtain
  \vspace{-2mm}
  \begin{align}
    \tilde \ell_{t}^{(n)}\!(\bm a_n)\!=&\frac{\mu }{2} \!  \sum_{\tau=P}^{t}\! \gamma^{t-\tau}  \big ( y_n^2[\tau]\!+\!\bm a_n^\top \bm g[\tau] \bm g^\top\![\tau] \bm a_n\!\!
    -\!2 y_n[\tau] \bm g^\top \![\tau]  \bm a_n \big ) \nonumber \\
     =& \frac{1}{2}\bm a_n^\top \bm \Phi[t] \bm a_n \!-\!\bm r_n^\top[t]  \bm a_n \!+\! \frac{\mu}{2}  \sum_{\tau=P}^{t}  \gamma^{t-\tau} y_n^2[\tau], \label{eq:recloss3}
  \end{align}
  where
    \begin{subequations}
    \begin{equation} \textstyle
      \label{eq:tirsocov}
      \bm \Phi[t]\triangleq \mu\sum_{\tau=P}^{t}  \,  \gamma^{t-\tau} \bm g[\tau] \, \bm g^\top[\tau],
    \end{equation}  
    \begin{equation} \label{eq:tirsocrosscov}
    \textstyle
      \bm r_n[t]\triangleq  \mu \sum_{\tau=P}^{t} \,  \gamma^{t-\tau} y_n[\tau] \,  \bm g[\tau].
    \end{equation}
\end{subequations}
    The variables $\bm \Phi[t]$ and $\bm r_n[t]$ can be respectively thought of as a weighted sample
    autocorrelation matrix and a weighted sample cross-correlation
    vector. The key observation here is that, as occurs in RLS, these
    quantities can be updated recursively as $\bm \Phi[t]= \gamma \,
    \bm \Phi[t-1]+\mu \,\bm 	g[t] \, \bm g^\top[t]$ and $\bm
    r_n[t]= \gamma \, \bm r_n[t-1]+\mu\, y_n[t]\, \bm g[t]$.
  \myitem \cmt{TIRSO update}Noting that
  \vspace{-1mm}
      \begin{equation} \label{eq:tirsograd}
      	\nabla \tilde \ell_{t}^{(n)}(\bm a_n)= \bm \Phi[t] \bm a_n-\bm r_n[t],
      \end{equation}
  and letting $ \tbm v_n[t]  \triangleq [\tbm v_{n,1}^\top[t],
  \ldots, \tbm v_{n,N}^\top[t] ]^\top \!\define\! \nabla \tilde
  \ell_{t}^{(n)}(\bm a_n[t])$, \rev{the estimate ${\tbm a}_n[t+1]$ after receiving $\{\bm
  y[\tau]\}_{\tau=0}^t$ becomes
  \begin{align} \textstyle \label{eq:TIRSOupdateexp}
  {\tbm a}_n[t+1]&=\underset{\tbm a_n }{\arg \min} ~ \tilde J\rev{_{t}^{(n)}}(\tbm a_n),
    \vspace{-2mm}
  \end{align}
  \vspace{-2mm}
  where
  \begin{multline}\label{eq:tirsoupdateobj}
  \tilde J\rev{_{t}^{(n)}}(\tbm a_n)\define \tbm v_n^\top[t] (\tbm a_n-\tbm a_n[t])+  \frac{1}{2\alpha_t} \left \lVert \tbm a_n- \tbm a_n[t] \right \rVert_2^2  \\
   +\lambda\textstyle \sum_{\substack { n'=1, n' \neq n}}^{N} \left \lVert \tbm a_{n,n'}\right\rVert_2.		
  \end{multline}
}
   Proceeding similarly to Sec.~\ref{ss:tiso}
yields the update
\begin{equation}  \label {eq:comidgroupsoltirso}
      {\tbm a}_{n,n'}[t+1]= \tbm a^\text{f}_{n,n'}[t ]\bigg [1-\frac{\alpha_t \lambda ~ \mathds 1 \{ n \neq n'\}}{\left \lVert\tbm a^\text{f}_{n,n'}[t ] \right\rVert_2}\bigg ]_+,
    \end{equation}
    where $\tbm a^\text{f}_{n,n'}[t] \!\triangleq\! \tbm
    a_{n,n'}[t]\!-\! \alpha_t \tbm v_{n,n'}[t]$.
\cmt{Overall algorithm}Due to the recursive
    nature of the updates for $\bm \Phi[t]$ and $\bm r_n[t]$, the
    resulting algorithm is termed \emph{Topology Identification via Recursive Sparse Online optimization} (TIRSO) and tabulated as \textbf{Procedure~\ref{alg:TIRSO}}.

\rev{
The choice of the step size affects the convergence properties of TIRSO, as analyzed in Sec.~\ref{sec:theoresults}. 
 Regarding step size selection, natural choices include
  \begin{myitemize}%
    \myitem \cmt{Constant}(i) constant step size, which is convenient in
    dynamic setups where changes in the coefficients
    $\mathcal{A}$ need to be tracked over time (see \thref{th:dynamicregretbound})
    but also gives rise to performance guarantees in static scenarios (\thref{prop:regrettiso} and \thref{prop:regrettirso} in the supplementary material);
    \myitem \cmt{Diminishing}(ii) diminishing step size, commonly in
    the form of  $\mathcal{O}(1/\sqrt{t})$ or $\mathcal{O}(1/t)$ (see \thref{th:strongconvexitytirso}); 
    \myitem \cmt{Adaptive}or (iii) an adaptive step size that depends
    on the data, as discussed at the end of
 Sec.~\ref{sec:dynamicregret}. 
  \end{myitemize}%
}
\end{myitemize}%

\begin{myitemize}%
    \myitem\cmt{corr. updates}Observe that  $\bm \Phi [t]$ only needs to be updated once per observed sample $t$, \rev{whereas the vector  $\bm r_n[t]$ need to be updated for each $n$ at every $t$.}
    \myitem \cmt{Computational Complexity}The computational complexity
    is dominated by step 7, which is $\mathcal O(N^3P^2)$ 
    operations per $t$.  \rev{However, exploiting the group-sparse structure of $ {\tbm a}_n[t]$ may reduce the computation by disregarding the columns of $\bm \Phi [t]$  corresponding to the zero entries of $\tbm a_{n}[t]$. If, for instance, the number of edges is $\mathcal O(N)$, then the complexity of TIRSO becomes  $\mathcal O(N^2P^2)$ per $t$. }%
Regarding memory complexity, TIRSO requires $N^2P^2$
memory positions to store $\bm \Phi[t]$ and $N^2P$ positions to store        $\{\bm r_n[t]\}_{n=1}^N$.
  \begin{algorithm}[t]
    \caption{Topology Identification via Recursive Sparse Online optimization (TIRSO)}\label{alg:TIRSO}
    \textbf{Input:} $\gamma, \mu, P, \lambda, \sigma^2, \{\alpha_t\}_t, \{\bm y[\tau]\}_{\tau =0}^{P-1} $ \\
    \textbf{Output:} $\{  {\tbm a}_n[t]\}_{n=1}^N, t=P+1,...$ \\
    \textbf{Initialization:} $  {\tbm a}_n[P]=\bm 0_{NP}, n=\!1,...,N, \, \bm \Phi[P-1]\!= \!\sigma ^2 \bm I_{NP}\\  \bm r_n[t]=\bm 0_{NP}, n=1,...,N$
    \begin{algorithmic}[1] 
      \For {$t=P,P+1, \ldots $} 
      \State {Receive data vector  $\bm y[t]$}
      \State {Form $\bm g[t] $} via \eqref{eq:g} 
      \State $\bm \Phi[t]= \gamma \, \bm \Phi[t-1]+\mu \,\bm g[t] \, \bm g^\top[t]$ 
      \For {$n=1, \ldots,N $} 
      \State $\bm r_n[t]= \gamma \, \bm r_n[t-1]+\mu\, y_n[t]\, \bm g[t]$ 
      \State $\tbm v_n[t]= \bm \Phi[t] \, {\tbm a}_n[t] -\bm r_n[t]$ 
      \For {$n'=1,2, \ldots ,N $ } 
      \State  $\tbm a^\text{f}_{n,n'}[t] = \tbm a_{n,n'}[t]- \alpha_t \tbm v_{n,n'}[t]$
      \State {Compute $ {\tbm a}_{n,n'}[t+1]$ via \eqref {eq:comidgroupsoltirso} }
      \EndFor
      \State \textbf{end for}
      \EndFor
      \State \textbf{end for}
      \EndFor
      \State \textbf{end for}
    \end{algorithmic}
  \end{algorithm}
\end{myitemize}%
\section{Theoretical Results} \label{sec:theoresults}
\rev{
In this section, the performance of TISO and TIRSO is analyzed.
The upcoming results will make use of one or more of the following assumptions:
\begin{enumerate}[{A}1.]
	\item {\textit{Bounded samples:} There exists $\energybound\!>0\!$
	such that $|y_n[t]|^2 \leq \energybound ~\forall \, n, t$. } \label{as:boundedprocess}
	\item {\textit{Bounded minimum eigenvalue of $\bm \Phi[t]$:}
	There exists $\strongcvxparf~ >~0$ such that  $\lambda_{\mathrm{min}}(\bm \Phi[t]) \geq \strongcvxparf , ~\forall \, t \geq P$.} \label{as:mineig}
	\item {\textit{Bounded maximum eigenvalue of $\bm \Phi[t]$:}
	There exists  $L~>~0$ such that $\lambda_{\mathrm{max}}(\bm \Phi[t]) \leq L, ~\forall \, t \geq P$. }\label{as:maxeig}
	\item {\textit{Asymptotically invertible sample covariance:} There exists~$T_m$ and $ \beta$ such that
		\vspace{-2mm}
		\begin{equation} \label{eq:tisocov}
		\textstyle
		\lambda_{\mathrm{min}} \left (\frac{1}{t-P}  \sum_{\tau=P}^{t}  \bm g[\tau] \, \bm g^\top[\tau] \right ) \geq  \beta~~ \forall \, t \geq T_m.
		\end{equation}%
	\label{as:tisocov}}%
\end{enumerate}%
}%
\vspace{-4mm}
\rev{
\cmt{Hypotheses}Note that A\ref{as:boundedprocess}
 entails no loss of generality  in real-world
applications, where
\begin{myitemize}%
  \myitem\cmt{bounded}
  \rev{data are bounded and thus $\energybound$
   necessarily exists.  \myitem\cmt{covariance}A\ref{as:mineig}
  usually holds in practice unless the data is redundant, meaning that
  some time series can be obtained as a linear combination of the
  others. In general, the latter will not be the case e.g.  if the
  data $\{\bm y[t]\}_t$ adheres to a continuous probability
  distribution, in which case  $\bm \Phi [t]$ is positive definite for
  all $t\geq P$ with probability 1.
  A\ref{as:maxeig} will also hold in practice since it can be shown
  that it is implied by A\ref{as:boundedprocess}. In particular, if
  A\ref{as:boundedprocess} holds, then    A\ref{as:maxeig} holds with
  $L = PN\energybound$. Similarly, A\ref{as:tisocov} will also
  generally hold since it is a weaker version of A\ref{as:mineig}. 
} 
\end{myitemize}%
}%
\par 
\rev{
Next, the asymptotic equivalence of the batch solutions for TISO and TIRSO is established.
}%
\vspace{-2mm}
\subsection{Asymptotic Equivalence between TISO and TIRSO}
\cmt{Asymptotic equivalence TISO-TIRSO}To complement the \rev {arguments given in Sec.~\ref{ss:tirso}} to support the decision of
setting $ \rev {\ftn}(\bm a_n)\!\! = \!\!\ltn(\bm a_n)$, which laid
the grounds to develop TIRSO, we establish that the batch problems that TISO and TIRSO implicitly solve become asymptotically equivalent as $T\!\rightarrow\!\infty$.
\begin{myitemize}%
  \myitem \cmt{Defs.}To this end, let  $\bm a_n^{*}[T]$ denote the hindsight solution for TISO, which is given by
   \vspace{-2mm}
  \begin{eqnarray} \label{eq:minimizertiso}
    \bm a_n^{*}[T]\!=\!\underset{\bm
    a_n}{\arg\min}~\tisohindsightobj_T(\bm a_n),
  \end{eqnarray}
  \vspace{-4mm}
  where
  \vspace{2mm}
  \begin{eqnarray}\label{eq:tisohindsightobjdef}
  \textstyle
    \tisohindsightobj_T(\bm a_n)\!\define\!	\frac{1}{T-P}\!\sum_{t=P}^{T-1} \!\Big[ \ell_{t}^{(n)}(\bm a_n)\!+ \!\!\lambda \!\sum_{\substack { n'=1, n' \neq n}}^{N} \!\left \lVert \bm a_{n,n'}\right\rVert_2 \!\Big ].
  \end{eqnarray}
 Observe that \eqref{eq:tisohindsightobjdef} is identical to the
 objective in the batch criterion \eqref{eq:prob-sep}. Likewise,
let $\tbm a^*_n[T]$ denote the hindsight solution of TIRSO, which is
given  by
  \vspace{-2mm}
  \begin{eqnarray} \label {eq:minimizertirso}
    \tbm a^*_n[T]=\underset{\bm a_n}{\arg\min}
    ~~ \tirsohindsightobj_T(\bm a_n)
      \vspace{-2mm}
  \end{eqnarray}%
\vspace{-2mm}with
 \vspace{2mm}
  \begin{eqnarray}    \label{eq:tirsohindsightobjdef}
  \textstyle
    \tirsohindsightobj_T(\bm a_n)\!\define\!\!\frac{1}{T-P}\!\sum_{t=P}^{T-1}\! \Big   [\tilde \ell_{t}^{(n)}(\bm a_n)\!+ \! \lambda\!\sum_{\substack { n'=1, n' \neq n}}^{N} \!\left \lVert \bm a_{n,n'}\right\rVert_2 \! \Big ].
  \end{eqnarray}
\myitem\cmt{relevance}In this case, \eqref{eq:tirsohindsightobjdef} no longer coincides
  with the objective in \eqref{eq:prob-sep}. Therefore, one can argue that the TIRSO algorithm is not pursuing the estimates that
  minimize the batch criterion \eqref{eq:prob-sep}. This idea is
  dispelled next by establishing the asymptotic equivalence between
  minimizing $ \tirsohindsightobj_T(\bm a_n)$ and minimizing
  $ \tisohindsightobj_T(\bm a_n)$, since the latter is identical to
  \eqref{eq:prob-sep}.

\myitem\cmt{result}%
  \begin{theorem}
                          \thlabel{prop:asymptoticequivalence}
    \cmt{hypothesis}%
     \rev {Under assumption A\ref{as:boundedprocess}}:
    \begin{enumerate}%
    \item\cmt{pointw. convergence}It	holds for all $\bm a_n$ that
      $ \displaystyle \label{eq:pointwiseconvergence}
        \lim_{T\rightarrow\infty}|\tisohindsightobj_T(\bm
        a_n) - \tirsohindsightobj_T(\bm a_n)| = 0.
      $
    \item\cmt{convergence of minima}It holds that
      $  \displaystyle
        \lim_{T\rightarrow\infty}\big |\inf_{\bm a_n}\tisohindsightobj_T(\bm	a_n) - \inf_{\bm a_n}\tirsohindsightobj_T(\bm a_n)\big| = 0.
      $
    \item\cmt{convergence of minimizers}If, additionally, \rev{assumption A\ref{as:mineig} holds,}
       then
       $
        \lim_{T\rightarrow\infty}\|
        \bm a_n^{*}[T] -\tbm a_n^{*}[T] \|_2 = 0.
      $
    \end{enumerate}
  \end{theorem}
  \begin{IEEEproof}
See Appendix~\ref{sec:proof:asymptoticequivalence} in the supplementary material.
  \end{IEEEproof}
\cmt{Interpretation}%
\begin{myitemize}%
\myitem\cmt{conclusions}\thref{prop:asymptoticequivalence} essentially
establishes not only that the TISO and TIRSO hindsight objectives
are asymptotically the same but also that their minima and minimizers
asymptotically coincide. Since the TISO hindsight objective equals the batch
objective  \eqref{eq:prob-sep}, it follows that the TIRSO hindsight
objective asymptotically approaches the batch objective
\eqref{eq:prob-sep}. This observation is very important since  the
regret analysis from Sec.~\ref{sec:staticregret} will establish that the
TISO and TIRSO estimates asymptotically match their hindsight
counterparts.
\end{myitemize}%
\end{myitemize}%
\vspace{-1mm}
\subsection{Static Regret Analysis} \label{sec:staticregret}
\cmt{overview}This section characterizes the performance of TISO and TIRSO analytically. Specifically, it is shown that the sequences of estimates produced by these algorithms yield a sublinear static regret, which is a basic requirement in online optimization; see
Sec.~\ref{sec:onlineopt}. Broadly speaking, this property means that, on average and asymptotically, the online estimates perform as well  as their hindsight counterparts.
\par
\cmt{regret}\rev{A general definition of the regret metric is given
in \eqref{eq:regret_def}. Since the problem at hand is separable
across nodes, it is natural to separately quantify the regret for each node. The
total regret will be the sum of the regret for all nodes. Applying
this idea
and shifting the time index to simplify notation, one can
 replace $R_{\rev{s}}[T_0]$  in \eqref{eq:regret_def} with $R_s^{(n)}[T_0+P-1]$, function $h_t$ with
$h_{t+P}^{(n)}$, and $T_0$ with $T-P+1$ to  write the regret of TISO
for the $n$-th node at time $T$ as
	\vspace{-1mm}
\begin{equation} \label{eq:defstaticregretTISO}
R_s^{(n)}[T]\triangleq \sum_{t=P}^{T}\big [ h_t^{(n)}(\bm a_n[t])- h_t^{(n)}(\bm a_n^*[T])\big  ],
\end{equation}
 where $h_t^{(n)}(\cdot) = \rev{ \ell_t^{(n)} }(\cdot)+\rev
 {\nReg}(\cdot)$ and $\bm a_n^*[T]$
 is defined in \eqref {eq:minimizertiso}. For TIRSO, the regret for the $n$-th node is given by
 \vspace{-1mm}
\begin{equation} \label{eq:defstaticregretTIRSO}
\tilde R_s^{(n)}[T]\triangleq \sum_{t=P}^{T}\big [ \tilde h_t^{(n)}(\tbm a_n[t])- \tilde h_t^{(n)}(\tbm a_n^*[T])\big  ],
\end{equation}
  where $\tilde h_t^{(n)}(\cdot) = \rev{ \ltn }(\cdot)+\rev {\nReg}(\cdot)$
and $\tbm a_n^*[T]$ is defined in \eqref {eq:minimizertirso}.}%
\par 
\rev{
Since constant step size sequences allow tracking time-varying
topologies, one could think of seeking a sublinear bound for the
regret. However, it is easy to see (cf. \eqref{eq:comidsol2}
and \eqref{eq:comidgroupsol} in the case of TISO) that the sequences
of estimates in this case are generally noisy, unless the innovation
process $\bm u[t]$ in \eqref{eq:model} is $\bm 0_N$. For this reason, a
sublinear regret bound cannot be obtained for a constant
$\alpha_t$. However, it is possible to establish sublinear regret when
the step size is ``asymptotically constant,''  as described next.}

\rev{The idea is to  run the selected algorithm in time windows of exponentially
increasing length with a step size that differs across windows but is
constant within each one.  Specifically, let the $(m+1)$-th window,
$m=1,\ldots, M$, comprise the time indices $t_0 2^{m-1} < t \leq t_0
2^m$ for some user-selected $t_0\geq P$. Set
$\alpha_t= \alpha\winnot{m}$ for those $t$ satisfying $t_0 2^{m-1} <
t \leq t_0 2^m$.  The following result proves sublinear regret
 for TISO.}
\rev{
\begin{theorem} \thlabel{cor:doublingtricktiso}
	Let $\{\bm a_n[t]\}_{t=P}^T$ be generated by applying TISO
	(\textbf{Procedure~\ref{alg:TISO}}) 
	with step size $\alpha_t=\alpha\winnot{m}=\mathcal O(1/\sqrt{t_0
	2^{m-1}})$ in the window $t_0 2^{m-1} < t \leq t_0
2^m$, $m=1,2,\ldots$
        Then, the regret of TISO under assumptions A\ref{as:boundedprocess} and A\ref{as:tisocov} is
        \vspace{-2mm}
	\begin{equation}\label{eq:sublinearafterdoublingtiso}
	R_s^{(n)}[T] = \mathcal O\left(PN\energybound B_{\bm a}^2 \sqrt{T} \right ),
	\end{equation}
\rev {where $B_{\bm a}=1/\beta  (\energybound \sqrt{PN} + \sqrt{\energybound^2PN+\beta \energybound})$. }
\end{theorem}
\begin{IEEEproof} See Appendix~\ref{proof:doublingtricktiso} in the
        supplementary material.
\end{IEEEproof}
}

\par
 \cmt{Regret Bound}Similarly, the regret of TIRSO is characterized as follows:
\rev {
	\begin{theorem} \thlabel{cor:doublingtricktirso}
		Let $\{\tbm a_n[t]\}_{t=P}^T$ be generated by applying TIRSO
		(\textbf{Procedure~\ref{alg:TIRSO}}) 
	with step size $\alpha_t=\alpha\winnot{m}=\mathcal O(1/\sqrt{t_0
	2^{m-1}})$ in the window $t_0 2^{m-1} < t \leq t_0
2^m$, $m=1,2,\ldots$
Then, the regret of TIRSO under assumptions A\ref{as:boundedprocess}, A\ref{as:mineig}, and A\ref{as:maxeig}, is
		\begin{equation}\label{eq:sublinearafterdoublingtirso}
		R_s^{(n)}[T] = \mathcal O\left( L B_{\tbm a}^2 \sqrt{T} \right ), 
		\end{equation}
                \rev {where $ B_{\tbm a}\define1/\strongcvxparf  (\energybound \sqrt{PN} + \sqrt{\energybound^2PN+\strongcvxparf \energybound})$. }
	\end{theorem}
\begin{IEEEproof}
See Appendix~\ref{proof:doublingtricktirso} in the supplementary material.
\end{IEEEproof}
} 
\cmt{Interpretation}\thref{cor:doublingtricktirso} has the same form as
\thref{cor:doublingtricktiso} with the exception of  \eqref{eq:regrettirso},
where the constant term multiplying $\sqrt{T}$ differs from the one in
\eqref{eq:sublinearafterdoublingtiso}. However,   \rev{it  can be readily shown
that  $L \leq PN\energybound$, which implies that TIRSO 
also satisfies \eqref{eq:sublinearafterdoublingtiso}.}

\cmt{Summary}To sum up, both TISO and TIRSO behave asymptotically in
the same fashion and provide, on average, the same performance as the
hindsight solution of TISO, which coincides with the batch solution in
\eqref{eq:prob-sep}. The difference between TISO and TIRSO is,
therefore, in the non-asymptotic regime, where TIRSO can
track changes in the estimated graph more swiftly than TISO. This
is at the expense of a slight increase in the number of arithmetic
operations and required memory. Note, however, that TIRSO offers an
additional degree of freedom through the selection of the forgetting
factor $\gamma$. This enables the user to select the desired point in
the trade-off between adaptability to changes and low variability in
the estimates.
\par
\rev {
\cmt{Logarithmic regret bounds for TIRSO}As demonstrated next, tighter regret bounds can be
obtained when a diminishing
step size sequence is adopted. Such sequences are of special interest
when 
the VAR coefficients 
 do not change over time. Even in this
scenario, the application of
online algorithms such as TISO or TIRSO is well-motivated when the
number or dimension of the data vectors is prohibitively large to
tackle with a batch algorithm.}
\rev{
\begin{theorem} \thlabel{th:strongconvexitytirso}
	\cmt{Hypotheses}%
	\begin{myitemize}%
		\myitem \cmt{Assumptions}Under assumptions A\ref{as:boundedprocess}, A\ref{as:mineig}, and A\ref{as:maxeig},
		\myitem \cmt{TIRSO }let $\{\tbm a_n[t]\}_{t=P}^{T}$ be generated by TIRSO (\textbf{Procedure~\ref{alg:TIRSO}}) with
		\myitem \cmt{decreasing step-size}$\alpha_t=1/(\strongcvxparf t)$.
	\end{myitemize}%
Then, the static regret of TIRSO satisfies
\vspace{-2mm}
\begin{equation}
	\tilde R_s^{(n)}[T]\leq  \frac{\gradboundf^2}{2\strongcvxparf} \left (\mathrm{log}(T-P+1)+1\right ) +\frac{1}{2\alpha_{P-1}} B_{\tbm a}^2,
\end{equation}
where $\gradboundf \triangleq
(1+ \kappa_{\bm \Phi}) \sqrt{PN}\energybound$ with
$\kappa_{\bm \Phi}=L/\strongcvxparf$ and $B_{\tbm a}$ is defined in
\thref{cor:doublingtricktirso}.
\end{theorem}
\begin{IEEEproof}
	See Appendix~\ref{appendix:proof{th:strongconvexitytirso}} in the supplementary material.
\end{IEEEproof}
}
\rev{Next, we analyze the performance of TIRSO in dynamic environments.}
	\vspace{-2mm}
\rev{
\subsection{Dynamic Regret Analysis of TIRSO} \label{sec:dynamicregret}
\cmt{Dynamic regret}%
In this section, the performance of TIRSO is analyzed in dynamic settings. Specifically, a dynamic regret bound is derived for TIRSO, and its steady-state tracking error in dynamic scenarios is also discussed.}
\begin{myitemize}%
	\rev{
	\myitem \cmt{Definition of dynamic regret}To characterize the performance of TIRSO in dynamic setups, the dynamic regret is defined as:
	\begin{equation} \label{eq:defdynamicregret}
	\textstyle
		\tilde R_d^{(n)}[T]\triangleq  \sum_{t=P}^{T}\big [ \tilde h_t^{(n)}(\tbm a_n[t])- \tilde h_t^{(n)}( \timevarhindsight[t])\big  ],
	\end{equation}
	where
	\begin{myitemize}%
		\myitem \cmt{estimate}$\tbm a_n[t]$ is the TIRSO estimate
		\myitem \cmt{optimal solution}and  $ \timevarhindsight[t] = \arg \min _{\tbm a_n} \tilde h_t^{(n)}(\tbm a_n)$. 
	\end{myitemize}}%
	\begin{myitemize}%
			\rev{
		\myitem \cmt{Relation to static regret}
The dynamic regret in \eqref{eq:defdynamicregret} compares the
estimate $\tbm a_n[t]$ with $ \timevarhindsight[t] $ in terms of the
metric  $\tilde h_t^{(n)}(\cdot)$. As opposed to
	$ \timevarhindsight[t] $, estimate  $\tbm a_n[t]$ does
	not ``know'' $\tilde h_t^{(n)}(\cdot)$ since $\tbm a_n[t]$ is
	obtained  from $\{\bm y[\tau]\}_{\tau<t}$ whereas   $\tilde
	h_t^{(n)}(\cdot)$  depends on both  $\{\bm y[\tau]\}_{\tau<t}$
	and $\bm y[t]$. This means that the dynamic regret captures
	the ability of an algorithm to attain small \emph{future}
	residuals. Furthermore, note that comparing with
	$ \timevarhindsight[t]$ is highly meaningful in the present
	case since, by definition, $ \timevarhindsight[t] = \arg \min
	_{\tbm
	a_n} \mu\sum_{\tau=P}^{t}  \gamma^{t-\tau}\ell_{\tau}^{(n)}(\tbm
	a_n) + \lambda \textstyle \sum_{\substack { n'=1,
  n' \neq n}}^{N} \left \lVert \tbm a_{n,n'}\right\rVert_2 $, which
	therefore minimizes 
	a version of the batch \eqref{eq:prob-sep} or hindsight \eqref {eq:minimizertirso} objectives where
	the more recent residuals are weighted more heavily. Thus,
	$ \timevarhindsight[t]$ constitutes a significant estimator in
	dynamic setups and therefore the dynamic regret also
	quantifies the ability of an estimator to track changes.}
\par 
\rev{
		It can be easily shown that the static regret is
		upper-bounded by the dynamic regret. The dynamic
		regret in \eqref {eq:defdynamicregret}  would coincide
		with the static regret if  $ \timevarhindsight[t]$
		were replaced with  $\arg \min _{\tbm a_n} \sum
		_{t=P}^T \tilde h_t^{(n)}(\tbm a_n)$.
                Attaining a low dynamic regret is therefore more
		challenging because the estimator under consideration
		is compared with a \emph{time-varying} reference. 
		\myitem \cmt{Path length and dynamic regret bound}
                This implies that a sublinear dynamic regret may not
		be attained if this time-varying reference changes too
		rapidly, which generally occurs when the tracked
		parameters vary too quickly.
For this reason, the dynamic regret  is commonly upper-bounded in
		terms of the cumulative distance between two
		consecutive instantaneous optimal solutions, known as \emph {path length}:
		\begin{equation} \label {eq:pathlength}
		\textstyle
			W^{(n)}[T] \triangleq \sum_{t=P+1}^T \left \lVert \timevarhindsight[t] - \timevarhindsight[t-1] \right \rVert_2.
		\end{equation}
	}
		\end{myitemize}%
	\rev{
	Next, we  bound  the dynamic regret of TIRSO.}
\end{myitemize}%
\rev{
\cmt{Theorem}%
\begin{myitemize}%
	\myitem \cmt{Theorem statement}%
	\begin{theorem} \thlabel{th:dynamicregretbound}
	    	\begin{myitemize}%
			\myitem \cmt{assumptions}Under assumptions A\ref{as:boundedprocess}, A\ref{as:mineig}, and A\ref{as:maxeig},
			\myitem \cmt{TIRSO }let $\{\tbm a_n[t]\}_{t=P}^{T}$ be generated by TIRSO (\textbf{Procedure~\ref{alg:TIRSO}}) with
			\myitem \cmt{constant step size}a constant step size $\alpha \in  (0,1/L]$.
			\myitem \cmt{bounded variations}If there exists $\sigma$ such that
			\vspace{-2mm}
				\begin{equation} \label {eq:boundedvariationsassump1}
					\left \lVert \timevarhindsight[t] - \timevarhindsight[t-1] \right \rVert_2 \leq \sigma,~ \forall\,t\geq P+1,
				\end{equation}
	    	\end{myitemize}%
		then the dynamic regret of TIRSO satisfies:
		\vspace{-2mm}
		\begin{align}\label{eq:regretbound}
		&	\tilde R_d^{(n)}[T]\leq \\ & \frac{1}{\alpha \strongcvxparf}\left ( \left (1+\kappa_{\bm \Phi}\right )
			\sqrt{PN} \energybound+ \lambda N \right )  \big (\lVert 
			 \timevarhindsight[P] \rVert_2 + W^{(n)}[T]\big ), \nonumber
		\end{align}
		where $\kappa_{\bm \Phi}\define L/\strongcvxparf$.
	\end{theorem}
\begin{IEEEproof}
	See Appendix~\ref{appendix:proof{th:dynamicregretbound}} in the supplementary material.
\end{IEEEproof}
	\myitem \cmt{Interpretations and remarks on the dynamic regret bound}
	\revv{The derivation of the dynamic regret bound above relies on the strong convexity of \eqref{eq:recloss3}, and thus cannot be done for TISO.}  \revv{
	The derivation of a different dynamic regret bound in \cite{fosson2018onlineoptimization} relies on the strong convexity of the quadratic term of an elastic-net regularizer, which is not necessary here.}
	Several remarks about \thref{th:dynamicregretbound} are in order.
	\begin{myitemize}%
		\myitem \cmt{Path length}If the path length
	$W^{(n)}[T]$ is sublinear in $T$, then the dynamic regret is
	also sublinear in $T$. 
	\end{myitemize}%
\end{myitemize}}%
\rev {
\cmt{Bounding the error of each iterate}When the path length is not
	sublinear, the dynamic regret may not be sublinear, but we can still bound the \emph{steady-state error} under certain conditions:
\begin{myitemize}%
	\myitem \cmt{Theorem}%
	\begin{theorem} \thlabel {th:boundingerror}
		\begin{myitemize}%
                \myitem              \cmt{assumptions} Under assumptions A\ref{as:boundedprocess}, A\ref{as:mineig}, and A\ref{as:maxeig},
			\myitem \cmt{TIRSO }%
			let $\{\tbm a_n[t]\}_{t=P}^{T}$ be generated by TIRSO (\textbf{Procedure~\ref{alg:TIRSO}}) with
			\myitem \cmt{constant step size}a constant step size $\alpha \in  (0,1/L]$.
			\myitem 
			\myitem \cmt{bounded variations}If there exists $\sigma$ such that \eqref{eq:boundedvariationsassump1} holds,
	    	\end{myitemize}%
		then 
		\vspace{-2mm}
		\begin{equation} \label{eq:boundingerror}
		\limsup_{t \rightarrow \infty}  \left \lVert \tbm a_n[t]-\timevarhindsight [t] \right \rVert_2 \leq \frac{ \sigma}{ \alpha \strongcvxparf}.
		\end{equation}
	\end{theorem}
	\myitem \cmt{Proof}%
	\begin{IEEEproof}
		Following similar arguments as in the proof of \thref{th:dynamicregretbound}, \eqref {eq:boundingerror} follows by applying \cite[Lemma 4]{dixit2019onlineproximal}.
	\end{IEEEproof}
	\myitem \cmt{Interpretations}%
	\begin{myitemize}%
		\myitem \cmt{Steady state error}This theorem establishes that the steady-state error incurred by TIRSO with $\alpha \in  (0,1/L]$  in dynamic scenarios eventually becomes bounded, which shows its
		tracking capability in time-varying environments.
		\myitem \cmt{dependence on condition number }If $\alpha=1/L$, then the upper bound on the steady-state error becomes $\sigma \kappa_{\bm \Phi}$, where $\kappa_{\bm \Phi} \define L/ \strongcvxparf$  is an upper bound on the condition number of $\bm \Phi[t], \, t \geq P$. This clearly agrees with intuition.
	\end{myitemize}%
        \myitem\cmt{Adaptive step size}In practice, one may not know
        the value of $L$ and therefore selecting an $\alpha$
        guaranteed to be in $(0,1/L]$ would not be possible. In those cases,
        it makes sense to compute a running approximation of $L$ given
        by $\hat L_t = \max_{P\leq \tau \leq
        t} \lambda_{\mathrm{max}}(\bm \Phi[\tau])$ and adopt the
        approximately constant step size $\alpha_t = c/\hat L_t$,
        where $c\in(0,1]$. However, in setups where the true VAR parameters change over time, the $\max$ operation may lead the algorithm to use an overly pessimistic approximation of $L$. Thus, it may be preferable to directly adopt the \emph{adaptive step size} 
$\alpha_t = c/ \lambda_{\mathrm{max}}(\bm \Phi[t])$, as  analyzed in Sec.~\ref{sec:numericalresults}. 
\end{myitemize}%
}
\\
\rev{
\textbf{Remark.} None of the algorithms and analytical results in this
paper require any probabilistic assumption or mention to probability
theory\revv{, making our results fully compatible with the deterministic interpretation of the estimator at hand.}  This is because these results establish performance guarantees
for the proposed online algorithms relative to the batch estimator or
hindsight solutions. If one wished to obtain performance guarantees in
terms of \emph{probabilistic} metrics, such as consistency of the
estimators, probabilistic assumptions would of course be required.
For example, when $\lambda=0$, the batch estimator
in \eqref{eq:probl1} boils down to the ordinary least squares
estimator, which is consistent if the VAR process is stable and the
noise is standard white \cite[Lemma 3.1]{lutkepohl2005}. When
$\lambda>0$, consistency of \eqref{eq:probl1} is discussed in
\cite{bolstad2011groupsparse}. Remarkably, consistency of the VAR
coefficient estimates is not enough to ensure the correct
identification of the true graph.  Theorem 1 in
\cite{bolstad2011groupsparse} provides conditions that depend on the
true VAR parameters that guarantee that the graph is successfully
recovered.
}
\vspace{-3mm}
\section{Numerical Results and Analysis} \label{sec:numericalresults}	

	 \begin{myitemize}%
	 \myitem 	 \cmt{Introduction}Simulation tests for the
         proposed algorithms are performed on both synthetic and
         real data. \rev{All code will be made public at the authors' websites.}
		\begin{figure*}[t]
	\includegraphics[width=\linewidth]{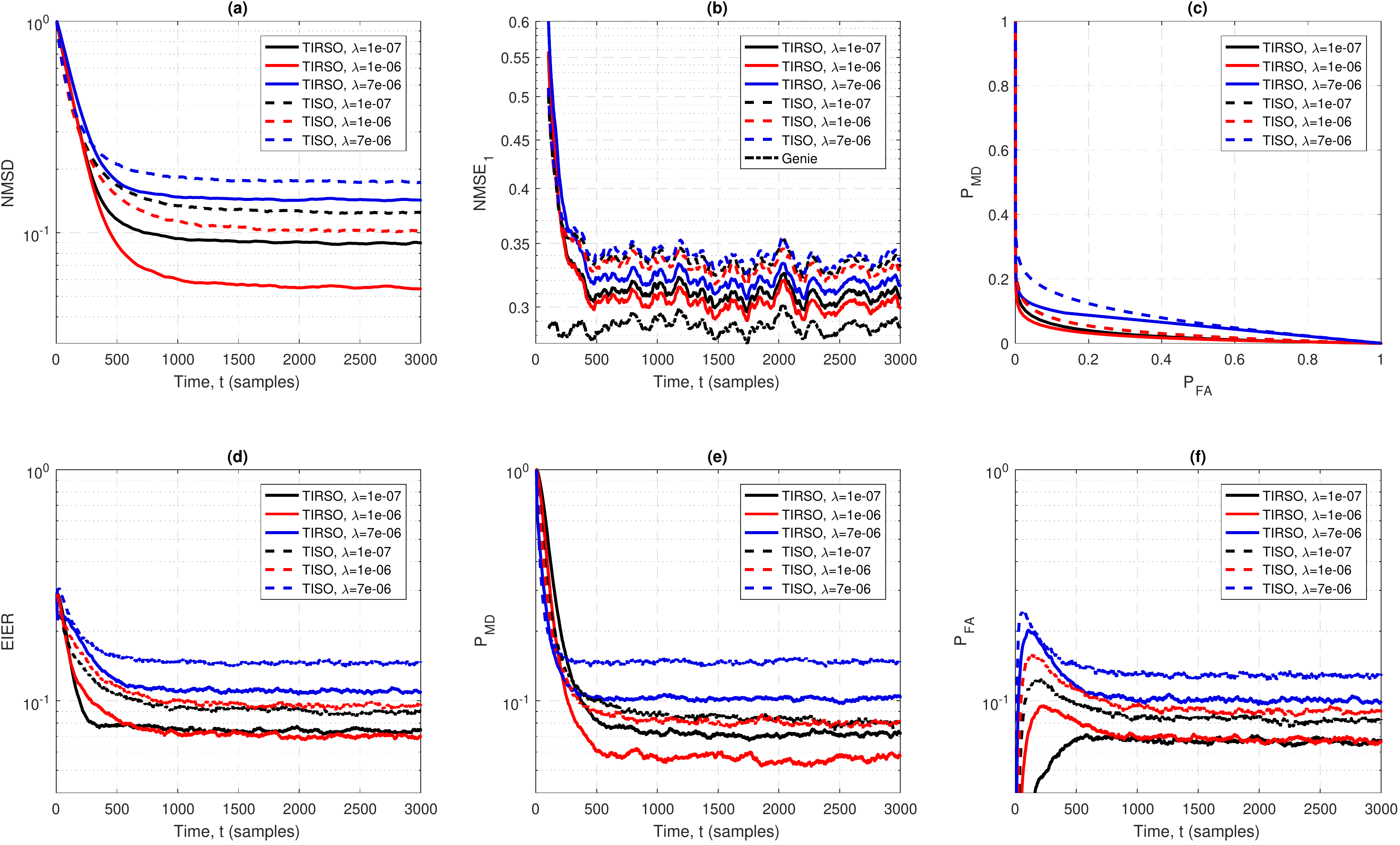}
	\caption{\small Performance of TISO and TIRSO on stationary
		time series for different degrees of sparsity-promoting
		regularization ($N = 12$, $P = 2$,
		$p_e=0.2$, $\sigma_u=0.005$, $\gamma= 0.99$, $T=3000$, 
		$T_1=500$, $T_2=3000$,  300  Monte Carlo
		runs).  
	}
	\label{fig:stationarysetting}
\end{figure*}
\begin{figure}
	\centering
	\includegraphics[width=0.9\linewidth]{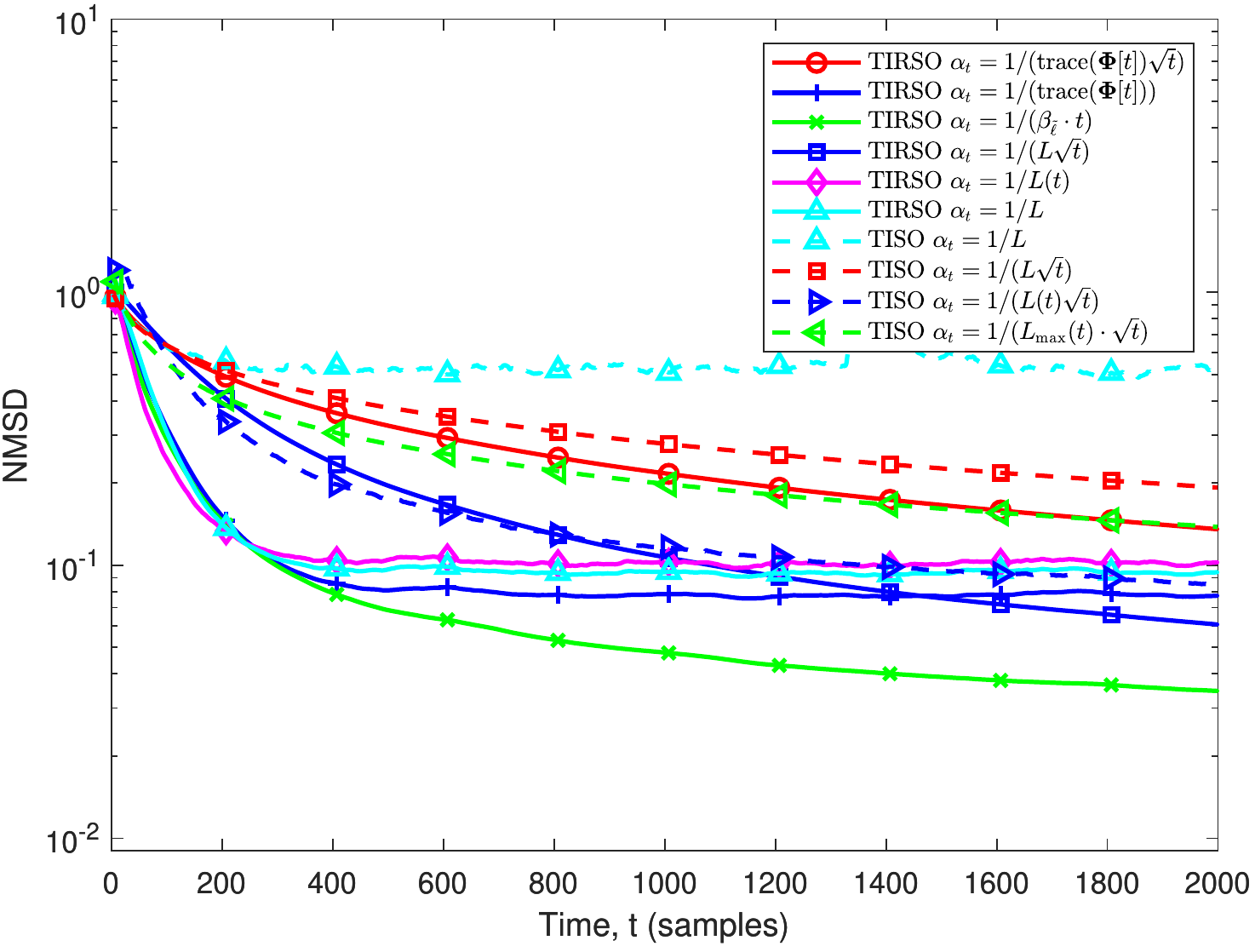}
	\caption{\rev {\small NMSD vs. time: comparison of TISO and TIRSO for various options of step sizes ($N = 10$, $P = 3$,
		$p_e=0.2$, $\sigma_u=0.1$, $\gamma= 0.99$, $\lambda=8\times 10^{-4}$,  $T=2000$,  50  Monte Carlo
		runs). Moreover, $ L_\mathrm{max}(t) := \mathrm{max}_{\tau = 1}^t L(\tau)$.}}%
	\label{fig:fig-stepsizetisotirso}%
\end{figure}%
\begin{figure}%
	\centering
	\includegraphics[width=0.9\linewidth]{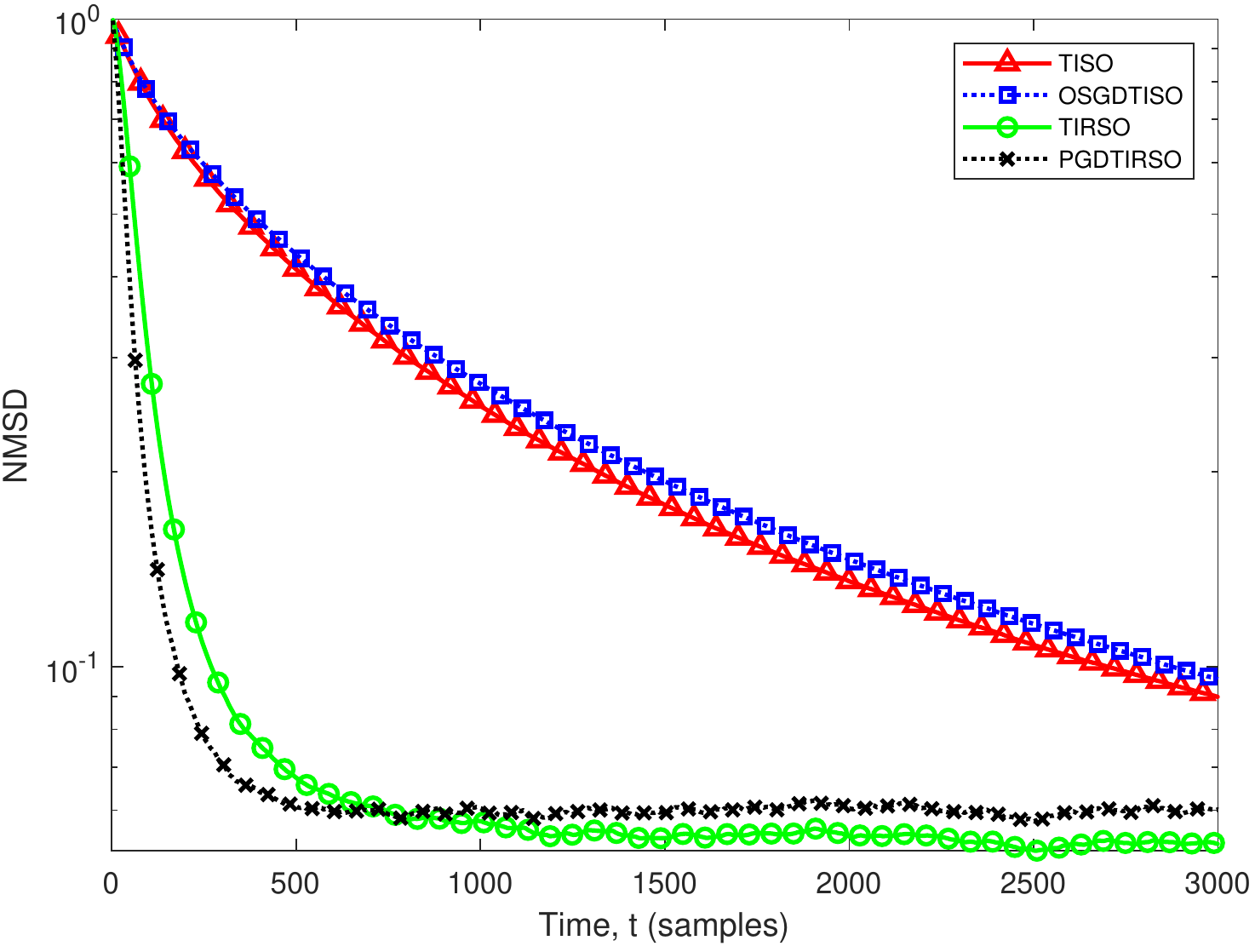}
	\caption{\rev{\small NMSD vs. time: comparison of TISO and TIRSO  with other algorithms. ($N = 10$, $P = 2$,
		$p_e=0.2$, $\sigma_u=0.01$, $\alpha_t=0.1/L$, $\gamma= 0.99$, $T=3000$, $K_\text{PGD}=5$, 200  Monte Carlo runs). The parameter $\lambda$ for each algorithm is selected based on minimum NMSD.}}
	\label{fig:comparisonosgpgd}
\end{figure}%
\begin{figure}
	\centering
	\includegraphics[width=\linewidth]{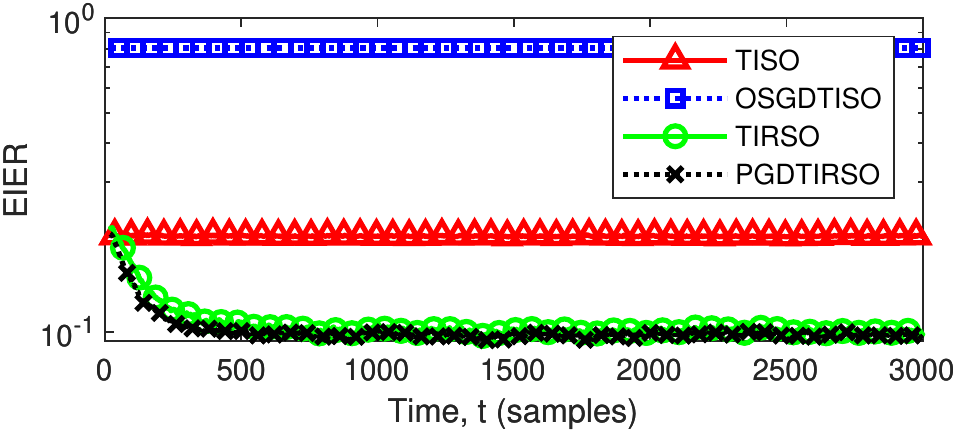}
	\caption{\rev{\small EIER vs time for $\delta=0$, same parameters as of Fig. \ref {fig:comparisonosgpgd}. The parameter $\lambda$ for each algorithm is selected based on minimum EIER. }}
	\label{fig:comparisonEIERTISOOSGD}%
\end{figure}%
\newcommand{\true}{^\text{true}}
         \\
\rev{The proposed algorithms are evaluated based on
         the performance metrics described next, where expectations are approximated by the Monte Carlo method.}
	 \myitem \cmt{performance metrics}%
	\begin{myitemize}%
            \myitem\cmt{synthetic data}For synthetic-data experiments,
          \begin{myitemize}%
          \myitem \cmt{NMSD}the normalized
          mean square deviation 
          \vspace{-2mm}
		\begin{equation} \label{eq:nmsd}
		\text{NMSD}[t] \define  \frac{\mathbb{E}\big [\textstyle\sum_{n=1}^N \lVert  \hat{\bm{a}}_n[t] - \bm{a}_n\true[t] \lVert_2^2\big ]}{\mathbb{E}\big [\textstyle\sum_{n=1}^N\lVert\bm{a}_n\true[t] \lVert_2^2\big]}
		\end{equation}
                measures the difference between the estimates $\{\hbm a_n[t]\}_{t}$ and the (possibly time-varying) true VAR coefficients $\{\bm a_n\true[t]\}_{t}$. 
		\myitem \cmt{Prob. of miss detection:}The ability
                to detect edges of the true VAR-causality graph is
                assessed using the probability of miss detection
                \vspace{-2mm}
                \begin{equation*} \label{eq:probMD}
                   P_\text{MD}[t] \define \frac{\sum_{n\neq n'}
                   \mathbb{E}\left[
                   \mathds{1}\{ \left \lVert\hbm a_{n,n'}[t]\right \rVert_2 <  \delta  \}
                   \mathds{1}\{ \left \lVert\bm a_{n,n'}\right \rVert_2 \neq 0  \}\right] }{\sum_{n\neq n'}
                     \mathbb{E}\left[
                   \mathds{1}\{ \left \lVert\bm a_{n,n'}\right \rVert_2 \neq 0  \} \right]  }
                \end{equation*}
for a given threshold $\delta$, which is the probability of not  identifying an edge that
actually exists,
		\myitem\cmt{Prob. of false alarm:}and the probability
                of false alarm 
\begin{align*} 
P_\text{FA}[t] \define \frac{\sum_{n\neq n'}
	\mathbb{E}\left[
	\mathds{1}\{ \left \lVert\hbm a_{n,n'}[t]\right \rVert_2 \geq  \delta  \}
	\mathds{1}\{ \left \lVert\bm a_{n,n'}\right \rVert_2 = 0  \}\right] }{\sum_{n\neq n'}
  \mathbb{E}\left[
	\mathds{1}\{ \left \lVert\bm a_{n,n'}\right \rVert_2 = 0  \} \right]  },
\end{align*}
which is the probability of detecting an edge that does not exist.
		\myitem \cmt{EIER:}Another relevant metric is the
                \emph{edge identification error rate} (EIER), which measures how
                many edges are misidentified relative to the number
                of possible edges \cite{giannakis2018nonlineartopologyid}:
 \vspace{-1mm}
		\begin{multline} \label{eq:EIER}
		\text {EIER}[t ]=
%
                   \textstyle \frac{1}{N(N-1)}   \textstyle\sum_{n'\neq n}\mathbb{E}
                  \big [ \big|\mathds{1}\{ \left \lVert\hbm a_{n,n'}[t]\right \rVert_2 \geq \delta  \} \\
                   \quad  - \mathds{1}\{ \left \lVert \bm a_{n,n'}\right \rVert _2 \neq 0   \}\big|\big ].
		\end{multline}
Note that self-loops are excluded in these metrics.
	\end{myitemize}%
\end{myitemize}%
\myitem\cmt{real data tests}%
\begin{myitemize}%
		\myitem \cmt{NMSE or $h$-step normalized prediction
                  error:}To quantify the forecasting  performance,
                define recursively the $h$-step ahead predictor given
                $\{\bm y[\tau]\}_{\tau\leq t}$ as:
                \begin{equation}
\label{eq:predictor}
 \textstyle
                                \hat {\bm y}[t+h|t] \triangleq \sum_{p=1}^{P}  \mathbf  {\hat{A}}_p[t] \hat {\bm y}\left[t+h-p\vert t\right],
                \end{equation}
                where $\{\mathbf  {\hat{A}}_p[t]\}_{p=1}^P$ are the estimated VAR coefficients at time $t$ and $\hat {\bm y}[t+j|t]=\bm y[t+j]$ for $j\leq 0$.
                The  $h$-step normalized mean square error is given by
                \vspace{-2mm}
		\begin{align} \label{eq:NMSE}
			\text{NMSE}_h[t] &=\frac{\mathbb{E}\big[\left  \lVert \bm y[t+h] - \hat{\bm y}[t+h|t] \right \rVert_2^2 \big]}{\mathbb{E}\left [\lVert \bm y[t+h]\lVert_2^2\right ]}.
		\end{align}
          \end{myitemize}%
\end{myitemize}%
The values of all parameters involved in the experiments are listed in
the captions and legends of the figures.
\vspace{-4mm}
	\subsection{Synthetic Data Tests} \label{sec:results}
Throughout this section, unless otherwise stated, the expectations in \eqref{eq:nmsd}
to \eqref{eq:NMSE} are taken with respect to realizations of the
graph, VAR parameters, and innovation process $\bm u[t]$. Similarly,
the  step size is set to $\alpha_t= 1/(4 \lambda_{\max} (\bm \Phi [t]))$; \rev{see Sec.~\ref{sec:dynamicregret}. The regularization parameter is selected to approximately minimize NMSD.}
        \subsubsection{Stationary VAR Processes}
\label{sec:stationary}
		 \begin{myitemize}%
		 	\myitem  \cmt{Data Generation}%
		 \begin{myitemize}%
		 \myitem \cmt{Network via Erd\H{o}s-R\'enyi model}An
                Erd\H{o}s-R\'enyi random graph is generated with edge probability $p_e$ and
                 self-loop probability 1.
                 This graph determines which entries of
the matrices $\{\bm A_p\}_{p=0}^P$ are zero. 
		 \myitem \cmt{VAR coefficients from Gaussian  distribution}The rest of entries are drawn i.i.d. from a standard normal distribution.
         Matrices $\{\bm
                 A_p\}_{p=0}^P$ are  scaled down afterwards by a constant
                 that ensures that the VAR process is stable \cite{lutkepohl2005}.
		 \myitem \cmt{Time series data generation}The
                 innovation process samples are drawn independently as
                  $\bm{u}[t]\sim\mathcal{N}(\bm{0}, \sigma_u^2\bm
                  I_{N})$. 
		 \end{myitemize}%

\myitem	\cmt{Stationary synthetic data results}%
	\begin{myitemize}%
		 \myitem \cmt{Figure: Stationary setting }The first
                 experiment analyzes TISO and TIRSO
                  in a stationary setting.
		 \begin{myitemize}%
		 \myitem \cmt{Figure: NMSD and
                   NMSE}Figs.~\ref{fig:stationarysetting}(a) and
                 \ref{fig:stationarysetting}(b) depict the NMSD  and
                 $\text{NMSE}_1$ for three different values of
                 $\lambda$. As a benchmark,
                 Fig.~\ref{fig:stationarysetting}(b) includes the
  $\text{NMSE}_1$ of the  \emph{genie-aided predictor}, obtained from
  \eqref{eq:predictor} after replacing $\mathbf {\hat{A}}_p$ with $ \bm A_p$.
It is observed that  $\lambda=10^{-6}$ yields a better NMSD  and
$\text{NMSE}_1$ than lower and higher values of $\lambda$. This
corroborates the importance of promoting sparse solutions, as done in TISO
and TIRSO.  Furthermore, as expected,
                  TIRSO generally converges faster than TISO.
		 \myitem \cmt{Figure
                   ROC}Fig.~\ref{fig:stationarysetting}(c) shows the receiver operating
                 characteristic (ROC) curve, composed of pairs $(\text {P}_{\text {FA}},\text {P}_{\text {MD}})$ for different values of the threshold $\delta$. The values of these pairs 
                 are obtained by respectively averaging $\text {P}_{\text {FA}}[t]$ and $\text {P}_{\text {MD}}[t]$ over time in the interval $[T_1,T_2]$. Remarkably, both TISO and
                 TIRSO can simultaneously attain $\text {P}_{\text
                   {FA}}$ and $\text {P}_{\text {MD}}$  below 10\%.
		 \myitem \cmt{Figures EIER, P-MD P-FA}This ability to
                 satisfactorily detect edges is further investigated
                 in
                 Figs.~\ref{fig:stationarysetting}(d-f), where
                 $\delta$ is set for each algorithm so that $\text{P}_{\text{FA}}[t]$ and
                 $\text {P}_{\text {MD}}[t]$ have the same  average over the
                 time interval $[T_1,T_2]$.
		\end{myitemize}%
	\par
	\rev{
		\myitem \cmt{Figure: different Step sizes }Fig. \ref {fig:fig-stepsizetisotirso} analyzes different step size sequences. Because the true VAR parameters remain constant, the diminishing sequence yields the best performance; see \thref{th:strongconvexitytirso}.
		\myitem \cmt{Figure: Comparison with PGD and OSGD}Besides, TISO and TIRSO are compared with benchmarks in Fig.~\ref {fig:comparisonosgpgd}, namely online subgradient descent (OSGD) and proximal gradient descent (PGD). The former obtains a minimizer for \eqref{eq:prob-sep} in an online fashion (labeled as OSGDTISO since it uses the same information as TISO at each iteration). The latter approximates $ \timevarhindsight[t] = \arg \min _{\tbm a_n} \tilde h_t^{(n)}(\tbm a_n)$ by using the (batch) algorithm PGD for $K_\text{PGD}$ iterations over $ \tilde h_t^{(n)}(\tbm a_n)$ (labeled as PGDTIRSO since it uses the same information as TIRSO at each iteration). Fig. \ref {fig:comparisonosgpgd} shows that TISO outperforms OSGDTISO in terms of NMSD, and TIRSO eventually attains better NMSD level than PGDTIRSO. Note that the computational complexity of PGDTIRSO is significantly larger than the complexity of TIRSO.  
		Although the NMSD of TISO in Fig.~\ref {fig:comparisonosgpgd} is close to that of OSGD, a more in-depth study reveals that the former yields sparse iterates without any thresholding; moreover, TIRSO offers a significantly improved edge-detection performance (EIER), see Fig. \ref{fig:comparisonEIERTISOOSGD}.}
		 \myitem \cmt{Figure: Estimated graphs via TIRSO and
                   TISO}Fig.~\ref{fig:syntheticgraph} compares  the
                 true (left) and recovered (right) graphs via TIRSO and TISO
		 \begin{myitemize}%
		\myitem \cmt{true and estimated graphs}%
		 \myitem \cmt{averaged graphs}by thresholding
                 the average of the estimated VAR coefficients 
               across the intervals $[k/(3T),(k+1)/(3T)]$,
                 $k=0,1,2$. 
		 \myitem \cmt{threshold for selecting and edge}The
                 threshold $\delta$  is selected to detect  $p_e(N^2-N)$ edges. 
       Note that this is displayed for a single graph and realization of the VAR process; in other words, this is not a Monte Carlo experiment.
		 \myitem \cmt{comparison of TIRSO and TISO}It is
                 observed that both TIRSO and TISO can identify the
                 true graph quite accurately and  approximate the true
                 VAR coefficients
                 soon afterwards.
		\end{myitemize}%
		\end{myitemize}%
\subsubsection{Non-stationary VAR Processes}
		 \begin{myitemize}%
		\myitem \cmt{data generation}The next experiment
                analyzes TISO and TIRSO when $\bm y[t]$ is a
		\begin{myitemize}%
		 \myitem \cmt{ST-VAR model}%
                 (non-stationary) smooth-transition VAR process
                 \cite[Ch. 18]{kilian2017}
		$ 	\bm y[t]=\sum_{p=1}^{P} \big(\bm A_{p}+ s_f[t](\bm B_p-\bm A_p ) \big)\bm y[t-p]+\bm u[t].
		$ 
		 \myitem \cmt{Two regimes}%
		 \myitem \cmt{transition function}The signal $s_f[t]$
                 determines the transition profile
from a VAR model with parameters  $ \{\bm A_p\}_p$ to a VAR model with
parameters $ \{\bm B_p\}_p  $. In this experiment, $
s_f[t]=1-\text{exp}( -\kappa  ([t-T_B]_+  )^2),$ where $\kappa >0$
controls the transition speed and $T_B$ denotes transition starting instant.
		 \myitem \cmt{Generating coefficients(A,B)}Over an  Erd\H{o}s-R\'enyi random graph,   $\{\bm A_p\}$ and $\{\bm B_p\}$ are
                 generated independently as in Sec.~\ref{sec:stationary}.
It is easy to show that the coefficients $\bm A_{p}+ s_f[t](\bm
B_p-\bm A_p )$  yield a  \emph{stable} VAR process for all~$t$.
		\end{myitemize}%
		 \par
		\cmt{Nonstationary results discussion}%
		\myitem  \cmt{Figure 1: time-varying}%
		\begin{myitemize}%
			\myitem \cmt{NMSD and NMSE}%
                        Figs.~\ref{fig:nmsenmsdstarvaryinggamma}(a) and
                         \ref{fig:nmsenmsdstarvaryinggamma}(b)
                         illustrate the influence
                        of the forgetting factor $\gamma$, of critical
                        importance in non-stationary setups.
			\myitem \cmt{Observations}TISO and TIRSO are
                        seen to  satisfactorily estimate and track the
                        model coefficients. As intuition predicts, the lower
                        $\gamma$ is, the more rapidly TIRSO can
                        adapt to changes, but after a sufficiently
                        long time after the transition, a higher $\gamma$ is preferred.
		\end{myitemize}%
		\par
		\myitem \cmt{Figure 2: time-varying}%
		\begin{myitemize}%
                  \myitem \cmt{description}Finally, to demonstrate
                  that TISO and TIRSO successfully leverage sparsity
                  to track \emph{time-varying} topologies,
                  Fig.~\ref{fig:nmsenmsdstarvaryinglambda} illustrates
                  an approximately optimal point in the trade-off  of selecting  $\lambda$. 
		\end{myitemize}%
		\end{myitemize}%
		\begin{figure*}
			\vspace{-1.3cm}
			\includegraphics[width=\linewidth]{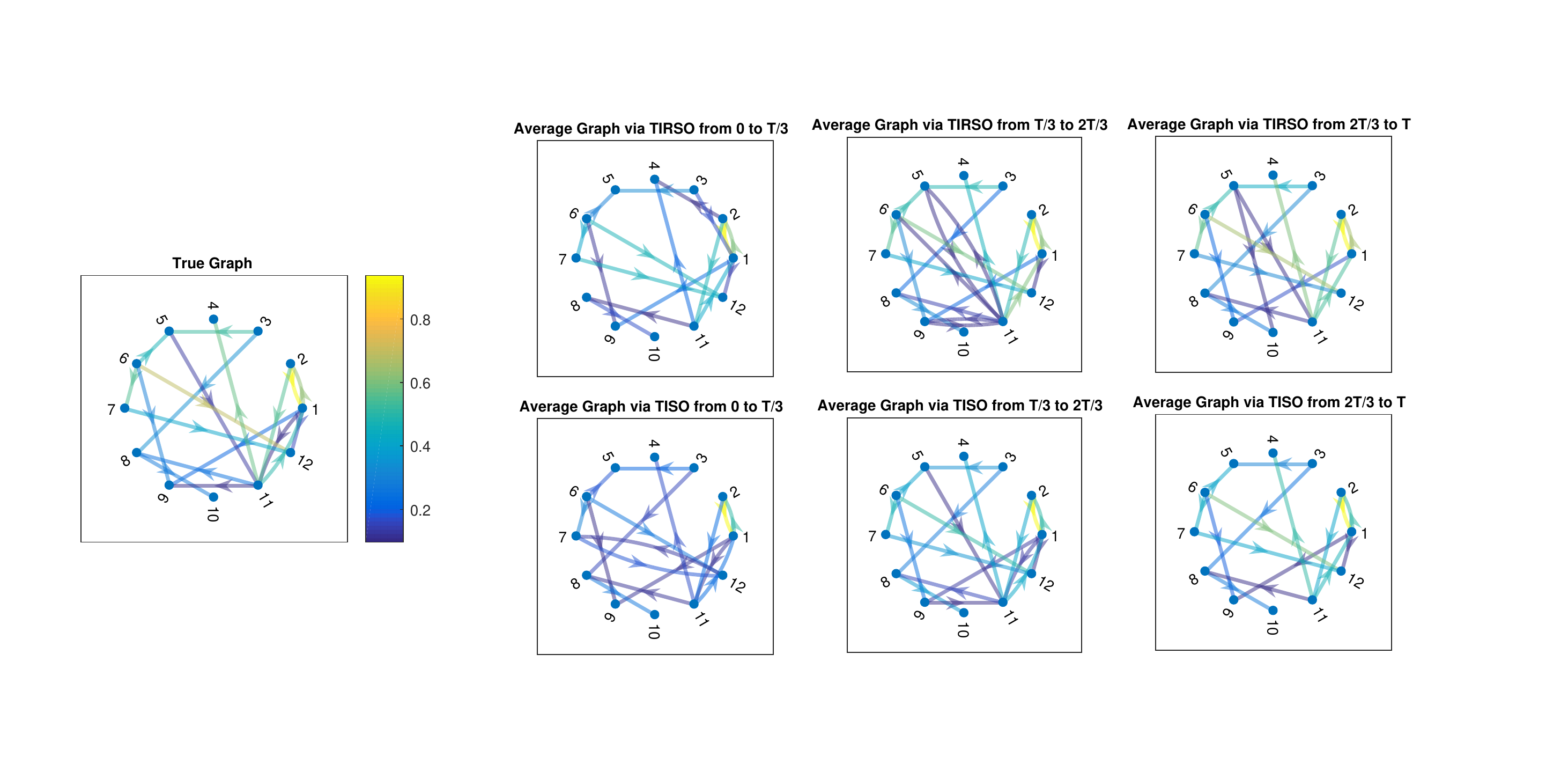}
			\vspace{-1.5cm}
			\caption{\small True and recovered graphs ($N = 12$, $P = 2$, $p_e=0.2$, $\sigma_u=0.005$, $\gamma= 0.98$, $\lambda=10^{-6}$, $T=600$).				 }
			\label{fig:syntheticgraph}
		\end{figure*}
\vspace{-4mm}
		\subsection{Real-Data Tests} \label{sec:realdatatests}
		 \cmt{Real Data Description}%
		 \begin{myitemize}%
                   \myitem \cmt{describing the data}The real data is taken from
                   Lundin's offshore oil and gas (O\&G) platform
                   Edvard-Grieg\footnote{\scriptsize
                     https://www.lundin-petroleum.com/operations/production/norway-edvard-grieg}. Each
                   node corresponds to a temperature, pressure, or
                   oil-level sensor placed in  the decantation system that separates oil,
                   gas, and water.  \myitem \cmt{describing possible
                     causal relations}The measured time series are
                   physically coupled due to the pipelines connecting
                   the system parts and due to the control
                   systems. Hence, causal relations among time series are
                   expected.
                   \myitem \cmt{motivation of
                     inferring dependencies in real data}Topology identification is motivated to forecast the short-term future state of the system and to unveil dependencies that cannot
                   be detected by simple inspection.  \myitem
                   \cmt{Pre-processing of the data}%
		 \begin{myitemize}%
		 	\myitem \cmt{time alignment}All time series
                        are resampled to a common  set of equally-spaced sampling
                        instants using linear interpolation.
                        Since the data was quantized and compressed using a lossy scheme, a significant amount of noise is expected.
		 	\myitem \cmt{zero-mean and unit magnitude}Each time
                        series is normalized to have  zero mean and
                        unit sample standard deviation.
		 \end{myitemize}%
		\end{myitemize}%
		\end{myitemize}%
\begin{figure}
	\centering
	\begin{subfigure}[b]{0.5\textwidth}
		\includegraphics[width=\textwidth]{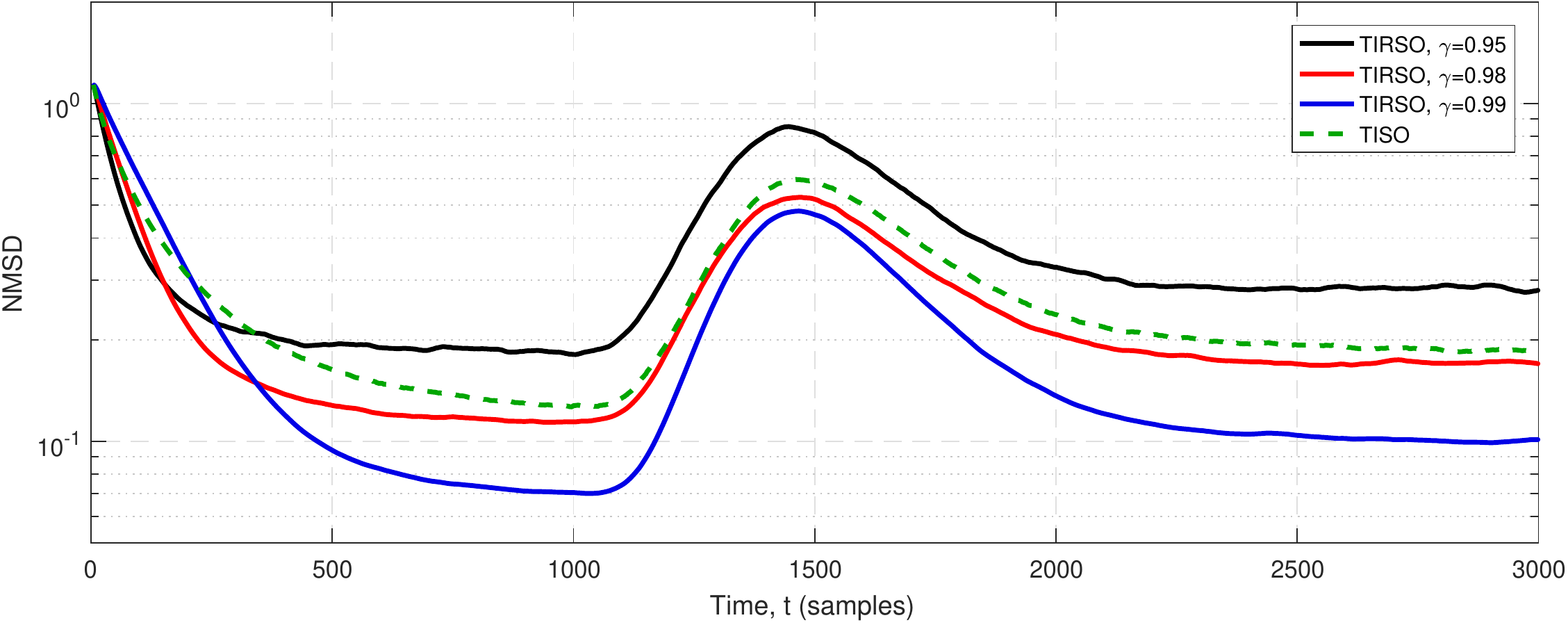}
		\caption{\small NMSD vs. time }
		\label{fig:NMSDgammaNonstationary}
	\end{subfigure}
	\begin{subfigure}[b]{0.5\textwidth}
		\includegraphics[width=\textwidth]{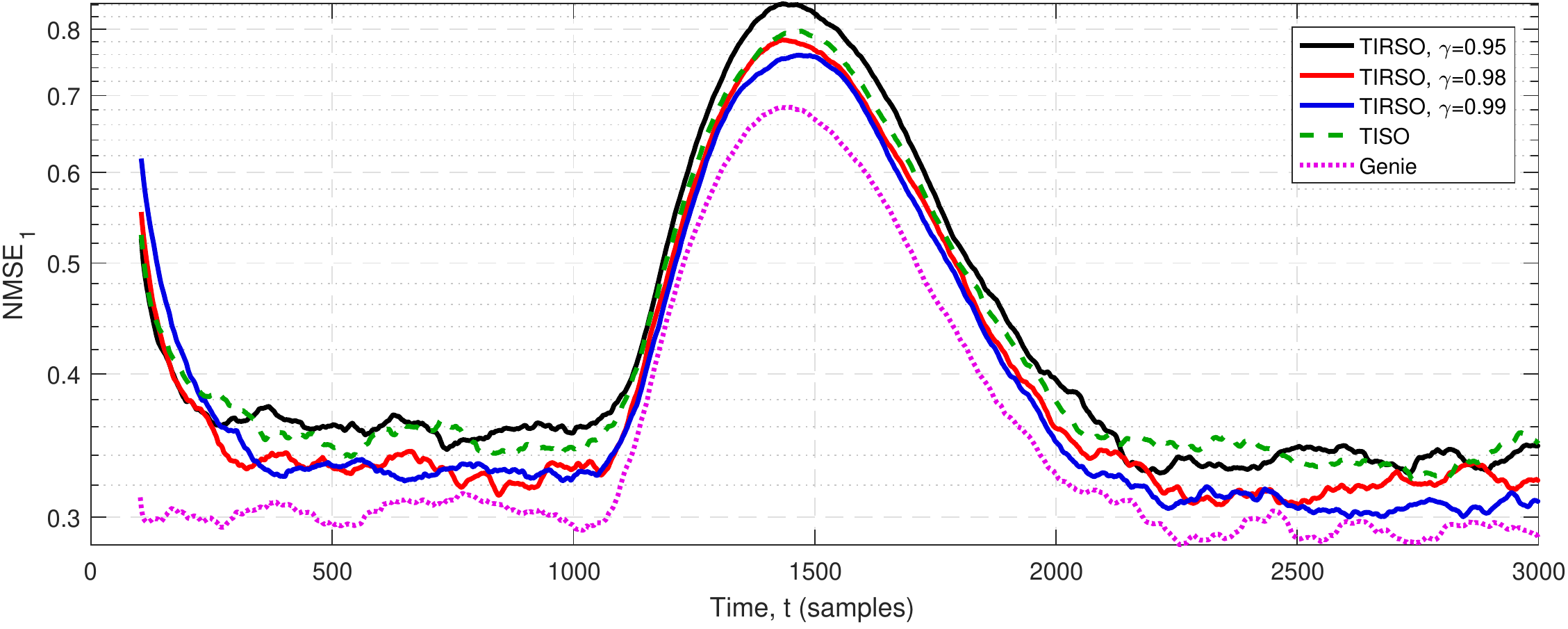}
		\caption{\small NMSE$_1$ vs. time}
		\label{fig:NMSEgammaNonstationary}
	\end{subfigure}
	\caption{\small Effect of the forgetting factor on the performance in a smooth-transition VAR model
          ($\kappa=0.99$, $T_B=1000$, $N=12$, $P = 2$, $p_e=0.2$, 300
          Monte Carlo runs).  
	}\label{fig:nmsenmsdstarvaryinggamma}
\end{figure}
\begin{figure}[h]
	\centering
	\begin{subfigure}[b]{0.5\textwidth}
		\includegraphics[width=\textwidth]{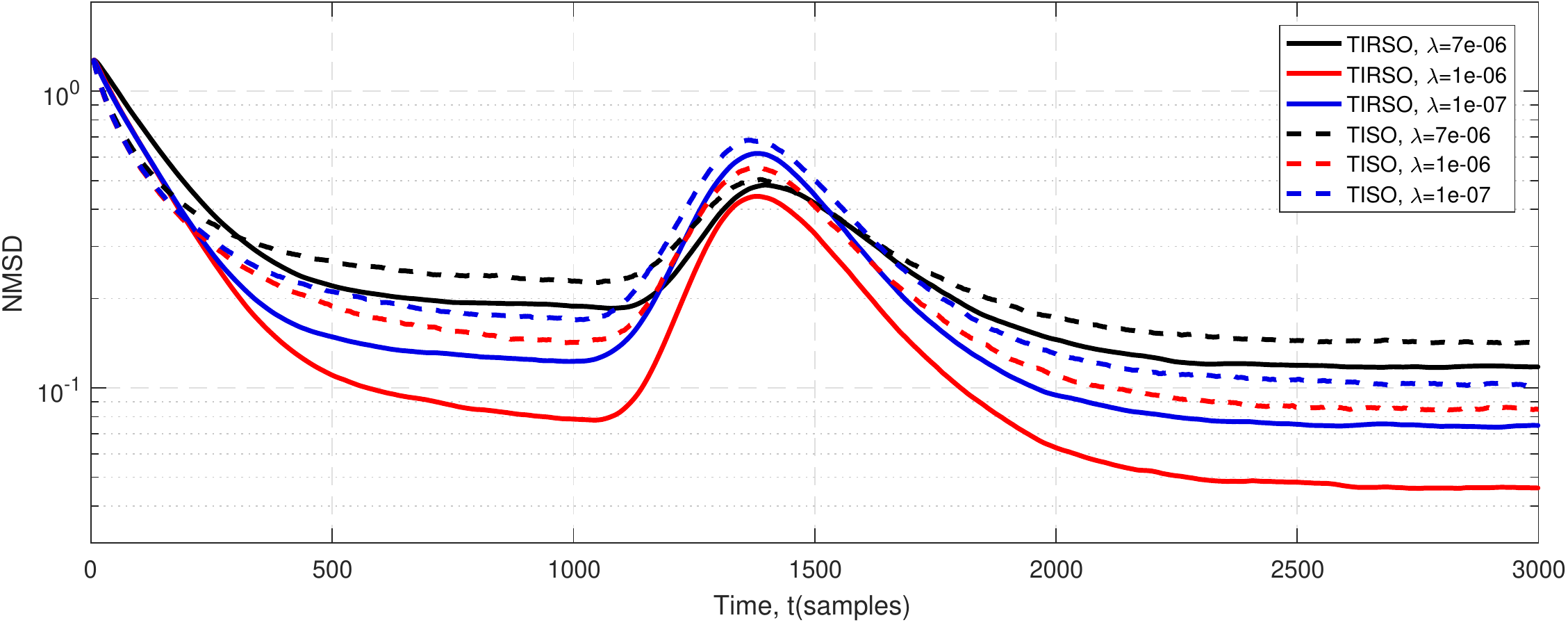}
		\caption{\small NMSD vs. time }
		\label{fig:NMSDlambdaNonstationary}
	\end{subfigure}
	\begin{subfigure}[b]{0.5\textwidth}
		\includegraphics[width=\textwidth]{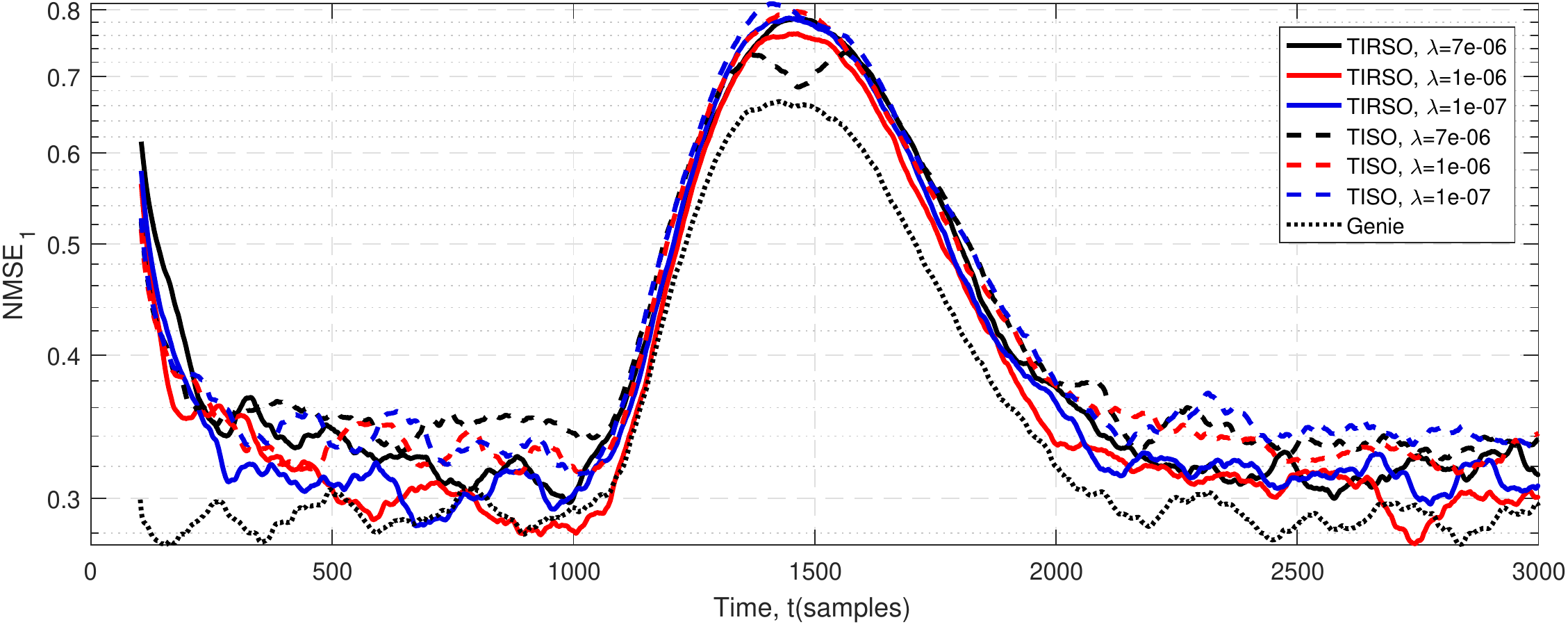}
		\caption{\small NMSE$_1$ vs. time}
		\label{fig:NMSElambdaNonstationary}
	\end{subfigure}
	\caption{\small {Effect of the regularization parameter on the
            performance in a smooth-transition VAR model ($\kappa=0.99$, $T_B=1000$, $N=12$, $T=3000$ $P = 2$, $p_e=0.2$, \rev{$\gamma=0.98$}, 200 Monte Carlo runs).}}  
	\label{fig:nmsenmsdstarvaryinglambda}
\end{figure}
		\begin{figure}
			\centering
			\includegraphics[width=\linewidth]{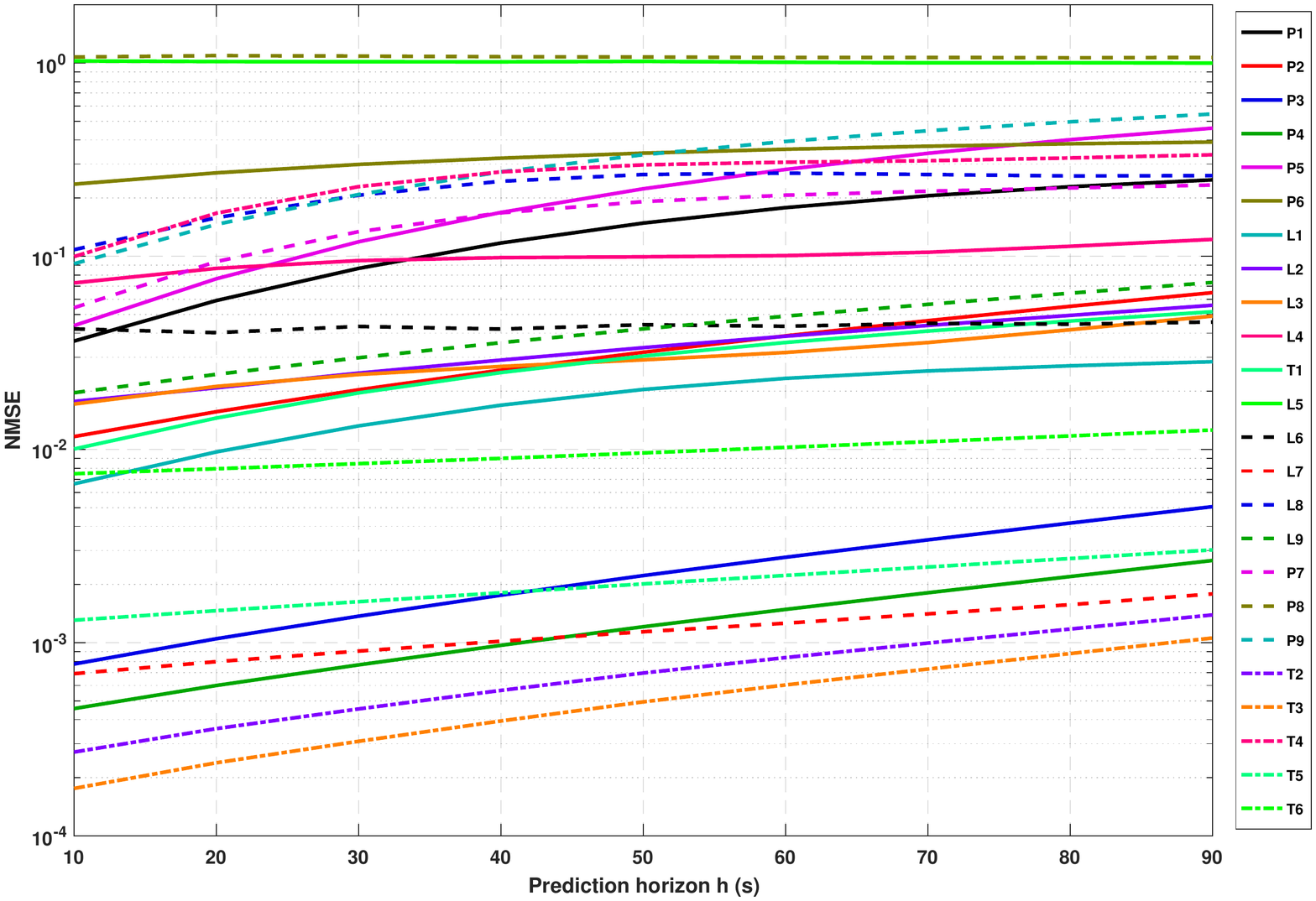}
			\caption{\small Prediction NMSE vs. prediction horizon for individual variables of oil, gas, and water separation system. TIRSO is used with $P=8$, $\gamma=0.9$, $T=4$ hours, sampling interval = 10 s. The  parameter $\lambda$ is selected based on minimum average NMSE.}
			\label{fig:individualerrors}
		\end{figure}
			\begin{figure}
			\centering
			\includegraphics[width=\linewidth]{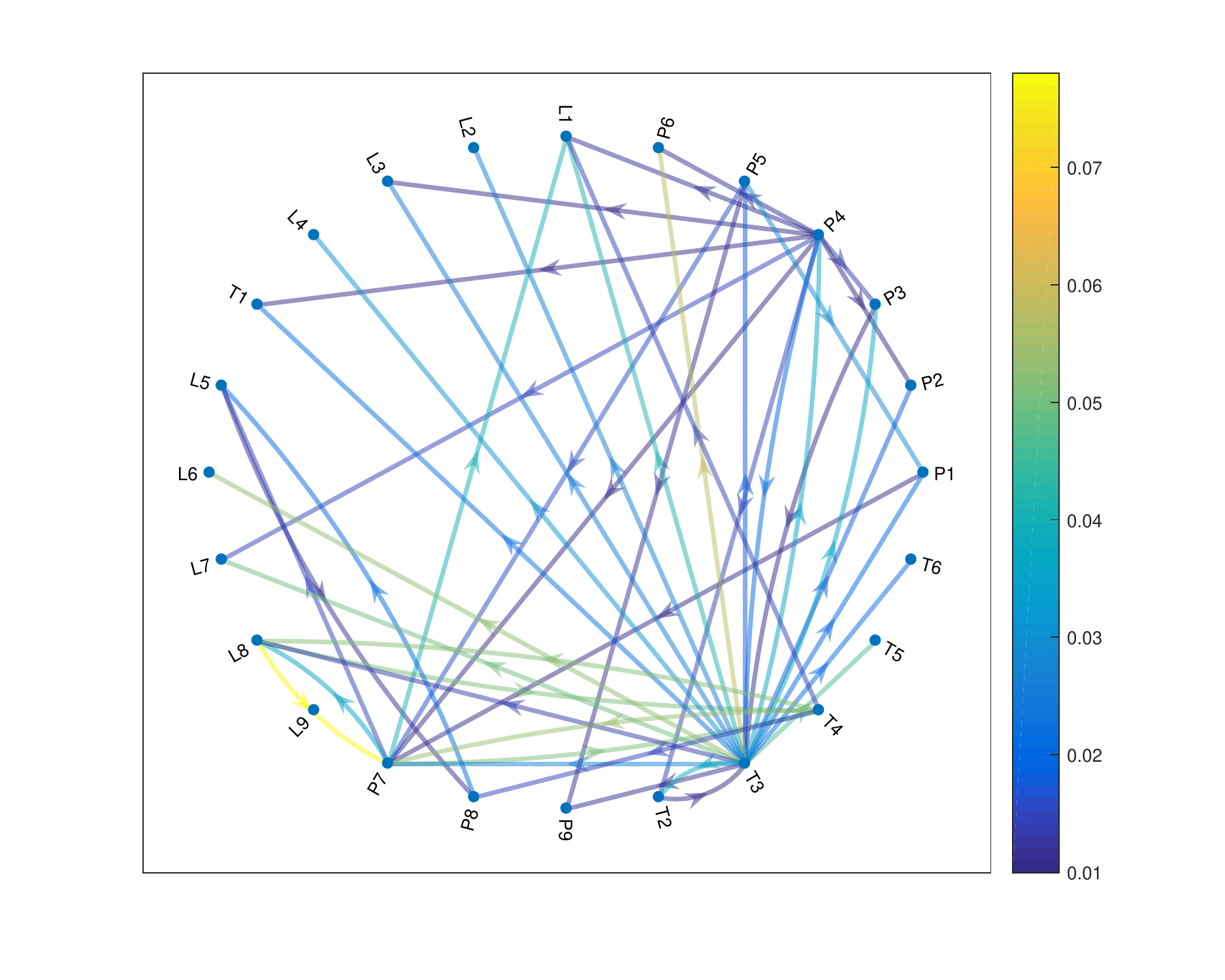}
			\caption{\small The estimated topology of a subset of the variables. The sampling interval is set to 10 seconds. The topology is  obtained via TIRSO with $\gamma=0.9$, $T=3$ hours,  and $P=8$. The  parameter $\lambda$ is selected based on minimum average NMSE.
			}
			\label{fig:estimatedGraph_colorbar}
		\end{figure}
        Here, the  step size is set to $\alpha_t\!= \!1/( \lambda_{\max} (\bm \Phi [t]))$ and the NMSE is defined as
$
			\text{NMSE}_h =1/(\sum_t\lVert \bm y[t+h]\lVert_2^2)\sum_t\left  \lVert \bm y[t+h] - \hat{\bm y}[t+h|t] \right \rVert_2^2 .
$

		\begin{myitemize}%
                  \myitem \cmt{First real data result:prediction for
                    the time series in the data
                    set}Fig.~\ref{fig:individualerrors} shows the
                  $\text{NMSE}_h$ vs. the \emph{prediction horizon}
                  $h$ for the time series in the data set. The
                  temperature, pressure, and oil level time series are
                  respectively denoted by T, P, and L and an
                  identifying index.  As expected, the prediction
                  error increases with $h$. The NMSE ranges from
                  $10^{-4}$ to $1$ due to the different   predictability of each time series.

		\myitem \cmt{3rd: estimated
                  graph}Fig.~\ref{fig:estimatedGraph_colorbar}
                presents the graph obtained by thresholding the average  coefficient estimates over a
                three-hour duration.  The
                threshold is such that the number of reported edges is $4N$. Self-loops are omitted for clarity, and arrow colors encode
                 edge weights. It is observed that most identified edges connect sensors within each subsystem.
		\end{myitemize}

\vspace{-3mm}
\section {Conclusions} \label{s:concl}
	Two online algorithms were proposed for identifying and
        tracking VAR-causality graphs from time series. These
        algorithms sequentially accommodate data and refine their
        sparse topology estimates accordingly. The proposed algorithms
        offer complementary benefits: whereas TISO is computationally
        simpler, TIRSO showcases improved tracking
        behavior. Performance is assessed theoretically and
        empirically. Asymptotic equivalence of the hindsight solutions
        of the proposed algorithms is established and sublinear regret
        bounds are derived.  Experiments with synthetic and real data
        validate the conclusions of the theoretical
        analysis.  \cmt{future
        directions}
        Future directions include explicitly modeling the variations
        in the VAR coefficients, possibly along the lines
        of~\cite{abramovich2005timevarying,
        abramovich2007orderestimation,
        lundbergh2003smoothtransition}, \myitem \cmt{low rank plus
        sparsity}\rev{as well as identifying topologies whose
        adjacency matrix has a low-rank plus sparse structure along the lines
        of \cite{kanada2018lowranksparse} to account for  clusters}.  

	\if\editmode1
	\onecolumn
	\printbibliography
	\else
	\bibliographystyle{IEEEtran}
	\bibliography{IEEEabrv,\bibfilenames}
	\fi

\clearpage
\clearpage 
\centering 
\appendices{\textbf{Supplementary Material}}
\section{Proof of  \thref{prop:asymptoticequivalence}}
\setcounter{equation}{44}
\label{sec:proof:asymptoticequivalence}
\begin{myitemize}%
	\myitem\cmt{hindisght obj diff}The first step is to rewrite \eqref{eq:tirsohindsightobjdef} to be able to obtain a simple expression for  $\tisohindsightobj_T(\bm a_n) - \tirsohindsightobj_T(\bm a_n)$.
	\begin{myitemize}%
		\myitem\cmt{Exchange sums}To this end, substitute  \eqref{eq:recloss} into \eqref{eq:tirsohindsightobjdef} and  exchange the order of the summations  to obtain
		\begin{align}
			\nonumber
			\tirsohindsightobj_T(\bm a_n)&= \textstyle \frac{1}{T-P} \displaystyle \sum_{\tau=P}^{T-1}\Big[ \sum_{t=\tau}^{T-1} \gamma^{t-\tau} \mu \, \ell_{\tau}^{(n)}(\bm  a_n)+ \lambda\sum_{\substack { n'=1\\ n' \neq n}}^{N} \left \lVert \bm a_{n,n'}\right\rVert_2 \Big ]\\
			&= \textstyle \frac{1}{T-P}\displaystyle \sum_{\tau=P}^{T-1} {\auxSgamma_{\tau,T}} \, \mu \, \ell_{\tau}^{(n)}(\bm a_n)
			+ \lambda\sum_{\substack { n'=1, n' \neq n}}^{N} \left \lVert \bm a_{n,n'}\right\rVert_2  ,\nonumber 
		\end{align}
		where  $\auxSgamma_{\tau,T} \define \sum_{t=\tau}^{T-1} \gamma^{t-\tau}$.
		\myitem\cmt{$\theta$ and $\mu$}From the geometric series~summation formula, which establishes that $\auxSgamma_{\tau,T}= ({1-\gamma^{T-\tau}})/({1-\gamma})$, and noting that $\mu=1-\gamma$, 
		the above equation becomes
		\begin{equation*} 
			\tirsohindsightobj_T(\bm a_n)=\textstyle \frac{1}{T-P} \displaystyle \sum_{\tau=P}^{T-1} (1-\gamma^{T-\tau}) \, \ell_{\tau}^{(n)}(\bm a_n)
			+ \lambda\sum_{\substack { n'=1\\ n' \neq n}}^{N} \left \lVert \bm a_{n,n'}\right\rVert_2  .
		\end{equation*}
		\myitem\cmt{diff}From \eqref{eq:tisohindsightobjdef} and 
		the equation above, the difference $d_T(\bm	a_n)\define \tisohindsightobj_T(\bm
		a_n) - \tirsohindsightobj_T(\bm a_n)$ between the TISO and TIRSO hindsight objectives is given by:
		\begin{align} \label{eq:difference}
			d_T(\bm
			a_n)&= \textstyle\frac{1}{T-P}\sum_{\tau=P}^{T-1} \gamma^{T-\tau} \, \ell_{\tau}^{(n)}(\bm a_n).
		\end{align}
	\end{myitemize}%
	\myitem\cmt{pointwise convergence}To prove part 1, it suffices  to show that $d_T(\bm	a_n)\rightarrow 0$ as $T\rightarrow \infty $ for all $\bm a_n$.
	\begin{myitemize}%
		\myitem\cmt{bound $\ell$}To this end, expand $
		\ell_t^{(n)}(\bm a_n)$
		\begin{eqnarray} \label{eq:expanded_loss_function}
			\ell_t^{(n)}(\bm a_n)\!= \!  \frac{1}{2} \left (y_n^2[t]+\bm a_n^\top \, \bm g[t] \, \bm g^\top[t] \, \bm a_n \!-\! 2 \,y_n[t] \, \bm g^\top [t] \,\bm a_n\right ),
		\end{eqnarray}
		and apply  \rev{Cauchy-Schwarz} inequality  to obtain
		\begin{align} \label{eq:bound_on_ell}
			\ell_t^{(n)}(\bm a_n) \!\leq \!\frac{1}{2}\left [ \left\lVert \bm a_n \right\rVert_2 \cdot \left\lVert \bm g[t]\right\rVert_2 \right]^2 \!+ \!\frac{1}{2}\energybound \!+\! \sqrt{\energybound}\left\lVert \bm g[t] \right\rVert_2 \cdot \left\lVert \bm a_n \right\rVert_2.
		\end{align}
		On the other hand, the hypothesis  $|y_n[t]|^2 \leq \energybound \forall n, t$ implies that $ \lVert \bm y[t]  \rVert_2^2\leq  N \energybound$, and hence
		\begin{equation*} 
			\left \lVert \bm g[t] \right \rVert_2^2 \!= \sum_{\tau=t-P}^{t-1} \left \lVert \bm y[\tau] \right \rVert_2^2\leq  P \!\underset{t-P \leq \tau \leq t-1} {\text {max}} \! \left \lVert \bm y[\tau] \right \rVert_2^2 \leq  PN \energybound.
		\end{equation*}
		Substituting the upper bound of $ \lVert \bm g[t]
		\rVert_2^2$  into
		\eqref{eq:bound_on_ell} yields
		\begin{align}
			\label{eq:boundonl}
			\ell_t^{(n)}(\bm a_n) &\leq \frac{1}{2} NP \energybound  \left\lVert \bm a_n \right\rVert_2^2+ \!\frac{1}{2} \energybound\!+ \!\sqrt{NP}\energybound\left \lVert \bm a_n \right\rVert_2\define G(\bm a_n) 
		\end{align}
		\myitem\cmt{bound $d_T$}Applying the latter bound  to \eqref{eq:difference} results in
		\begin{align}
			d_T(\bm a_n)& \leq \frac{1}{T-P}\sum_{\tau=P}^{T-1} \gamma^{T-\tau} \, G(\bm a_n)  \nonumber \\
			&= \frac{G(\bm a_n)  \gamma^T}{T-P}\sum_{\tau=P}^{T-1} \gamma^{-\tau}
			=\frac{G(\bm a_n) \left ( 1-\gamma^{T-P}\right )}{(T-P)(\gamma^{-1} -1)} \label{eq:proofa5}.
		\end{align}
		Taking the limit of the right-hand side clearly yields
		\begin{equation}
			\lim\limits_{T \rightarrow \infty} \frac{G(\bm a_n) \left ( 1-\gamma^{T-P}\right )}{(T-P)(\gamma^{-1} -1)}=0.
		\end{equation}
		\myitem\cmt{limit $d_T$}Noting  from \eqref{eq:difference} that $d_T(\bm a_n)\geq 0$, it follows that $\lim_{T\rightarrow\infty} d_T(\bm a_n)=0$, which concludes the proof of part 1.
	\end{myitemize}%
	\par
	\myitem\cmt{convergence of minima}To prove part 2,
	\begin{myitemize}%
		\myitem\cmt{TISO$\ge $TIRSO}note from \eqref{eq:difference}
		that $d_T(\bm	a_n)\geq 0$, which in turn implies that
		\begin{equation} \label{eq:tirso-lessthan-tiso}
			\tirsohindsightobj_T(\bm a_n) \leq \tisohindsightobj_T(\bm a_n),
		\end{equation}
		for all $\bm a_n$ and $T>P$.
		\myitem\cmt{bound inf diff}On the other hand, it follows from
		\eqref{eq:minimizertirso} that
		\begin{equation}  \label {eq:infTIRSO}
			\tirsohindsightobj_T(\tbm a_n^*[T]) \leq \tirsohindsightobj_T(\bm a_n^*[T]) .
		\end{equation}
		Thus, by combining \eqref{eq:tirso-lessthan-tiso} and \eqref{eq:infTIRSO},
		\begin{equation} \label{eq:tiso-rel-tirso}
			\tirsohindsightobj_T(\tbm a_n^*[T])  \leq  \tisohindsightobj_T(\bm a_n^*[T]).
		\end{equation}
		Similarly, from \eqref{eq:minimizertiso}, it holds that
		$\tisohindsightobj_T(\bm a_n^*[T]) \leq
		\tisohindsightobj_T(\tbm a_n^*[T]) $. Subtracting
		$\tirsohindsightobj_T(\tbm a_n^*[T])$ from both sides of the
		latter inequality yields
		\begin{eqnarray} \label{eq:diffsequence}
			\tisohindsightobj_T(\bm a_n^*[T]) - \tirsohindsightobj_T(\tbm a_n^*[T]) \leq \tisohindsightobj_T(\tbm a_n^*[T]) - \tirsohindsightobj_T(\tbm a_n^*[T])\nonumber \\
			= d_T(\tbm a_n^*[T]).
		\end{eqnarray}
		By combining \eqref{eq:tiso-rel-tirso} and
		\eqref{eq:diffsequence}, it holds that
		\begin{equation} \label{eq:sandwichsequence}
			0\leq \tisohindsightobj_T(\bm a_n^*[T]) - \tirsohindsightobj_T(\tbm a_n^*[T]) \leq d_T(\tbm a_n^*[T]).
		\end{equation}
		\myitem\cmt{lim inf diff}Since $\lim_{T\rightarrow\infty} d_T(\tbm a_n^*[T])=0$, \eqref{eq:sandwichsequence} implies that
		\begin{equation} \label {eq:minimaconvergence}
			\lim_{T\rightarrow\infty}\tisohindsightobj_T(\bm a_n^*[T]) - \tirsohindsightobj_T(\tbm a_n^*[T])=0.
		\end{equation}
	\end{myitemize}
	\myitem\cmt{convergence of minimizers}Finally, to establish part 3, \rev {note that it follows from assumption A\ref{as:mineig}, \eqref{eq:recloss3} and \eqref{eq:tirsohindsightobjdef} that  $\tirsohindsightobj_T$ is $\tilde \beta$-strongly convex for some $\tilde \beta>0, \forall\, T$. } Thus, from \eqref{eq:minimizertirso}, one finds that
	\begin{equation} \label {eq:tirsostrongconvexity}
		\tirsohindsightobj_T(\bm a_n^*[T]) \geq \tirsohindsightobj_T(\tbm a_n^*[T])+ \frac{\rev{\tilde \beta}}{2} \left\lVert \bm a_n^*[T]- \tbm a_n^*[T] \right\rVert_2^2.
	\end{equation}
	By combining \eqref{eq:tirso-lessthan-tiso} and
	\eqref{eq:tirsostrongconvexity}, it follows that
	\begin{equation} \label {eq:tirsosandtiso}
		\tisohindsightobj_T(\bm a_n^*[T])  \geq  \tirsohindsightobj_T(\tbm a_n^*[T])+\frac{\rev{\tilde \beta}}{2} \left\lVert \bm a_n^*[T]- \tbm a_n^*[T] \right\rVert_2^2,
	\end{equation}
	or, equivalently,
	\begin{equation} \label {eq:sandwichminmizers}
		\tisohindsightobj_T(\bm a_n^*[T]) - \tirsohindsightobj_T(\tbm a_n^*[T]) \geq  \frac{\rev{\tilde \beta}}{2} \left\lVert \bm a_n^*[T]- \tbm a_n^*[T] \right\rVert_2^2 \geq 0.
	\end{equation}
	Taking limits gives rise to
	\begin{multline}
		\hspace{-5mm}
		\lim_{T\rightarrow\infty}\big [\tisohindsightobj_T(\bm a_n^*[T]) - \tirsohindsightobj_T(\tbm a_n^*[T]) \big ] \geq 	\lim_{T\rightarrow\infty} \Big [ \frac{\rev{\tilde \beta}}{2} \lVert \bm a_n^*[T] \\
		- \tbm a_n^*[T] \rVert_2^2 \Big ]\geq 0. \label {eq:sandwichlimit}
	\end{multline}
	From \eqref{eq:minimaconvergence} and the sandwich theorem applied to \eqref{eq:sandwichlimit}, 
	we have
	\begin{equation}
		\lim_{T\rightarrow\infty} \Big [ \frac{\rev{\tilde \beta}}{2} \lVert \bm a_n^*[T]- \tbm a_n^*[T] \rVert_2^2 \Big ]=0,
	\end{equation}
	which concludes the proof.
\end{myitemize}%

\section{Proof of \thref{cor:doublingtricktiso}}
\label{proof:doublingtricktiso}
\rev{
	\cmt{regret of TISO}Consider first the  regret
	of TISO with constant step size.}
\begin{mylemma}
	\thlabel{prop:regrettiso}
	\begin{myitemize}%
		\myitem\cmt{let}Let $\{\bm
		a_n[t]\}_{t=P}^T$ be generated by TISO
		(\textbf{Procedure~\ref{alg:TISO}}) with
		constant step size
		$\alpha_t\!=\alpha\!=\!\mathcal O\big
		(1/\sqrt{T}\big )$. 
		\myitem\cmt{hypothesis}\rev{Under assumptions A\ref{as:boundedprocess} and A\ref{as:tisocov},
			we have}
		\begin{align}
			\label{eq:regrettiso}
			R\rev{_{s}^{(n)}[T]}=\mathcal{O}\left (  PN\energybound \, \rev{B_{\bm
					a}^2} \,\sqrt{T}\right ).
		\end{align}
	\end{myitemize}%
\end{mylemma}%
\begin{IEEEproof} \rev{
		See Appendix~\ref{appendix:proof{prop:regrettiso}}.}
\end{IEEEproof}%
\cmt{Interpretation}
\begin{myitemize}%
	\rev {
		\myitem \cmt{Comparison with the conditions for online subgradient
			descent methods to have sublinear regret}%
	}%
	\myitem\cmt{constant stepsize}Observe that the step size
	in \thref{prop:regrettiso} depends on $T$ and
	therefore \eqref{eq:regrettiso} cannot be interpreted as directly
	establishing sublinear regret for TISO. To understand this result,
	consider a sequence of copies of TISO, each one for a value of
	$T$. Each copy has a (potentially) different step size, but uses the
	same step size for all $t$. \rev {Expression \eqref{eq:regrettiso}
		bounds the regret of the $T$-th copy at time $T$.}
	\cmt{how to select $\alpha$:the doubling
		trick}\rev{However, \thref{prop:regrettiso} can be used next to establish
		sublinear regret for step size sequences that remain constant over
		windows of exponentially increasing length; see the \emph{doubling
			trick} \cite{shalev2011online}. 
	}
	
\end{myitemize}%
\rev{To this end, let the regret in the window
	$[t_1,t_2]$ be 
	\begin{equation} \label{eq:regretwindow}
		R_s^{(n)}[t_1,t_2] \define  \sum_{t=t_1}^{t_2}
		h_t^{(n)} (\bm a_n[t]) -  h_t^{(n)} (\bm a_n^*[t_1,t_2]),
	\end{equation}
	where  $\{\bm
	a_n[t]\}_t\subset \rfield^{NP}$ is   an arbitrary sequence  and
	\begin{equation} \label{eq:defhindsightforwindow}
		\bm a_n^*[t_1,t_2] \define \argmin_{\bm a_n} \sum_{t=t_1}^{t_2} h_t^{(n)} (\bm a_n).
	\end{equation}
}
\rev{
	The next result establishes a bound on the static regret given the
	regret at each window.}
\rev {
	\begin{mylemma} \thlabel{lemma:individualregrets}
		For $T= t_0 2^M$ and for an arbitrary sequence $\{\bm
		a_n[t]\}_t\subset \rfield^{NP}$, the regret in \eqref {eq:defstaticregretTISO} is bounded as: 
		\begin{align}
			\label{eq:boundedwindowregret}
			&R_s^{(n)}[T] \leq  R_s^{(n)}[P,t_0] + \sum_{m=1}^M
			R_s^{(n)}[t_0 2^{m-1}\!\!+\!1, t_0 2^m ].
		\end{align}
	\end{mylemma}
	\begin{IEEEproof}
		For $T= t_0 2^M$,  expression  \eqref {eq:defstaticregretTISO} can be written as:
		\begin{align} \label{eq:regretfort02powM}
			R_s^{(n)}[T] & = \sum_{t=P}^{t_0 2^M}  h_t^{(n)}(\bm a_n[t])- \sum_{t=P}^{t_0 2^M} h_t^{(n)}\left ( \bm a_n^*[T]\right).  
		\end{align}
		On the other hand, it follows 
		from \eqref{eq:regretwindow}
		that \eqref{eq:boundedwindowregret} is equivalent to         
		\begin{align}
			&R_s^{(n)}[T] \leq \sum_{t=P}^{t_0} \left[ h_t^{(n)}(\bm a_n[t])- h_t^{(n)}\left ( \bm a_n^*[P,t_0]\right) \right] \nonumber \\
			& + \sum_{m=1}^M \sum_{t=t_0 2^{m-1}+1}^{t_0 2^m} \!\!\!\!\left[ h_t^{(n)}\!(\bm a_n [t]) \!-\! h_t^{(n)} \left ( \bm a_n^*\left [t_0 2^{m-1}\!\!+\!1, t_0 2^m\right ] \right)
			\right], \label{eq:indvidualhindsightsregret}
		\end{align}
		The inequality in \eqref{eq:indvidualhindsightsregret} can also be rewritten as
		\begin{align} 
			R_s^{(n)}[T]& \leq \sum_{t=P}^{t_0 2^M}  h_t^{(n)}(\bm a_n[t]) - \bigg[ \sum_{t=P}^{t_0} h_t^{(n)}\left ( \bm a_n^*[P,t_0]\right) \nonumber \\ 
			& \quad + \sum_{m=1}^M \sum_{t=t_0 2^{m-1}+1}^{t_0 2^m} h_t^{(n)} \left ( \bm a_n^*\left [t_0 2^{m-1}\!\!+\!1, t_0 2^m\right ] \right) \bigg]. \label{eq:regretindividualhindsights}
		\end{align}
		By comparing \eqref {eq:regretfort02powM} and \eqref{eq:regretindividualhindsights}, proving \eqref{eq:boundedwindowregret} is equivalent to showing that  
		\begin{multline}
			\sum_{t=P}^{t_0 2^M} h_t^{(n)}\left ( \bm a_n^*[T]\right) \geq   \bigg[ \sum_{t=P}^{t_0} h_t^{(n)}\left ( \bm a_n^*[P,t_0]\right) + \\ \sum_{m=1}^M \sum_{t=t_0 2^{m-1}+1}^{t_0 2^m} h_t^{(n)} \left ( \bm a_n^*\left [t_0 2^{m-1}\!\!+\!1, t_0 2^m\right ] \right) \bigg].
		\end{multline}
		From the definitions of $	\bm a_n^*[T]$ in \eqref{eq:minimizertiso}  and $\bm a_n^*[t_1,t_2]$ in \eqref{eq:defhindsightforwindow}, the above inequality holds since $\inf_{\bm x, \bm y} f(\bm x, \bm y) \leq \inf_{\bm x= \bm y} f(\bm x, \bm y) $.
	\end{IEEEproof}
	The next step is to bound
	the regret at each window
	using \thref{prop:regrettiso}. To this end, one must set
	$\alpha\winnot{m}$ as a function $\mathcal{O}(1/\sqrt{T_m})$,
	where $T_m\define t_02^m -t_02^{m-1}= t_02^{m-1}$ is the
	length of the $(m+1)$-th window, $m=1,\ldots, M$.
	Invoking \thref{prop:regrettiso}, the
	regret for the $(m+1)$-th window is given by $R_s^{(n)}[t_0
	2^{m-1}+1,t_0 2^m]= \mathcal O(PN\energybound B_{\bm
		a}^2 \sqrt{2^{m-1}})$. By
	\thref{lemma:individualregrets}, the regret of TISO
	becomes
	\begin{align} 
		R_s^{(n)}[T] &= \mathcal O\left(PN\energybound B_{\bm a}^2 \sqrt{t_0-P+1}\right) \nonumber \\
		& \quad + \sum_{m=1}^M \mathcal O\left(PN\energybound B_{\bm a}^2 \sqrt{t_0 2^{m-1}}\right) \nonumber \\
		& =  \mathcal O\left(PN\energybound B_{\bm a}^2 \sum_{m=1}^M \sqrt{t_0 2^{m-1}}\right) \nonumber \\
		& = \mathcal O\left(PN\energybound B_{\bm a}^2 (\sqrt{2})^M \right ) \nonumber \\
		& = \mathcal O\left(PN\energybound B_{\bm a}^2 \left (2^{\log_2 \frac{T}{t_0} } \right )^{\frac{1}{2}} \right ) \nonumber \\
		& = \mathcal O\left(PN\energybound B_{\bm a}^2 \sqrt{T} \right ) \nonumber ,
	\end{align}
	which concludes the proof.
}

\vspace{8mm}
\section{Proof of \thref{prop:regrettiso}}
\label{appendix:proof{prop:regrettiso}}
\rev {
	First we present a lemma that establishes that the hindsight solution of TISO is bounded and then we will present the proof of  \thref {prop:regrettiso}.
	\begin{mylemma}\thlabel{lemma:boundedhindsightsolutionTISO}
		Under assumptions A\ref{as:boundedprocess}, A\ref{as:mineig}, and A\ref{as:tisocov}, the hindsight solution of TISO $\bm a_n^*[T]$ given in \eqref{eq:minimizertiso} is bounded as
		\begin{eqnarray} \label{eq:boundonatiso}
			\left \lVert \bm a_n^*[T] \right \rVert_2 \leq B_{\bm a}\triangleq \frac{1}{\beta} \left (\energybound \sqrt{PN} + \sqrt{\energybound^2PN+\beta \energybound} \right  ).
		\end{eqnarray}
	\end{mylemma}
	\vspace{5mm}
	\begin{IEEEproof}
		Note that $a_n^*[T]$ belongs to the sublevel set of TISO hindsight objective for $\bm a_n= \bm 0_{NP}$, given by
		\begin{equation}
			\mathcal S_T \triangleq \left \{\bm a_n: \tisohindsightobj_T(\bm a_n)\leq \tisohindsightobj_T(\bm 0_{NP}) \right \},
		\end{equation}
		where $\tisohindsightobj_T(\bm 0_{NP})$ is upper bounded by
		\begin{align*}
			\tisohindsightobj_T(\bm 0_{NP}) &= \frac{1}{T-P}  \sum_{t=P}^{T-1}\frac{1}{2}  y_n^2[t] \\
			& \leq  \frac{1}{2(T-P)}\sum_{t=P}^{T-1}\energybound =\frac{\energybound}{2}.
		\end{align*}
		This means that we can write:
		\begin{equation} \label{eq:sublevelsetTISO1}
			\mathcal  S_T \subset \left \{\bm a_n^*[T]:  \tisohindsightobj_T(\bm a_n^*[T])\leq \frac{\energybound}{2} \right \}.
		\end{equation}
		Next, we find a lower bound to $\tisohindsightobj_T(\bm a_n^*[T])$ that is an increasing function of $\lVert \bm a_n^*[T] \rVert _2$ as follows
		\begin{subequations}
			\begin{align*}
				\tisohindsightobj_T(\bm a_n^*[T]) &= \frac{1}{T-P}  \sum_{t=P}^{T-1} \Big [\frac{1}{2}(\bm a_n^*[T])^\top \bm g[t] \bm g^\top[t] \bm a_n^*[T]  \\
				&-y_n[t] \bm g^\top[t]  \bm a_n^*[T] + \frac{1}{2} y_n^2[t] + \nReg (\bm a_n^*[T])\Big ]\\
				& \geq \frac{1}{T-P}  \sum_{t=P}^{T-1} \Big [\frac{1}{2}(\bm a_n^*[T])^\top \bm g[t] \bm g^\top[t] \bm a_n^*[T] \\
				&-y_n[t] \bm g^\top[t]  \bm a_n^*[T]  \Big ] \\
				& \geq \frac{1}{2}  \lambda_{\mathrm{min}}\left (\frac{1}{T-P}\sum_{t=P}^{T-1}\bm g[t] \bm g^\top[t] \right ) \left \lVert \bm a_n^*[T] \right \rVert_2^2 \\
				& -\frac{1}{T-P}  \sum_{t=P}^{T-1} y_n[t] \left \lVert \bm g[t] \right \rVert_2 \cdot  \left \lVert \bm a_n^*[T] \right \rVert_2   \\
				& \geq \frac{1}{2}  \beta  \left \lVert \bm a_n^*[T] \right \rVert_2^2-\energybound \sqrt{PN}  \left \lVert \bm a_n^*[T] \right \rVert_2.
			\end{align*}
		\end{subequations}
		Therefore,
		\begin{eqnarray} \label{eq:sublevelsetTISO2}
			\mathcal  S_T\!\subset\! \left \{\bm a_n^*[T]\!: \!  \frac{1}{2} \beta \lVert \bm a_n^*[T] \rVert _2^2\!-\!\energybound \sqrt{PN} \lVert \bm a_n^*[T] \rVert _2 \leq \frac{\energybound}{2} \right \}.
		\end{eqnarray}
		Further, we can write
		\begin{eqnarray} \label{eq:sublevelsetTISO3}
			\mathcal  S_T \subset \left \{\bm a_n^*[T]:  \left \lVert \bm a_n^*[T] \right \rVert_2 \leq  B_{\bm a} \right \},
		\end{eqnarray}
		with $B_{\bm a}\triangleq 1/\beta(\energybound \sqrt{PN} + \sqrt{\energybound^2PN+\beta \energybound})$.
		Expression \eqref {eq:sublevelsetTISO3} implies that the TISO hindsight solution is bounded.
	\end{IEEEproof}
}
\rev{Now, we present the proof of \thref {prop:regrettiso}. This proof is based on the idea that if the inequality $ \lVert \nabla  \ell_{t}^{(n)}(\bm
	a_n) \rVert_2^2 \leq 2PN\energybound \,
	\ell_t^{(n)}(\bm a_n), \forall \, t,n$ holds and the strong convexity parameter of $\psi$ is 1, then it follows from \cite [Corollary 5]{duchi2010comid} that:
	\begin{align*}
		R\rev{_{s}^{(n)}[T]}&=\mathcal O\Big (\tfrac{1}{2} \, \rho\sqrt{T-P} \, \left \lVert \bm a_n^{*} [T]- \bm a_n[P] \right \rVert_2^2 \Big )\\
		&= \mathcal O\Big (\tfrac{1}{2} \,  \rho\sqrt{T} \, \left \lVert \bm a_n^{*}[T] \right \rVert_2^2 \Big ) \\
		&= \mathcal O\Big ( PN\energybound \sqrt{T}\left \lVert \bm a_n^{*} [T] \right \rVert_2^2 \Big ) \\
		&= \mathcal O\Big ( PN\energybound \sqrt{T}B_{\bm a}^2 \Big ),
	\end{align*}
	where $B_{\bm a}$ is defined in \eqref {eq:boundonatiso}. } We still need to show that the inequality $ \lVert \nabla  \ell_{t}^{(n)}(\bm
a_n) \rVert_2^2 ~\leq~ 2PN\energybound \,
\ell_t^{(n)}(\bm a_n)$, $\forall \, t,n$, holds. To this end, note from \eqref {eq:instgrad}
that:
\begin{align}
	\big \lVert \nabla  \ell_{t}^{(n)}(\bm a_n) \big \rVert_2^2
	&=\left \lVert \bm g[t] \left (\bm g^\top [t] \bm a_n - y_n[t] \right ) \right \rVert_2^2 \nonumber\\
	&= \left \lVert \bm g[t] \right \rVert_2^2 \cdot  \left \lvert y_n[t] - \bm g^\top [t] \, \bm a_n   \right \rvert^2. \label {eq:proof2}
\end{align}
On the other hand, the hypothesis  $|y_n[t]|^2 \leq \energybound ~\forall~ n, t$ implies that $\left \lVert \bm y[t] \right \rVert_2^2\leq  N \energybound$ and,  therefore:
\begin{equation} \label {eq:bound_on_g1}
	\left \lVert \bm g[t] \right \rVert_2^2 \!= \sum_{\tau=t-P}^{t-1} \left \lVert \bm y[\tau] \right \rVert_2^2\leq  P \!\underset{t-P \leq \tau \leq t-1} {\text {max}} \! \left \lVert \bm y[\tau] \right \rVert_2^2 \leq  PN \energybound.
\end{equation}
Combining   \eqref{eq:proof2} and \eqref{eq:bound_on_g1} yields
\begin{equation} \label{eq:bound1}
	\big \lVert \nabla  \ell_{t}^{(n)}(\bm a_n) \big \rVert_2^2 \leq PN\energybound \left \lvert y_n[t] - \bm g^\top [t] \, \bm a_n   \right \rvert^2.
\end{equation}
Thus, to satisfy
\begin{equation*} 
	\big \lVert \nabla  \ell_{t}^{(n)}(\bm a_n) \big \rVert_2^2  \leq \rho \, \ell_t^{(n)}(\bm a_n)= \rho \, \frac{1}{2} \left (y_n[t]- \bm g^\top[t] \, \bm a_n \right )^2,
\end{equation*}
it suffices to set
$\rho=2PN\energybound$.

\vspace{5mm}
\section{Proof of \thref{cor:doublingtricktirso}}
\label{proof:doublingtricktirso}
\rev{The first step is to obtain a bound for constant step size.}
\begin{mylemma} \thlabel{prop:regrettirso}
	\begin{myitemize}%
		\myitem\cmt{let}Let $\{\tbm a_n[t]\}_{t=P}^{T}$ be  generated
		by TIRSO (\textbf{Procedure~\ref{alg:TIRSO}}) with constant step
		size $\alpha_t\!=\!\alpha\!=\!\mathcal O\big (1/\sqrt{T}\big
		)$. 
		\myitem\cmt{hypothesis}\rev{Under assumptions A\ref{as:boundedprocess}, A\ref{as:mineig}, and A\ref{as:maxeig}, we have}
		\vspace{-2mm}
		\begin{align}
			\label{eq:regrettirso}
			\tilde R\rev{_{s}^{(n)}[T]}=\mathcal{O}\Big (   L \, \rev { B_{\tbm a}^2} \,\sqrt{T}\Big ).
		\end{align}
	\end{myitemize}%
\end{mylemma}
\begin{IEEEproof}
	See Appendix \ref
	{appendix:proof{prop:regrettirso}}.
\end{IEEEproof}
The rest of the proof proceeds along the lines of the proof of \thref{cor:doublingtricktiso}.

\vspace{5mm}

\section{Proof of \thref {prop:regrettirso}}
\label{appendix:proof{prop:regrettirso}}
\rev {
	First, we present a lemma that establishes that the hindsight solution of TIRSO is bounded. Then, we will present the proof of  \thref {prop:regrettirso}.
	\begin{mylemma}\thlabel{lemma:boundedhindsightsolutionTIRSO}
		Under the assumptions A\ref{as:boundedprocess} and A\ref{as:mineig}, the hindsight solution of TIRSO  $\tbm a_n^*[T]$  given in \eqref {eq:minimizertirso} is bounded as
		\begin{eqnarray} \label{eq:boundonatirso}
			\left \lVert \tbm a_n^*[T] \right \rVert_2 \leq B_{\tbm a}\triangleq \frac{1}{\strongcvxparf} \left (\energybound \sqrt{PN} + \sqrt{\energybound^2PN+\strongcvxparf \energybound} \right  ).
		\end{eqnarray}
	\end{mylemma}
	\vspace{5mm}
	\begin{IEEEproof}
		The proof follows similar steps to those of  \thref{lemma:boundedhindsightsolutionTISO}. Consider the sublevel set of TIRSO hindsight objective for $\tbm a_n^*[T]= \bm 0_{NP}$,
		\begin{equation}
			\mathcal {\tilde S}_T \triangleq \left \{\tbm a_n^*[T]: \tirsohindsightobj_T(\tbm a_n^*[T])\leq \tirsohindsightobj_T(\bm 0_{NP}) \right \},
		\end{equation}
		where $\tirsohindsightobj_T(\bm 0_{NP})$ is upper bounded as follows:
		\begin{subequations}
			\begin{align*}
				\tirsohindsightobj_T(\bm 0_{NP}) &= \frac{1}{T-P}  \sum_{t=P}^{T-1}\frac{\mu}{2} \sum_{\tau=P}^{t} \gamma^{t-\tau}  y_n^2[t] \\
				& \leq \frac{\energybound\mu}{2(T-P)}\sum_{t=P}^{T-1}\sum_{\tau=P}^{t} \gamma^{t-\tau}\\
				& =\frac{\energybound\mu}{2(T-P)}\sum_{t=P}^{T-1}\frac{1-\gamma ^{t-P+1}}{1-\gamma} \nonumber\\
				& \leq \frac{\energybound}{2(T-P)}\sum_{t=P}^{T-1}1 \nonumber =\frac{\energybound}{2}.
			\end{align*}
		\end{subequations}
		This implies that
		\begin{equation} \label{eq:sublevelsetTIRSO1}
			\mathcal {\tilde S}_T \subset \left \{\tbm a_n^*[T]:  \tirsohindsightobj_T(\tbm a_n^*[T])\leq \frac{\energybound}{2} \right \}.
		\end{equation}
		Next, we find a lower bound to $\tirsohindsightobj_T(\tbm a_n^*[T])$ that is an increasing function of $\lVert \tbm a_n^*[T] \rVert _2$ as follows
		\begin{subequations}
			\begin{align*}
				\tirsohindsightobj_T(\tbm a_n^*[T]) &= \!\frac{1}{T-P}  \sum_{t=P}^{T-1} \Big [\frac{1}{2}(\tbm a_n^*[T])^\top \bm \Phi[t] \tbm a_n^*[T] \!- \!\bm r_n^\top[t]  \tbm a_n^*[T] \\
				&+ \frac{\mu}{2}  \sum_{\tau=P}^{t}  \gamma^{t-\tau} y_n^2[t] + \nReg (\tbm a_n^*[T])\Big ]\\
				& \geq \frac{1}{T-P}  \sum_{t=P}^{T-1} \Big [\frac{1}{2}(\tbm a_n^*[T])^\top \bm \Phi[t] \tbm a_n^*[T]\! - \!\bm r_n^\top[t]  \tbm a_n^*[T] \\
				&+ \frac{\mu}{2}  \sum_{\tau=P}^{t}  \gamma^{t-\tau} y_n^2[t] \Big ] \\
				& \geq  \frac{1}{T-P}  \sum_{t=P}^{T-1} \Big [\frac{1}{2} \lambda_{\mathrm{min}}(\bm \Phi[t] )\lVert \tbm a_n^*[T] \rVert _2^2 \\
				& -\left \lVert \bm r_n[t]\right \rVert_2 \lVert \tbm a_n^*[T] \rVert _2  \Big ] \\
				& \geq  \frac{1}{T-P} \! \sum_{t=P}^{T-1} \! \left [\frac{1}{2} \strongcvxparf \lVert \tbm a_n^*[T] \rVert _2^2\!-\!\energybound \sqrt{PN} \lVert \tbm a_n^*[T] \rVert _2  \right ] \\
				& =   \frac{1}{2} \strongcvxparf \lVert \tbm a_n^*[T] \rVert _2^2-\energybound \sqrt{PN} \lVert \tbm a_n^*[T] \rVert _2.
			\end{align*}
		\end{subequations}
		Therefore,
		\begin{eqnarray} \label{eq:sublevelsetTIRSO2}
			\mathcal {\tilde S}_T \!\subset\! \left \{\tbm a_n^*[T]:   \frac{1}{2} \strongcvxparf \lVert \tbm a_n^*[T] \rVert _2^2\!-\! \energybound \sqrt{PN} \lVert \tbm a_n^*[T] \rVert _2\! \leq\! \frac{\energybound}{2} \right \}.
		\end{eqnarray}
		Further, we can write
		\begin{equation} \label{eq:sublevelsetTIRSO3}
			\textstyle 
			\mathcal {\tilde S}_T \!\subset\! \left \{\tbm a_n^*[T]\!: \! \left \lVert \tbm a_n^*[T] \right \rVert_2 \leq \frac{1}{\strongcvxparf} \left (\energybound \sqrt{PN} \!+\! \sqrt{\energybound^2PN\!+\!\strongcvxparf \energybound} \right  ) \right \}.
		\end{equation}
		Expression \eqref {eq:sublevelsetTIRSO3} implies that the TIRSO hindsight solution is bounded.
	\end{IEEEproof}
	Now, we present the proof of \thref {prop:regrettirso}. }
\begin{myitemize}%
	\myitem \cmt{Introducing $\trho$}\rev{The proof has two parts. The first step is to prove that there exists  $ \trho>0$ such that
		\begin{equation} \label{eq:cond_for_rec}
			\big \lVert \nabla \tilde{\ell}_t^{(n)}(\bm a_n) \big  \rVert_2^2  \leq \trho  \, \tilde\ell_t^{(n)}(\bm a_n), ~~ \forall~  t, \, n,
		\end{equation}
		holds for all $\bm a_n$. The second step is to apply the result of \cite[Corollary 5]{duchi2010comid} in the present case. To prove the first part,} from  \eqref{eq:recloss3} and  $    \nabla \tilde
	\ell_{t}^{(n)}(\bm a_n)= \bm \Phi[t] \bm a_n-\bm r_n[t]$, it follows that \eqref{eq:cond_for_rec} is equivalent to
	\begin{align} \label{eq:recineq1}
		\lVert \bm \Phi[t] \bm a_n-\bm r_n[t] \rVert_2^2 &\leq \trho \Big (\frac{1}{2}\bm a_n^\top \bm \Phi[t] \bm a_n -\bm r_n^\top[t]  \bm a_n \nonumber \\
		&+\frac{1}{2} \textstyle \sum_{\tau=P}^{t} \mu \gamma^{t-\tau} y_n^2[t] \Big ), ~\forall ~ t,\,n.
	\end{align}
	By expanding the left-hand side of \eqref{eq:recineq1}, rearranging terms, and introducing $Z_t(\bm a_n)$ as
	\begin{align} \label{eq:recineq4}
		Z_t(\bm a_n&) \define  \;\bm a_n^\top \Big ( \frac{\trho}{2} \bm \Phi[t]- \bm \Phi^\top[t] \bm \Phi[t] \Big) \bm a_n + (2\bm r_n^\top[t] \bm \Phi[t] \nonumber \\
		&-\trho \, \bm r_n^\top[t]  )\bm a_n+\frac{\trho \mu}{2} \sum_{\tau=P}^{t} \gamma^{t-\tau} y_n^2[t] - \bm r_n^\top[t] \bm r_n[t],
	\end{align}
	the condition in \eqref{eq:cond_for_rec} is equivalent to $Z_t(\bm a_n) \geq 0$.
	So the goal becomes finding
	$\trho $ such that $Z_t(\bm
	a_n) \geq 0$ for all $\bm a_n$
	and $t$.
	\myitem \cmt{Conditions}For this
	condition to hold, it is necessary
	that (a) $\inf_{\bm a_n} Z_t(\bm a_n)$ is
	finite for all $t$, and (b)
	$\inf_{\bm a_n} Z_t(\bm a_n)\geq 0$
	for all $t$.
	\myitem\cmt{inf is finite}It can be seen \cite[Appendix A.5]{boyd} that condition (a) holds iff (a1) the Hessian matrix $ \bm H Z_t(\bm a_n) = {\trho} \bm \Phi[t]- 2\bm \Phi^\top[t] \bm \Phi[t]$ is positive semidefinite, and (a2) $2\bm \Phi[t]\bm r_n[t] -\trho \bm r_n[t] \in \mathcal R  ( \bm H Z_t(\bm a_n)   )$,  where $\mathcal R(\bm A)$ denotes the span of the columns of a matrix $\bm A$.
	\begin{myitemize}%
		\myitem\cmt{(a1) Positive Semidefiniteness of $\bm \Phi$}The first step is to find $\trho$ such that (a1) holds. To this end, consider the eigenvalue decomposition of $\bm \Phi[t]= \bm U \bm \Lambda \bm U^\top$, where the index $t$ is omitted to simplify notation. Therefore,
		\begin{equation} \label{eq:hessianztevd}
			\bm HZ_t(\bm a_n)=\bm U \left ( \trho \bm \Lambda - 2 \bm \Lambda^2 \right) \bm U^\top   .
		\end{equation}
		Let $\lambda_{\textrm{max}}(\bm \Phi[t])$ denote the maximum eigenvalue of
		$\bm \Phi[t]$. It follows from
		\eqref{eq:hessianztevd} that $\bm H Z_t(\bm a_n)$
		is positive semidefinite if
		\begin{equation}\label{eq:rho_cond}
			\trho \geq 2\lambda_{\textrm{max}}(\bm \Phi[t]).
		\end{equation}
		\myitem\cmt{(a2) $b \in
			\mathcal R(A)$}It remains to
		be shown that there exists
		$\trho>0$ such that
		\eqref{eq:rho_cond}, (a2), and
		(b) simultaneously hold. To
		this end,  focus first on (a2), which can be rewritten as                                         \begin{align} \label{eq:inspan}
			2\bm \Phi[t]\bm r_n[t] -\trho \bm r_n[t] &\in \mathcal {R} \left ( {\trho} \bm \Phi[t]-2 \bm \Phi^\top[t] \bm \Phi[t]  \right )\\
			&= \mathcal {R} \left(\bm \Phi[t] \left( {\trho} \bm I -
			2\bm \Phi[t] \right
			) \right ).\nonumber
		\end{align}
		Clearly, if $					\trho >
		2\lambda_{\textrm{max}}(\bm \Phi[t])$, then $                                                {\trho} \bm I -
		2\bm \Phi[t] $ is
		invertible and, hence, $\mathcal {R}
		(\bm \Phi[t] (
		{\trho} \bm I -
		2\bm \Phi[t]
		)  )
		= \mathcal {R}
		(\bm \Phi[t]
		)
		$  \cite[Ch. 4]{meyer2000}.
		Thus, \eqref{eq:inspan} holds if $2\bm \Phi[t]\bm r_n[t] \in  \mathcal {R}
		(\bm \Phi[t]
		)$ and
		$\trho \bm r_n[t] \in  \mathcal {R}
		(\bm \Phi[t]
		)$. The former
		condition is
		trivial. To verify the
		latter, define
		\begin{subequations}
			\begin{align}
				\bm  y_n   &\define \left [y_n[P],\ldots, y_n[t]\right ]^\top \in \mathbb R^{t-P+1\times 1},\\
				\bm  G     &\define \left[\bm g[P], \ldots, \bm g[t] \right ] \in \mathbb R^{NP\times t-P+1},\\
				\bm \Gamma &\define \mathrm{diag}\left ( \mu[\gamma^{t-P},  \ldots, \gamma^0]\right )  \in \mathbb R^{t-P+1\times t-P+1},                                                                                                                                                                                                                          \end{align}
		\end{subequations}
		and $\bm B \define \bm G \bm \Gamma^{1/2}$; note that
		$\bm \Phi[t]=\bm
		G \bm \Gamma \bm G^\top=\bm
		B \bm B^\top $. It follows
		that  $\bm
		r_n[t]=\bm G \bm \Gamma \bm
		y_n= \bm B\bm \Gamma^{1/2} \bm y_n \in
		\mathcal R( \bm
		B)
		=\mathcal {R} ( \bm B \bm B^\top   )=
		\mathcal R (\bm \Phi [t] )
		$. Therefore, $\trho \bm r_n[t] \in  \mathcal {R}
		(\bm \Phi[t]
		)$ holds and,
		consequently, (a2)
		holds whenever $					\trho >
		2\lambda_{\textrm{max}}(\bm \Phi[t])$.

	\end{myitemize}%
	\myitem \cmt{inf $>0$}So far,
	this proof has established
	that, if $					\trho >
	2\lambda_{\textrm{max}}(\bm \Phi[t])$, then both (a1) and (a2)
	hold. The next step is to show that (b) also holds when  $					\trho >
	2\lambda_{\textrm{max}}(\bm \Phi[t])$.
	\begin{myitemize}%
		\myitem \cmt{obtain infimum}To this end,
		set the gradient of
		$Z_t(\bm a_n)$ equal
		to zero and use  $					\trho >
		2\lambda_{\textrm{max}}(\bm \Phi[t])$ to  obtain
		$ \bm \Phi^{\dagger}[t] \bm
		r_n[t] \in \underset{\bm
			a_n}{\arg \min }~ Z_t(\bm a_n)
		$, where the symbol $\dagger$
		denotes
		pseudo-inverse. From this
		expression and   \eqref{eq:recineq4},
		it follows that
		\begin{multline}                                                     
			\underset{\bm a_n}{\inf}~ Z_t(\bm a_n)=Z_t(\bm \Phi^{\dagger}[t] \bm r_n[t]) \nonumber\\                                                                     
			= \bm r_n^\top[t] \bm \Phi^{\dagger} [t]\Big ( \frac{\trho}{2}\bm \Phi[t]- \bm \Phi^\top[t] \bm \Phi[t]  \Big ) \bm \Phi ^{\dagger}[t]\bm r_n[t]+ \big ( 2\bm r_n^{\top}[t] \bm \Phi[t] \nonumber\\                                                                              - \trho \bm r_n^\top[t]\big ) \bm \Phi ^{\dagger}[t] \bm r_n[t] \nonumber + \frac{\trho \mu}{2} \textstyle \sum_{\tau=P}^{t}\gamma^{t-\tau} y_n^2[t] - \bm r_n^\top[t] \bm r_n[t].   
		\end{multline}                                                      
		Applying the properties of the pseudoinverse
		and simplifying results in                                                            \begin{equation} \textstyle \label {eq:condition2}
			\underset{\bm a_n}{\inf}~  Z_t(\bm a_n)=\frac{\trho \mu}{2}\sum_{\tau=P}^{t}\gamma^{t-\tau} y_n^2[t]- \frac{\trho }{2}\bm r_n^\top[t] \bm \Phi ^{\dagger}[t] \bm r_n[t].
		\end{equation}
		\myitem \cmt{Minimum $\geq
			0$}From this expression, note
		that the condition $\underset{\bm a_n}{\inf}~ Z_t(\bm a_n) \geq 0$ is equivalent to
		\begin{equation} \label{eq:cond6}
			\bm
			y_n^\top \bm \Gamma \bm
			y_n \geq \bm
			y_n^\top \bm \Gamma \bm
			G^\top \left ( \bm
			G \bm \Gamma \bm
			G^\top \right)^{\dagger} \bm
			G \bm \Gamma \bm y_n,                                                                                   \end{equation}     and, upon defining $\tilde {\bm y}_n \define \bm \Gamma^{1/2} \bm y_n$,
		\begin{equation} \label{eq:cond7}
			\tilde {\bm y}_n^\top \tilde {\bm y}_n \geq \tilde {\bm y}_n^\top \bm \Gamma^{1/2} \bm G^\top \left ( \bm G \bm \Gamma \bm G^\top \right)^{\dagger} \bm G \bm \Gamma^{1/2} \tilde {\bm y}_n^\top.
		\end{equation}
		This inequality trivially holds when $\tilde {\bm y}_n=\bm 0_{t-P+1}$. Thus,
		assume without loss of generality that $\tilde {\bm
			y}_n\neq \bm 0_{t-P+1}$.
		By  setting $\bm A\define\bm \Gamma^{1/2} \bm
		G^\top  ( \bm
		G \bm \Gamma \bm
		G^\top )^{\dagger} $,
		one obtains
		$\bm A \bm B = \bm \Gamma^{1/2} \bm G^\top  ( \bm G \bm \Gamma \bm G^\top )^{\dagger} \bm G \bm \Gamma^{1/2} $ and
		$\bm B \bm
		A= \bm \Phi[t] \bm \Phi^{\dagger}[t]
		$.
		
		Since the nonzero eigenvalues
		of $\bm A \bm B$ and $\bm
		B \bm A$ are the
		same~\cite[Sec. 3.2.11]{horn1985} and the maximum
		eigenvalue of $\bm B \bm A$ is
		1, then the maximum eigenvalue
		of $\bm A \bm B$ is also 1. Therefore
		\begin{equation}
			\frac{\tilde {\bm y}_n^\top \bm A \bm B\tilde {\bm y}_n}{\tilde {\bm y}_n^\top\tilde {\bm y}_n}=\frac{\tilde {\bm y}_n^\top \bm \Gamma^{1/2} \bm G^\top \left ( \bm G \bm \Gamma \bm G^\top \right)^{\dagger} \bm G \bm \Gamma^{1/2} \tilde {\bm y}_n}{\tilde {\bm y}_n^\top \tilde {\bm y}_n} \leq 1,
		\end{equation}
		and, hence,  \eqref{eq:cond7} holds. To sum up, conditions (a) and (b)
		hold if $					\trho >
		2\lambda_{\textrm{max}}(\bm \Phi[t])$. In other
		words, \eqref{eq:cond_for_rec}
		holds for any choice of
		$\trho$ such that $					\trho >
		2\lambda_{\textrm{max}}(\bm \Phi[t])$ for all $t$. \rev {This completes the first part of the proof.}
	\end{myitemize}%
	\myitem \cmt{Applying the
		Corollary}\rev{The second part of the proof consists of setting}
	$\trho=\underset{t}{\textrm
		{sup}}
	~\lambda_{\textrm{max}}(\bm \Phi[t])
	+
	\epsilon$ with $\epsilon>0$ an
	arbitrary constant,
	and invoking \cite[Corollary
	5]{duchi2010comid} to conclude that
	\begin{equation*}
		\tilde R\rev{_{s}^{(n)}[T]}=\mathcal{O}\left ( \trho \, \left\lVert \tbm a_n^{*}[T] \, \right\rVert_2^2 \,\sqrt{T}\right ).
	\end{equation*}
	\rev {Using assumption A\ref{as:maxeig} and substituting the upper bound on $\lVert \tbm a_n^{*}[T]  \rVert_2$ from \eqref {eq:boundonatirso} into the above expression completes the proof.}
\end{myitemize}%
\section {Proof of \thref {th:strongconvexitytirso}}
\label{appendix:proof{th:strongconvexitytirso}}
\rev {
	To prove \thref {th:strongconvexitytirso}, first we present two lemmas. Before presenting the result related to logarithmic regret of TIRSO, it is worth mentioning that a related result is presented in \cite[Th. 7]{duchi2010comid}, which is applicable to strongly convex regularization functions. Note that in TIRSO, the data-fitting function is strongly convex. It can be easily shown that COMID applied to a problem with strongly convex regularizer produces different iterates than COMID applied to a strongly convex data-fitting function.
	\begin{mylemma} \thlabel{lemma:strongcvxlemma}
		\begin{myitemize}%
			\myitem \cmt{strongly convex f}Under assumption A\ref{as:mineig},
			\myitem \cmt{TIRSO}let the sequence  $\{\tbm a_n[t]\}_{t=P}^{T}$ be generated by TIRSO (\textbf{Procedure~\ref{alg:TIRSO}}) with a step size $\alpha_t$,
			\myitem \cmt{hindsight solution}and let $\tbm a_n^{*}[T]$ be the hindsight solution for TIRSO at time $T$ defined in \eqref{eq:minimizertirso}.
		\end{myitemize}%
		Then
		\begin{multline} \label {eq:strongcvxlemmaresult}
			\ltn(\tbm a_n[t])+\nReg(\tbm a_n[t+1])-\ltn(\tbm a_n^{*}[T])- \nReg(\tbm a_n^{*}[T]) \leq \\
			\frac{1}{2\alpha_t}  (1- \alpha_t \strongcvxparf) \left \lVert \tbm a_n^{*}[T]- \tbm a_n[t] \right \rVert_2^2- \frac{1}{2\alpha_t} \left \lVert  \tbm a_n^{*}[T]- \tbm a_n[t+1] \right \rVert_2^2 \\+ \frac{\alpha_t}{2} \left \lVert  \gradf \right \rVert_2^2,
		\end{multline}
		for $P\leq t\leq T$, $\forall~ \gradf \in \partial (\ltn (\tbm a_n[t]))$.
	\end{mylemma}
	\begin{IEEEproof}
		For a strongly convex $\ltn$, by the subgradient inequality, we have
		\begin{multline} \label{eq:inequalityforf}
			\ltn (\tbm a_n^{*}[T]) \geq \ltn (\tbm a_n[t])+   (\tbm a_n^{*}[T] - \tbm a_n[t])^\top \gradf \\ +\frac{\strongcvxparf}{2} \left \lVert \tbm a_n^{*}[T] - \tbm a_n[t] \right \rVert_2^2,
		\end{multline}
		$\forall \gradf \in \partial (\ltn (\tbm a_n[t]))$. On the other hand, since $\nReg$ is convex,
		\begin{equation}\label{eq:inequalityforOmega}
			\nReg (\tbm a_n^{*}[T]) \geq \nReg (\tbm a_n[t+1]) +   \left(\tbm a_n^{*}[T] - \tbm a_n[t+1]\right )^\top \gradreg,
		\end{equation}
		$\forall ~\gradreg \in \partial (\nReg (\tbm a_n[t+1]))$. Adding \eqref {eq:inequalityforf} and \eqref {eq:inequalityforOmega}, scaling by $\alpha_t$, and rearranging terms,
		\begin{align}
			&\!\alpha_t \left (\ltn (\tbm a_n[t])\!+\!\nReg (\tbm a_n[t+1])\!-\!\ltn (\tbm a_n^{*}[T])\!-\!\nReg (\tbm a_n^{*}[T]) \right ) \nonumber \\
			&\leq  \alpha_t \Big( (\tbm a_n[t]- \tbm a_n^{*}[T])^\top \gradf + (\tbm a_n[t+1]- \tbm a_n^{*}[T])^\top \gradreg \nonumber\\
			& - \frac{\strongcvxparf}{2} \left \lVert \tbm a_n[t] - \tbm a_n^{*}[T] \right\rVert_2^2 \Big ) \nonumber \\
			\stackrel{(a)}{=} & \left(\tbm a_n^{*}[T] - \tbm a_n[t+1]\right )^\top \left(\tbm a_n[t] - \tbm a_n[t+1]- \alpha_t \gradf - \alpha_t \gradreg\right ) \nonumber \\
			& + \alpha_t \left(\tbm a_n[t]-\tbm a_n[t+1]\right )^\top \gradf -\frac{\alpha_t \strongcvxparf}{2} \left \lVert \tbm a_n[t] - \tbm a_n^{*}[T] \right\rVert_2^2 \nonumber \\
			&+\left(\tbm a_n^{*}[T] - \tbm a_n[t+1]\right)^\top \left(\tbm a_n[t+1]- \tbm a_n[t]\right ) \nonumber \\
			\stackrel{(b)}{\leq} & \alpha_t \left(\tbm a_n[t]-\tbm a_n[t+1]\right )^\top \gradf -\frac{\alpha_t \strongcvxparf}{2} \left \lVert \tbm a_n[t] - \tbm a_n^{*}[T] \right\rVert_2^2 \nonumber \\
			& +  \left(\tbm a_n^{*}[T] -	\tbm a_n[t+1]\right )^\top \left(\tbm a_n[t+1]- \tbm a_n[t]\right ) \nonumber \\
			\stackrel{(c)}{=} & \alpha_t \left \langle \frac{1}{\sqrt{\alpha_t}}(\tbm a_n[t]-\tbm a_n[t+1]), \sqrt{\alpha_t}\gradf \right \rangle \!-\!\frac{\alpha_t \strongcvxparf}{2} \left \lVert \tbm a_n[t] - \tbm a_n^{*}[T] \right\rVert_2^2 \nonumber \\
			&  + \frac{1}{2} \left \lVert \tbm a_n^{*}[T]-\tbm a_n[t]\right \rVert_2^2 - \frac{1}{2} \left \lVert \tbm a_n^{*}[T]-\tbm a_n[t+1]\right \rVert_2^2\nonumber\\
			&- \frac{1}{2} \left \lVert \tbm a_n[t+1]-\tbm a_n[t]\right \rVert_2^2 \nonumber \\
			\stackrel{(d)}{\leq} & \frac{1}{2}\left \lVert \tbm a_n[t]-\tbm a_n[t+1]\right \rVert_2^2+ \frac{\alpha_t^2}{2} \left \lVert \gradf \right \rVert_2^2  - \frac{1}{2} \left \lVert \tbm a_n[t+1]-\tbm a_n[t]\right \rVert_2^2  \nonumber \\
			& - \frac{1}{2} \left \lVert \tbm a_n^{*}[T]-\tbm a_n[t+1] \right \rVert_2^2  +\left(\frac{1}{2}-\frac{\alpha_t \strongcvxparf}{2}\right ) \left \lVert \tbm a_n[t] - \tbm a_n^{*}[T] \right\rVert_2^2\nonumber \\
			= & \frac{\alpha_t^2}{2} \left \lVert \gradf \right \rVert_2^2 +\left(\frac{1}{2}-\frac{\alpha_t \strongcvxparf}{2}\right) \left \lVert \tbm a_n[t] - \tbm a_n^{*}[T] \right\rVert_2^2 \nonumber
			\\
			&- \frac{1}{2} \left \lVert \tbm a_n^{*}[T]-\tbm a_n[t+1] \right \rVert_2^2, \label{eq:prooflemmaresult}
		\end{align}
		where (a) results from adding and subtracting the term $\tbm a_n^\top[t+1] \gradf + (\tbm a_n^{*}[T] - \tbm a_n[t+1])^\top (\tbm a_n[t]- \tbm a_n[t+1])$ followed by rearranging terms; in (b) the inequality $(\tbm a_n^{*}[T] - \tbm a_n[t+1])^\top (\tbm a_n[t] - \tbm a_n[t+1]- \alpha_t \gradf - \alpha_t \gradreg)\leq 0$ is used, which is implied by the optimality of $\tbm a_n[t+1]$ in \eqref {eq:TIRSOupdateexp}, i.e., $(\bm a_n - \tbm a_n[t+1])^\top (\tilde \nabla \tilde {J}_t^{(n)}(\tbm a_n[t+1]))\geq 0, \forall \, \bm a_n$; in (c) the Pythagorean theorem for Euclidean distance (i.e. $(\tbm a_n^{*}[T] -\tbm a_n[t+1])^\top (\tbm a_n[t+1]- \tbm a_n[t])=1/2  \lVert \tbm a_n^{*}[T]-\tbm a_n[t] \rVert_2^2 - 1/2  \lVert \tbm a_n^{*}[T]-\tbm a_n[t+1] \rVert_2^2- 1/2  \lVert \tbm a_n[t+1]-\tbm a_n[t] \rVert_2^2$) is used;  in (d) the inequality $\langle  \bm x, \bm y\rangle \leq 1/2(\Vert \bm x \rVert_2^2+ \lVert \bm y \rVert_2^2)$ is used. Dividing both sides of \eqref {eq:prooflemmaresult} by~$\alpha_t$ completes the proof.
\end{IEEEproof}}
\vspace{5mm}
\rev {
	Next, we establish that TIRSO estimates $\tbm a_n[t] $ are bounded and a bound on $ \lVert \nabla \ltn(\tbm a_n[t])  \rVert_2$  that depends on parameters of the algorithm, is derived. %
	\begin{mylemma} \thlabel{lemma:boundingestiamtes}
		\cmt{Hypotheses}%
		\begin{myitemize}%
			\myitem \cmt{assumptions}Under assumptions A\ref{as:boundedprocess} and A\ref{as:mineig},
			\myitem \cmt{TIRSO  estimates}and let the sequence of iterates $\{\tbm a_n[t]\}$ be generated by TIRSO (\textbf{Procedure~\ref{alg:TIRSO}}).
		\end{myitemize}%
		Then 	
		\begin{align} \label{eq:boundedestimates1}
			\left \lVert \tbm a_n[t+1] \right  \rVert_2 &  \leq \left ( 1 - \alpha_t\strongcvxparf \right )\left \lVert \tbm a_n[t] \right \rVert_2 + \alpha_t \sqrt{PN} \energybound.
		\end{align}
	\end{mylemma}
	\begin{IEEEproof}
		From the update expression of TIRSO, we have
		\begin{align}
			\left \lVert \tbm a_n[t+1] \right  \rVert_2 & \leq \left \lVert \tbm a_n^{\text{f}}[t+1] \right  \rVert_2 \nonumber \\
			& = \left \lVert \tbm a_n[t] - \alpha_t \bm v_n[t]\right  \rVert_2 \nonumber \\
			& = \left \lVert \tbm a_n[t] - \alpha_t \left (\bm \Phi [t] \tbm a_n[t] -\bm r_n[t] \right )\right  \rVert_2 \nonumber \\
			& =\left \lVert \left ( \bm I - \alpha_t \bm \Phi[t]\right)\tbm a_n[t] + \alpha_t \bm r_n[t] \right  \rVert_2 \nonumber \\
			& \leq  \lambda_{\mathrm{max}} \left ( \bm I - \alpha_t \bm \Phi[t]\right) \left \lVert \tbm a_n[t] \right \rVert_2+ \alpha_t \left \lVert \bm r_n[t] \right  \rVert_2 \nonumber\\
			& =\left ( 1 - \alpha_t \lambda_{\mathrm{min}} ( \bm \Phi[t]\right) \left \lVert \tbm a_n[t] \right \rVert_2 + \alpha_t \left \lVert \bm r_n[t] \right  \rVert_2 \nonumber \\
			&\leq \left ( 1 - \alpha_t\strongcvxparf \right )\left \lVert \tbm a_n[t] \right \rVert_2 + \alpha_t \left \lVert \bm r_n[t] \right  \rVert_2. \label{eq:boundonestimates}
		\end{align}
		Now, we derive an upper bound on $\lVert \bm r_n[t]  \rVert_2$. By the definition of $\bm r_n[t]$ in \eqref {eq:tirsocrosscov} and assumption A\ref{as:boundedprocess}, we have
		\begin{subequations}
			\begin{align}
				\lVert \bm r_n[t]  \rVert_2& = \left \lVert  \mu \sum_{\tau=P}^{t} \,  \gamma^{t-\tau} y_n[\tau] \,  \bm g[\tau] \right \rVert_2 \nonumber  \\
				& \leq \mu \left \lVert   \sum_{\tau=P}^{t} \,  \gamma^{t-\tau} \sqrt{\energybound} \sqrt{\energybound} \bm 1_{NP} \right \rVert_2   \\
				& =  \mu \energybound \sqrt{PN} \gamma^{t} \sum_{\tau=P}^{t} \left (\frac{1}{\gamma}\right )^\tau \nonumber  \\
				&=  \energybound \sqrt{PN} (1-\gamma^{t-P+1})\nonumber \\
				& \leq  \sqrt{PN} \energybound.  \label{eq:boundonr}
			\end{align}
		\end{subequations}
		Substituting the upper bound of $\bm r_n[t]$ from \eqref {eq:boundonr} into \eqref {eq:boundonestimates} completes the proof. 
	\end{IEEEproof}
}
\vspace{5mm}
\rev{ 
	\begin{mylemma} \thlabel{lemma:boundg}
		\cmt{Hypotheses}%
		\begin{myitemize}%
			\myitem \cmt{assumptions}Under assumptions A\ref{as:boundedprocess}, A\ref{as:mineig}, and A\ref{as:maxeig},
			\myitem \cmt{TIRSO  estimates}and let the sequence of iterates $\{\tbm a_n[t]\}$ be generated by TIRSO (\textbf{Procedure~\ref{alg:TIRSO}})
			\myitem \cmt{decreasing step size}with $\alpha_t=1/(\strongcvxparf t)$.
		\end{myitemize}%
		Then \begin{equation}\label{eq:boundonTIRSOiterate}
			\lVert \tbm a_n[t] \rVert_2 \leq 1/\strongcvxparf \sqrt{PN} \energybound, \forall\, t\geq P,
		\end{equation}
		\begin{eqnarray} \label {eq:boundonTIRSOgrad}
			\left \lVert \nabla \ltn(\tbm a_n[t]) \right \rVert_2\! \leq \gradboundf \triangleq \! \left (1+\frac{L}{\strongcvxparf}\right ) \!\sqrt{PN} \energybound, \forall\, t\geq P.
		\end{eqnarray}
	\end{mylemma}
}%
\vspace{5mm}
\rev{
	\begin{IEEEproof}%
		\begin{myitemize}%
			\myitem \cmt{proving bounded estimates}Invoking \thref {lemma:boundingestiamtes} and setting $\alpha_t= 1/(\strongcvxparf t)$ in \eqref {eq:boundedestimates1},
			\begin{subequations}
				\begin{align}
					\left \lVert \tbm a_n[t+1] \right  \rVert_2 & = \left ( 1 - \frac{1}{\strongcvxparf t}\strongcvxparf \right) \left \lVert \tbm a_n[t] \right \rVert_2 + \frac{1}{\strongcvxparf t} \sqrt{PN} \energybound \nonumber \\
					& \leq \left ( 1 - \frac{1}{t} \right) \left \lVert \tbm a_n[t] \right \rVert_2 + \frac{1}{\strongcvxparf t} \sqrt{PN} \energybound \label {eq:boundonarecursive}\\
					& \leq \left ( 1 - \frac{1}{t} \right) \Bigg [ \left ( 1 - \frac{1}{t-1} \right)\left \lVert \tbm a_n[t-1] \right \rVert_2  \nonumber \\
					& + \frac{1}{\strongcvxparf (t-1)} \sqrt{PN} \energybound \Bigg ]+ \frac{1}{\strongcvxparf t} \sqrt{PN} \energybound \nonumber\\
					& \leq \left ( \frac{t-2}{t} \right) \left \lVert \tbm a_n[t-1] \right \rVert_2  + \frac{2}{\strongcvxparf t} \sqrt{PN} \energybound.
				\end{align}
			\end{subequations}
			Substituting the upper bound of $\lVert \tbm a_n[t-1]  \rVert_2$ using \eqref {eq:boundonarecursive}, we have
			\begin{align*}
				\left \lVert \tbm a_n[t+1] \right  \rVert_2 \leq \left ( \frac{t-3}{t} \right) \left \lVert \tbm a_n[t-2] \right \rVert_2  + \frac{3}{\strongcvxparf t} \sqrt{PN} \energybound.
			\end{align*}
			After $k$ substitutions, the above bound can be written in terms of $k$ as follows
			\begin{align*}
				\left \lVert \tbm a_n[t+1] \right  \rVert_2 & \leq \left ( \frac{t-k}{t} \right) \left \lVert \tbm a_n[t-k+1] \right \rVert_2  + \frac{k}{\strongcvxparf t} \sqrt{PN} \energybound,
			\end{align*}
			$1\leq k \leq t-P+1$. The bound on $\lVert \tbm a_n[t+1] \rVert_2$ in terms of the initial estimate $\lVert \tbm a_n[P]  \rVert_2$ is obtained for $k=t-P+1$ in the above inequality, given by
			\begin{subequations}
				\begin{align*}
					\left \lVert \tbm a_n[t+1] \right  \rVert_2 & \leq \left ( \frac{P-1}{t} \right) \left \lVert \tbm a_n[P] \right \rVert_2  + \frac{t-P+1}{\strongcvxparf t} \sqrt{PN} \energybound \\
					& = \frac{\sqrt{PN} \energybound }{\strongcvxparf} - \frac{P-1}{\strongcvxparf t} \sqrt{PN} \energybound \\
					& \leq \frac{\sqrt{PN} \energybound }{\strongcvxparf}, ~ t\geq P.
				\end{align*}
			\end{subequations}
			This completes the proof of \eqref{eq:boundonTIRSOiterate}, the first part of the lemma.
			\myitem \cmt{Proving the bound}To prove the second part of the lemma, by taking the value of the gradient in \eqref {eq:tirsograd}, and by the triangular inequality,
			\begin{subequations}
				\begin{align}
					\left \lVert \nabla \ltn (\tbm a_n[t]) \right \rVert_2 &= \left \lVert \bm \Phi[t] \tbm a_n[t]- \bm r_n[t] \right \rVert_2 \nonumber\\
					& \leq \left \lVert \bm \Phi[t] \tbm a_n[t] \right \rVert_2 + \left \lVert  \bm r_n[t] \right \rVert_2 \nonumber \\
					& \leq \lambda_{\mathrm{max}}(\bm \Phi[t])\left \lVert  \tbm a_n[t] \right \rVert_2 +  \left \lVert  \bm r_n[t] \right \rVert_2 \label{eq:generalgradbound} \\
					& \leq L\frac{\sqrt{PN} \energybound }{\strongcvxparf} +  \sqrt{PN} \energybound \nonumber \\
					& \leq \left (1+\frac{L}{\strongcvxparf}\right ) \sqrt{PN} \energybound.  \nonumber
				\end{align}
			\end{subequations}
		\end{myitemize}%
\end{IEEEproof}}
\rev{
	Now, we are ready to prove \thref{th:strongconvexitytirso}.
	We start from the result presented in \thref{lemma:strongcvxlemma}.
	Summing both sides of \eqref {eq:strongcvxlemmaresult} from $t=P$ to $T$ results in 
	\begin{align} 
		&\hspace{-8mm}\sum_{t=P}^{T} \! \left (\ltn (\tbm a_n[t])\!+\!\nReg (\tbm a_n[t\!+\!1])\!-\!\ltn (\tbm a_n^{*}[T])\!-\!\nReg (\tbm a_n^{*}[T]) \right ) \nonumber \\
		& \hspace{-4mm}\leq \sum_{t=P}^{T}   \Big(\frac{\alpha_t}{2} \left \lVert \gradf \right \rVert_2^2  +  \Big (\frac{1}{2\alpha_t}-\frac{ \strongcvxparf}{2}\Big ) \left \lVert \tbm a_n[t] - \tbm a_n^{*}[T] \right\rVert_2^2 \nonumber \\
		&  - \frac{1}{2\alpha_t} \left \lVert \tbm a_n^{*}[T]-\tbm a_n[t+1] \right \rVert_2^2  \Big ) \label{eq:proof1TIRSO} \\
		= &  \frac{1}{2}  \sum_{t=P}^{T} \left (\frac{1}{\alpha_t}- \strongcvxparf \right ) \left \lVert \tbm a_n^{*}[T]-\tbm a_n[t]\right \rVert_2^2
		- \frac{1}{2}  \sum_{t=P}^{T} \frac{1}{\alpha_t}  \lVert \tbm a_n^{*}[T] \nonumber \\
		& -\tbm a_n[t+1]  \rVert_2^2  +\frac{1}{2} \sum_{t=P}^{T} \alpha_t \left \lVert \gradf \right \rVert_2^2   \nonumber\\
		= &  \frac{1}{2}  \sum_{k=P-1}^{T-1} \left(\frac{1}{\alpha_{k+1}}- \strongcvxparf \right) \left \lVert \tbm a_n^{*}[T]-\tbm a_n[k+1]\right \rVert_2^2 \nonumber \\
		& - \frac{1}{2}  \sum_{t=P}^{T} \frac{1}{\alpha_t} \left \lVert \tbm a_n^{*}[T]-\tbm a_n[t+1] \right \rVert_2^2 +\frac{1}{2} \sum_{t=P}^{T} \alpha_t \left \lVert \gradf \right \rVert_2^2   \nonumber \\
		= &  \frac{1}{2}  \sum_{k=P-1}^{T-1} \frac{1}{\alpha_{k+1}} \left \lVert \tbm a_n^{*}[T]-\tbm a_n[k+1]\right \rVert_2^2
		+ \frac{1}{2} \sum_{t=P}^{T} \alpha_t \left \lVert \gradf \right \rVert_2^2 \nonumber \\
		& - \frac{1}{2}  \sum_{t=P-1}^{T-1} \frac{1}{\alpha_t} \left \lVert \tbm a_n^{*}[T]-\tbm a_n[t+1] \right \rVert_2^2 +\frac{1}{2\alpha_{P-1}}  \lVert \tbm a_n^{*}[T] \nonumber \\
		&-\tbm a_n[P]  \rVert_2^2  - \frac{1}{2\alpha_T} \left \lVert \tbm a_n^{*}[T]-\tbm a_n[T+1] \right \rVert_2^2   \nonumber \\
		\stackrel{(a)}{\leq} &  \frac{1}{2}  \sum_{t=P-1}^{T-1}  \left \lVert \tbm a_n^{*}[T]-\tbm a_n[t+1]\right \rVert_2^2 \left (\frac{1}{\alpha_{t+1}}- \frac{1}{\alpha_{t}}-\strongcvxparf \right) \nonumber \\
		&+\frac{1}{2} \sum_{t=P}^{T} \alpha_t \left \lVert \gradf \right \rVert_2^2 +\frac{1}{2\alpha_{P-1}} \left \lVert \tbm a_n^{*}[T]-\tbm a_n[P] \right \rVert_2^2, \label {eq:TIRSORegretProof1}
	\end{align}
	where the inequality in (a) results from ignoring the term $1/(2\alpha_T)  \lVert \tbm a_n^{*}[T]-\tbm a_n[T+1]  \rVert_2^2$ and combining similar terms. To relate the l.h.s. of  \eqref {eq:proof1TIRSO} and the static regret in this case, consider the definition of the static regret for TIRSO in \eqref {eq:defstaticregretTIRSO}
	\begin{align}
		\tilde R_s^{(n)}[T] &= \sum_{t=P}^{T}\left [ \ltn(\tbm a_n[t])- \ltn(\tbm a_n^*[T]) -\nReg (\tbm a_n^*[T])\right ] \nonumber \\
		&+ \sum_{t=P}^{T} \nReg (\tbm a_n[t]) \nonumber \\
		& = \sum_{t=P}^{T}\left [ \ltn(\tbm a_n[t])- \ltn(\tbm a_n^*[T]) -\nReg (\tbm a_n^*[T])\right ] \nonumber \\
		&+ \sum_{t=P-1}^{T-1} \nReg (\tbm a_n[t+1]). \label {eq:TIRSOregretdef1}
	\end{align}
	Adding and subtracting the term $\nReg (\tbm a_n[T+1])$ to the r.h.s. of \eqref {eq:TIRSOregretdef1} and rearranging of terms results in
	\begin{align}
		\tilde R_s^{(n)}[T] &= \sum_{t=P}^{T}\left [ \ltn(\tbm a_n[t])- \ltn(\tbm a_n^*[T]) -\nReg (\tbm a_n^*[T])\right ] \nonumber \\
		& + \!\sum_{t=P}^{T} \nReg (\tbm a_n[t+1])\!+ \!\nReg (\tbm a_n[P])\! - \! \nReg (\tbm a_n[T\!+\!1]) \nonumber  \\
		& \leq \sum_{t=P}^{T} \big [\ltn (\tbm a_n[t])+\nReg (\tbm a_n[t+1])-\ltn (\tbm a_n^{*}[T]) \nonumber \\
		& -\nReg (\tbm a_n^{*}[T]) \big ] \label{eq:TIRSOregret3},
	\end{align}
	where $\nReg (\tbm a_n[P])=0$ and $\nReg (\tbm a_n[T+1]) \geq 0$ are used in the above inequality. Observe that the r.h.s. of the above inequality coincides with the l.h.s. of \eqref {eq:proof1TIRSO}. Therefore, from \eqref {eq:TIRSORegretProof1} and \eqref {eq:TIRSOregret3}, we have
	\begin{align*}
		\tilde R_s^{(n)} [T] &\leq \frac{1}{2}  \sum_{t=P-1}^{T-1}  \left \lVert \tbm a_n^{*}[T]-\tbm a_n[t+1]\right \rVert_2^2 \left (\frac{1}{\alpha_{t+1}}- \frac{1}{\alpha_{t}}-\strongcvxparf \right) \\
		& +\frac{1}{2} \sum_{t=P}^{T} \alpha_t \left \lVert \gradf \right \rVert_2^2 +\frac{1}{2\alpha_{P-1}} \left \lVert \tbm a_n^{*}[T]-\tbm a_n[P] \right \rVert_2^2.
	\end{align*}
	Setting $\alpha_t= 1/(\strongcvxparf t)$ in the above inequality yields
	\begin{align}
		\tilde R_s^{(n)} [T] \leq &  \frac{1}{2}  \sum_{t=P}^{T-1}  \left \lVert \tbm a_n^{*}[T]-\tbm a_n[t+1]\right \rVert_2^2 \left(\strongcvxparf (t+1)-\strongcvxparf t- \strongcvxparf\right) \nonumber \\
		& +\frac{1}{2} \sum_{t=P}^{T}\frac{1}{\strongcvxparf t} \left \lVert \gradf \right \rVert_2^2 +\frac{1}{2\alpha_{P-1}} \left \lVert \tbm a_n^{*}[T]-\tbm a_n[P] \right \rVert_2^2 \nonumber \\
		& =\frac{1}{2\strongcvxparf} \sum_{t=P}^{T}\frac{1}{ t} \left \lVert \gradf \right \rVert_2^2 +\frac{1}{2\alpha_{P-1}} \left \lVert \tbm a_n^{*}[T]-\tbm a_n[P] \right \rVert_2^2 \nonumber \\
		\stackrel{(a)}{\leq}&  \frac{\gradboundf^2}{2 \strongcvxparf} \sum_{t=P}^{T}\frac{1}{ t}+\frac{1}{2\alpha_{P-1}} \left \lVert \tbm a_n^{*}[T]-\tbm a_n[P] \right \rVert_2^2  \nonumber \\
		\stackrel{(b)}{\leq}&  \frac{\gradboundf^2}{2 \strongcvxparf} \left (\mathrm{log}(T-P+1\right ) +1) +\frac{1}{2\alpha_{P-1}} \left \lVert \tbm a_n^{*}[T] \right \rVert_2^2 \nonumber\\
		\stackrel{(c)}{\leq}&  \frac{\gradboundf^2}{2 \strongcvxparf} \left (\mathrm{log}(T-P+1\right ) +1) +\frac{1}{2\alpha_{P-1}} B_{\tbm a}^2 \nonumber,
	\end{align}
	where in (a) the bound on the gradient given in \eqref {eq:boundonTIRSOgrad} is used; in (b) the inequality $\sum_{t=1}^{T}1/t\leq \mathrm{log}(T) +1$ and the fact $\tbm a_n[P]=\bm 0_{NP}$ is used, and (c) is obtained by using the bound from \eqref {eq:boundonatirso}.
}

\section{Proof of \thref {th:dynamicregretbound}}
\label {appendix:proof{th:dynamicregretbound}}
\rev {
	\begin{myitemize}%
		\myitem \cmt{Dynamic regret result}We derive the dynamic regret of TIRSO. To this end, since $\tilde h_t$ is convex, we have by definition
		\begin{equation} \label{eq:firstordercvxcond}
			\tilde h_t^{(n)}(\timevarhindsight[t]) \geq \tilde h_t^{(n)}(\tbm a_n[t]) + \left(\tilde \nabla \tilde h_t^{(n)}(\tbm a_n[t])\right )^\top\left(  \timevarhindsight[t] - \tbm a_n[t]\right )
		\end{equation}
		$\forall \, \timevarhindsight[t], \tbm a_n[t]$, where $\tilde \nabla \tilde h_t^{(n)}(\tbm a_n[t])= \nabla \ltn(\tbm a_n[t]) + \bm u_t$ with $ \bm u_t \in \partial  \Omega^{(n)}(\tbm a_n[t])$. Rearranging \eqref {eq:firstordercvxcond} and summing both sides of the inequality from $t=P$ to $T$ results in:
		\begin{multline*}
			\sum_{t=P}^{T}\left [ \tilde h_t^{(n)}(\tbm a_n[t])- \tilde h_t^{(n)}(\timevarhindsight[t])\right  ] \leq \sum_{t=P}^{T} \left(\tilde \nabla \tilde h_t^{(n)}(\tbm a_n[t])\right )^\top \\
			\cdot \left( \tbm a_n[t] -\timevarhindsight[t]\right ).
		\end{multline*}
		By applying the Cauchy–Schwarz inequality on each term of the summation in the r.h.s. of the above inequality, we obtain
		\begin{multline} \label{eq:dynamicregretinequality}
			\sum_{t=P}^{T}\left [ \tilde h_t^{(n)}(\tbm a_n[t])- \tilde h_t^{(n)}(\timevarhindsight[t])\right  ]\leq \sum_{t=P}^{T} \left \lVert \tilde \nabla \tilde h_t^{(n)}( \tbm a_n[t]) \right \rVert_2 \\
			\cdot \left\lVert \tbm a_n[t] -\timevarhindsight[t]\right \rVert_2.
		\end{multline}
		The next step is to derive an upper bound on $\lVert \tilde \nabla \tilde h_t^{(n)}( \tbm a_n[t]) \rVert_2$. From the definition of $ \tilde \nabla \tilde h_t^{(n)}( \tbm a_n[t]) $ and by the triangular inequality, we have
		\begin{equation} \label{eq:boundonsubgrad}
			\lVert \tilde \nabla \tilde h_t^{(n)}( \tbm a_n[t]) \rVert_2 \leq \lVert  \nabla  \ltn( \tbm a_n[t]) \rVert_2 + \left \lVert \bm u_t\right \rVert_2.
		\end{equation}
		To bound $\lVert  \nabla  \ltn( \tbm a_n[t]) \rVert_2$, we invoke \thref {lemma:boundingestiamtes} and set $\alpha_t=\alpha$ to obtain
		\begin{subequations} 
			\begin{align} 
				\left \lVert \tbm a_n[t+1] \right  \rVert_2 &  \leq \left ( 1 - \alpha\strongcvxparf \right )\left \lVert \tbm a_n[t] \right \rVert_2 + \alpha \sqrt{PN} \energybound\\
				& = \delta \left \lVert \tbm a_n[t] \right \rVert_2 + \alpha \sqrt{PN} \energybound, \label{eq:boundedestimatesconsttalpha}
			\end{align}
		\end{subequations} 
		where $\delta \triangleq 1-\alpha \strongcvxparf$. Observe that for $0<\alpha\leq 1/L$, we have $0<\delta<1$. Substituting \eqref {eq:boundedestimatesconsttalpha} recursively, we obtain
		\begin{subequations}
			\begin{align*} \label{eq:boundedestimatesconsttalpha2}
				\left \lVert \tbm a_n[t+1] \right  \rVert_2 &  \leq \delta \left ( \delta \left \lVert \tbm a_n[t-1] \right \rVert_2 + \alpha \sqrt{PN} \energybound  \right)  + \alpha \sqrt{PN} \energybound \\
				&=\delta^2   \left \lVert \tbm a_n[t-1] \right \rVert_2 + \delta \alpha \sqrt{PN} \energybound    + \alpha \sqrt{PN} \energybound \\
				& \leq \delta^3   \left \lVert \tbm a_n[t-2] \right \rVert_2 + \delta^2 \alpha \sqrt{PN} \energybound  + \delta \alpha \sqrt{PN} \energybound  \\
				& \quad + \alpha \sqrt{PN} \energybound \leq \ldots  \\
				& \leq \delta^k   \left \lVert \tbm a_n[t-k+1] \right \rVert_2 +  \alpha \sqrt{PN} \energybound \sum_{i=0}^{k-1} \delta ^i ,
			\end{align*}
		\end{subequations}
		where $ 1\leq k \leq t-P+1$. For $k=t-P+1$, the above inequality becomes
		\begin{subequations}
			\begin{align*}
				\left \lVert \tbm a_n[t+1] \right  \rVert_2 & \leq \delta^{t-P+1}   \left \lVert \tbm a_n[P] \right \rVert_2 +  \alpha \sqrt{PN} \energybound \sum_{i=0}^{t-P} \delta^i \\
				& = \frac{\alpha \sqrt{PN} \energybound\left( 1-\delta ^{t-P+1}\right)}{1-\delta} 	\\
				& \leq \frac{\alpha \sqrt{PN} \energybound}{1-(1- \alpha \strongcvxparf)}= \frac{1}{\strongcvxparf} \sqrt{PN} \energybound,
			\end{align*}
		\end{subequations}
		which implies that $ \lVert \nabla \ltn (\tbm a_n[t])  \rVert_2 \leq (1+L/\strongcvxparf) \sqrt{PN} \energybound$, as in the proof of  \thref {lemma:boundg} by following the same arguments as in \eqref {eq:generalgradbound}.
		Next, we need to find an upper bound on $\lVert \bm u_t \rVert_2$ in \eqref {eq:boundonsubgrad}. To this end, we apply the result in \cite[Lemma 2.6]{shalev2011online} to $\nReg$, which establishes that all the subgradients of $\nReg$ are bounded by its Lipschitz continuity parameter $L_{\Omega^{(n)}}$. In the following, we show that $L_{\Omega^{(n)}}=\lambda \sqrt{N}$. 			
		\myitem \cmt{Proving regularization function is Lipschitz continuous}Lipschitz smoothness of $\nReg$ means that there exists $L_{\nReg}$ such that
		\begin{equation} \label {eq:lipschitzcontinuousassump}
			\left \lvert \nReg(\bm a)- \nReg(\bm b) \right\rvert \leq L_{\nReg} \left\lVert \bm a -\bm b\right\rVert_2,
		\end{equation}
		for all $\bm a, \bm b$. 
		By definition, we have $\nReg(\bm x_n)
		=  \lambda \sum_{\substack { n'=1,
				n' \neq n}}^{N}\left \lVert \bm x_{n,n'} \right\rVert_2$ with $\bm x_n=[\bm x_{n,1}^\top,...,\bm x_{n,N}^\top]^\top, \bm x_{n,n'}\in \mathbb{R}^{P},n'=1,...,N$. Let $\bm z_n=[\bm z_{n,1}^\top,...,\bm z_{n,N}^\top]^\top, \bm z_{n,n'}\in \mathbb{R}^{P},n'=1,...,N$ and by taking the l.h.s. of \eqref {eq:lipschitzcontinuousassump}, we have
		\begin{subequations}
			\begin{align}
				\left \lvert \nReg(\bm x_n) - \nReg(\bm z_n) \right \rvert
				& = \lambda \left \lvert \sum_{\substack { n'=1\\
						n' \neq n}}^{N} \left\lVert \bm x_{n,n'} \right \rVert_2 - \sum_{\substack { n'=1\\
						n' \neq n}}^{N} \left \lVert \bm z_{n,n'} \right \rVert_2  \right\rvert \nonumber  \\
				& = \lambda \left \lvert \sum_{\substack { n'=1\\	n' \neq n}}^{N} \left [\left\lVert \bm x_{n,n'} \right \rVert_2 -  \left \lVert \bm z_{n,n'} \right \rVert_2 \right ] \right\rvert \nonumber  \\
				&\leq \lambda  \sum_{\substack { n'=1\\
						n' \neq n}}^{N} \left \lvert \lVert \bm x_{n,n'} \rVert_2 -  \lVert \bm z_{n,n'} \rVert_2 \right \rvert \label {eq:ineq1}\\
				& \leq \lambda  \sum_{\substack { n'=1\\
						n' \neq n}}^{N} \left\lVert \bm x_{n,n'} - \bm z_{n,n'} \right \rVert_2 \label{eq:ineq2}\\
				& \leq \lambda  \sum_{n'=1}^{N} \left\lVert \bm x_{n,n'} - \bm z_{n,n'} \right \rVert_2 \nonumber 
				\\
				& \leq \lambda \sqrt{N} \left \lVert \bm x_n- \bm z_n \right \rVert_2, \label{eq:Lipschitzinequaltiy}
			\end{align}
		\end{subequations}
		where the inequality in \eqref{eq:ineq1} holds due to the triangle inequality for scalars ($\lVert \bm x_{n,n'} \rVert_2 -  \lVert \bm y_{n,n'} \rVert_2$ as scalars); \eqref{eq:ineq2} holds due to the reverse triangle inequality (given by $\lvert \lVert \bm x_1  \rVert_2 -  \lVert \bm x_2 \rVert_2  \rvert  \leq   \lVert \bm x_1 - \bm x_2 \rVert_2$); 
		and \eqref {eq:Lipschitzinequaltiy} follows from the inequality $\lVert \bm b  \rVert_1 \leq \sqrt{N} \lVert \bm b \rVert_2$ with $\bm b \in \mathbb{R}^N$ \cite[Sec. 2.2.2]{golub1996}. The inequality in \eqref {eq:Lipschitzinequaltiy} implies that \eqref{eq:lipschitzcontinuousassump} is satisfied with $L_{\nReg}= \lambda \sqrt{N}$, i.e., $\nReg$ is $\lambda \sqrt{N}$-Lipschitz continuous.
		Thus, we have $\lVert \tilde \nabla \tilde h_t^{(n)}( \tbm a_n[t]) \rVert_2 \leq (1+L/\strongcvxparf) \sqrt{PN} \energybound+\lambda \sqrt{N}$. Substituting this bound in \eqref {eq:dynamicregretinequality} leads to:
		\begin{align} \label{eq:boundTIRSO}
			\!\!\sum_{t=P}^{T}\!&\left[ \tilde h_t^{(n)}(\tbm a_n[t])- \tilde h_t^{(n)}(\timevarhindsight[t])\right  ]\! \leq\! \sum_{t=P}^{T} \!\Bigg [ \left(1+\frac{L}{\strongcvxparf}\right ) \sqrt{PN} \energybound \nonumber \\&+ \lambda \sqrt{N}\Bigg ]\left \lVert \tbm a_n[t] -\timevarhindsight[t]\right \rVert_2.
		\end{align}
		\myitem \cmt{TIRSO as OPGD}Next, we show that TIRSO for a constant step size can alternatively be derived by applying online proximal gradient descent to minimize $\ltn + \nReg$. With $\ltn$  given by \eqref{eq:recloss} and $\Omega^{(n)}$ is given by \eqref{eq:comidlossdefreg}, applying the online proximal gradient algorithm with a constant step size $\alpha$ yields:
		\begin{equation}
			\tbm a_n[t+1] = \textbf{prox}_{\Omega^{(n)}}^{\alpha} \left ( \tbm a_n[t] - \alpha \nabla \ltn(\tbm a_n[t])   \right ),
		\end{equation}
		where
		\cmt{Proximal operator}the proximal operator of a  function $\Psi$ at point $\bm v$ is defined by \cite {parikh2014proximal}:
		\begin{equation} \label {eq:defproximaloperator}
			\textbf{prox}_{ \Psi}^{\eta}(\bm v)\triangleq \underset{\bm x \in \text{dom }\Psi}{\arg\min}\left [\Psi(\bm x)+\frac{1}{2\eta}\left \lVert \bm x-\bm v\right \rVert_2^2\right].
		\end{equation}
		\cmt{regularization parameter}The parameter $\eta$ controls the trade-off between minimizing $\Psi(\cdot)$ and being close to $\bm v$.
		According to the definition in Sec.~\ref{ss:tirso}, $\tbm a_n^{\text{f}} [t] \triangleq  \tbm a_n[t] - \alpha \nabla \ltn(\tbm a_n[t])$,  and $\tbm a^\text{f}_{n}[t]= [(\tbm a^\text{f}_{n,1}[t])^\top, \ldots, (\tbm a^\text{f}_{n,N}[t])^\top ]^\top$, which enables us to write the above update expression as
		\begin{align*}
			\tbm a_n[t+1]& =  \textbf{prox}_{ \Omega^{(n)}}^{\alpha} \left ( \tbm a_n^{\text{f}} [t]   \right )\\
			& = \underset{\bm z_n }{\arg\min} \left ( \Omega ^{(n)} (\bm z_n) + \frac{1}{2\alpha} \left \lVert \bm z_n - \tbm a_n^{\text{f}} [t] \right \rVert_2^2 \right ) \\
			& = \underset{\{\bm z_{n,n'}\}_{n'=1}^N  }{\arg\min} \Bigg ( \lambda \sum_{n'=1}^{N} \mathds 1 \{ n \neq n'\}  \left\lVert \bm z_{n,n'}  \right\rVert_2\\
			& + \frac{1}{2\alpha} \sum_{n'=1}^{N} \left \lVert \bm z_{n,n'} - \tbm a_{n,n'}^{\text{f}} [t] \right \rVert_2^2 \Bigg ).
		\end{align*}
		Observe that the above problem is separable and the solution to the $n'$-th problem is given by:
		\begin{align}
			\tbm a_{n,n'}[t+1] &= \underset{\bm z_{n,n'} }{\arg\min} \Bigg [ \mathds 1 \{ n \neq n'\}  \left\lVert \bm z_{n,n'}  \right \rVert_2 \nonumber \\ &\quad \quad \quad \quad \quad + \frac{1}{2\alpha \lambda} \left \lVert \bm z_{n,n'} - \tbm a_{n,n'}^{\text{f}} [t] \right \rVert_2^2 \Bigg] \nonumber \\
			&= \tbm a^\text{f}_{n,n'}[t ]\left [1-\frac{\alpha \lambda  \mathds 1 \{ n \neq n'\}}{\left \lVert \tbm a^\text{f}_{n,n'}[t ] \right\rVert_2}\right]_+, \label {eq:updateproxonline}
		\end{align}
		which is the same as \eqref {eq:comidgroupsoltirso} with a constant step size $\alpha$. Therefore, TIRSO can be equivalently derived by applying online proximal gradient descent method.	
		Next, we apply Lemma 2 in \cite{dixit2019onlineproximal} in order to bound $\sum_{t=P}^T\lVert \tbm a_n[t] -\timevarhindsight[t] \rVert_2$ in \eqref {eq:boundTIRSO}. The hypotheses of Lemma 2 are Lipschitz smoothness of $\ltn$, Lipschitz continuity of $\nReg$, and strong convexity of $\ltn $. Lipschitz continuity of $\nReg$ is proved in \eqref {eq:Lipschitzinequaltiy} whereas strong convexity of $\ltn $ is implied by the assumption A\ref{as:mineig}.
		\myitem \cmt{proving Lipschitz smoothness of loss function $f$}So we need to verify that $\rev{ \ltn }$ is Lipschitz-smooth, which means that there is $L'$ such that
		\begin{equation}\label{eq:lipscsmoothnessassump}
			\left \lVert\nabla \rev{ \ltn }(\bm a)- \nabla \rev{ \ltn }(\bm b) \right\rVert_2 \leq L' \left\lVert \bm a -\bm b\right\rVert_2,
		\end{equation}
		for all $\bm a, \bm b$. To this end, taking the l.h.s. of \eqref{eq:lipscsmoothnessassump} and substituting the value of the gradient of $\rev{ \ltn }$ from \eqref {eq:tirsograd} results in:
		\begin{align*}
			\left \lVert\bm \Phi[t] \bm a-\bm r_n[t]- \bm \Phi[t] \bm b+\bm r_n[t] \right \rVert_2 &= \left \lVert \bm \Phi [t](\bm a- \bm b)\right \rVert_2 \nonumber \\
			& \leq \lambda_{\mathrm{max}}(\bm \Phi [t])\left\lVert\bm a- \bm b \right\rVert_2, \label{ineq:lipschitzsmoothness}
		\end{align*}
		where $\lambda_{\mathrm{max}}(\cdot)$ denotes the maximum eigenvalue of the input matrix. Due to assumption A\ref{as:maxeig}, the inequality in \eqref {eq:lipscsmoothnessassump} holds with $L'=L$.
		\myitem \cmt{Applying Lemma 2}To apply Lemma 2 in \cite {dixit2019onlineproximal}, one can set $K$ in \cite {dixit2019onlineproximal} as $T-P+1$, $g_k$ as $\Omega^{(n)}$, and $f_k$ as $\tilde\ell_{P+k-1}^{(n)}$, it follows that $\bm x_k$ in \cite {dixit2019onlineproximal} equals $\tbm a_n[P+k-1]$ and $\bm x_k^\circ$ equals $\tbm a_n^\circ[P+k-1]$. Then, since we have already shown above that the hypotheses of Lemma 2 in \cite {dixit2019onlineproximal} hold in our case, applying it to bound $\lVert \tbm a_n[t] -\timevarhindsight[t]\rVert_2$ in \eqref {eq:boundTIRSO} yields:
		\begin{align}
			&\sum_{t=P}^{T}\left [ \tilde h_t^{(n)}(\tbm a_n[t])- \tilde h_t^{(n)}(\timevarhindsight[t])\right  ] \leq \frac{1}{\alpha \strongcvxparf}\Bigg[ \left  (1+\frac{L}{\strongcvxparf}\right ) \sqrt{PN} \energybound\nonumber \\
			&+ \lambda \sqrt{N}\Bigg]  \left(\lVert \tbm a_n[P]- \timevarhindsight[P] \rVert_2 + W^{(n)}[T]\right ).
		\end{align}
	\end{myitemize}%
	Noting that $\tbm a_n[P]=\bm 0_{NP}$ concludes the proof.
}

\end{document}